\documentclass{JHEP}

\usepackage{epsfig}
\usepackage{amsmath}
\usepackage{amssymb,amsfonts}
\usepackage{graphicx}

\makeatletter
\renewcommand{\@makecaption}[2]{%
{ #2}%
}

\makeatother

\newcommand{\ie}{{\em i.e. }}

\newcommand{\R}{\mathbb{R}}
\newcommand{\Co}{\mathbb{C}}
\newcommand{\Cs}{\mathbb{C}^*}

\newcommand{\CP}{\mathbb{C}\mathbb{P}}
\newcommand{\RP}{\mathbb{R}\mathbb{P}}
\newcommand{\Z}{\mathbb{Z}}

\newcommand{\C}{{\cal C}}
\newcommand{\G}{{\cal G}}
\newcommand{\M}{{\cal M}}
\newcommand{\N}{{\cal N}}

\newcommand{\W}{{\cal W}}

\newcommand{\Ka}{K{\"a}hler }

\newcommand{\I}{{\cal I}}

\newcommand{\pamatrix}[1]{\begin{pmatrix} #1 \end{pmatrix}}

\newcommand\fverb{\setbox\pippobox=\hbox\bgroup\verb}
\newcommand\fverbdo{\egroup\medskip\noindent%
                        \fbox{\unhbox\pippobox}\ }
\newcommand\fverbit{\egroup\item[\fbox{\unhbox\pippobox}]}
\newbox\pippobox

\title{Geometric Transitions, Brane Dynamics and Gauge Theories}

\author{A. Giveon$^1$, A. Kehagias$^2$ and H. Partouche$^3$ \\

$^1$ Racah Institute of Physics, The Hebrew Univ., Jerusalem 91904, ISRAEL \\
E-mail: \email{giveon@vms.huji.ac.il}\\
\vskip .1 in
$^2$ Physics Department, NTUA, 15773 Zografou, Athens, GREECE\\
E-mail: \email{kehagias@mail.cern.ch}\\
\vskip .1 in
$^3$ Centre de Physique Th{\'e}orique, Ecole Polytechnique\\
F-91128 Palaiseau cedex, FRANCE\\
E-mail: \email{Herve.Partouche@cpht.polytechnique.fr}}
\preprint{\hepth{0110115}\\ CPHT-S040.1001\\ RI-10-01}

\abstract{We consider the interplay between brane constructions
and type IIA, IIB or M-theory geometries on Calabi-Yau (CY) and $G_2$
holonomy manifolds. This is related to $\N=1$ (and $\N=2$)
gauge theories in four dimensions. We first discuss
simple geometric transitions corresponding to brane set ups
involving orthogonal (or parallel) Neveu-Schwarz branes
that approach each other.
This is related to confinement and Seiberg duality in SQCD.
In particular, we argue that in type
IIA, a $\CP^1$ of singularities and one unit of Ramond-Ramond (RR)
flux is dual to a
D6 brane wrapped on a Lens space, describing the UV and IR
of $\N=1$ Super-Yang-Mills (SYM), respectively. Also, in the large $N_c$
duality that relates D6 branes on $S^3$ to an $S^2$ with
RR flux, we implement the presence of $N_f$ flavors of
quarks.  We then compactify
M-theory on $(T^*(S^3)\times S^1)/\Z_2$ and observe that one
phase describes $SO(4)$ SYM in the UV and two others describe
confinement. Moreover, we consider compact 7-spaces (CY$\times
S^1)/\Z_2$. We describe transitions where disconnected $S^3$'s
approach and connect each other before they vanish.
These effects correspond to non-Abelian Higgs mechanism and
confinement. The similar transitions involving $S^2\times S^1$'s are
also considered. Finally, we present transitions at finite
distance in moduli space, where the first Betti number $b_1$ of
3-cycles of singularities changes.
}

\begin{document}

\maketitle 

\setlength{\baselineskip}{1.2\baselineskip}

\section{Introduction}

The interplay between gauge theories, brane dynamics
and geometry is a useful tool to learn about
non-trivial phenomena in one of these frameworks since
sometimes it is manifest in one of the other descriptions
(see \cite{gk}  for a review on brane dynamics and gauge theory,
and \cite{geometry} for reviews on geometrical engineering).
For instance, various properties of brane dynamics can be deduced
by known properties of the gauge theory living on the branes at low
energies and vice versa. On the other hand,
non-trivial dualities in gauge theory --
like the Seiberg's duality \cite{Seiberg:1995pq}
in Supersymmetric Yang-Mills (SYM) --
are manifested as deformations in the space of brane
configurations \cite{egk},
providing a ``unification'' of the dualities for different gauge groups
with various matter content and in various dimensions
in a single framework.

There are however subtleties in the study of gauge dynamics using branes.
The ``rules'' governing the behavior of branes, especially in transitions
involving coincident branes, are not always clear.
In particular, the transitions due to coincident Neveu-Schwarz (NS)
fivebranes involve the non-trivial theory on their worldvolume.
Hence, generically, one is not guaranteed to have a smooth transition
when parallel NS branes approach each other.
On the other hand, the crossing of orthogonal NS branes is leading
to the $\N=1$ electric-magnetic duality in four dimensions, and thus is
expected to be smooth.
An extensive discussion of transitions due to
intersecting NS branes appears in \cite{gk}, section IX B2.

Confinement can also be described by two
orthogonal flat NS branes intersecting in $1+3$ dimensions,
in the presence of RR flux.
Quantum gauge theory effects will turn
out to bend the branes classically.

In many cases, known properties in certain
gauge theories allow to set  rules of brane dynamics, which can
then be used in more complicated systems. Sometimes, such rules can
be confirmed by standard perturbative worldsheet considerations \cite{egkrs}.
Alternatively, known properties in geometry can also be used
\cite{Ooguri:1997ih}
to shed light on the kinematics and dynamics of brane constructions
as well as gauge dynamics. In this work we focus on the latter.

The frameworks employed here are M-theory on $G_2$ holonomy manifolds
and orientifolds of type IIA on Calabi-Yau (CY) threefolds.  The low
energy physics is four dimensional and is  $\N=1$ supersymmetric.
The geometry on CY threefolds is accompanied by D-branes wrapped on
non-trivial cycles and/or RR fluxes.
On the contrary, the description in the framework of $G_2$ 7-manifolds
is purely geometric.

Since there is no non-Abelian structure in eleven
dimensions to start with, compactifications on smooth 7-manifolds give
rise to Abelian vector multiplets. However, it is known
that the existence of singularities in the internal space leads to
non-Abelian  gauge theories. In the M-theory framework,
it has been shown in
\cite{Sen:1997kz} that,
in certain cases, as 2-cycles collapse in the internal space,
 an $A_n$ singularity is developed giving rise to
 an enhancement of the gauge group.  This is the mechanism employed in
\cite{Ferrara:1998vf} for the study of $\N=1$ six dimensional  gauge
theories in the
AdS/CFT correspondence. From the type IIA point of view, the enhancement of
the gauge group is due to coincident D6 branes, the description of which
is purely geometric from the M-theory point of view \cite{Townsend:1995kk}.

In this work, we argue that smooth transitions corresponding
to intersecting NS branes translate in geometry into topology changes
at finite distance in moduli space,
and vice versa. To investigate this interplay,
we begin in Section 2.1 with a review of relatively simple brane
configurations describing $\N=1$ SQCD and $\N=2$ SYM, and
geometries obtained from them by performing
various chains of T-dualities. In particular, the two possible small
resolutions of the conifold by a 2-sphere $S^2$ or $\tilde{S}^2$ give
rise to a flop transition $S^2\to 0\to \tilde{S}^2$. In type IIB, when
D5-branes are wrapped on them, this transition is
associated to electric-magnetic duality. In Section 2.2, we
review various geometric constructions involving special Lagrangian
(SLAG) 3-cycles on which D6 branes are wrapped in type IIA. In
particular, it has been
argued that a stack of $N_c$ D6 branes wrapped on $S^3$ is dual to an $S^2$
with $N_c$ units of Ramond-Ramond (RR) flux through it
\cite{Vafa:2000wi}. We shall see
that in this geometric set up, which describes confinement of pure $SU(N_c)$
SYM, one can include $N_f$ massless flavors of quarks. Section 2.3 is
devoted to the lift of such type IIA descriptions to M-theory.  First,
the case of orbifolds of the non-compact $G_2$ holonomy manifold
$Spin(S^3)$ is recalled \cite{Acharya:2000gb, Atiyah:2000zz}. We then
treat the case of $(T^*(S^3)\times S^1)/\Z_2$, where $\Z_2$ acts
antiholomorphically on  $T^*(S^3)$ and as an inversion on  $S^1$. We
shall see that there are three distinct phases in this model. The first
one is associated to $SO(4)$ SYM in the UV, while the other two
describe confinement. The situation is similar to the three phases
occuring in the models based on $Spin(S^3)$ \cite{Atiyah:2001qf}.
Also, we propose a new duality conjecture. It will be argued that
in type IIA on a CY threefold, a $\CP^1$ of $A_{N_c-1}$
singularity and one unit of RR charge through it is dual to a Lens space
$S^3/\Z_{N_c}$ with one D6 brane wrapped on it. These two phases
describe the UV and IR physics of the pure $\N=1$ $SU(N_c)$ gauge
theory, respectively. \footnote{And similarily for $D$ and $E$ Lie groups.}
Finally, an M-theory background $Spin(S^3)/(\Z_{N_f-N_c}\times \Z_{N_f})$ is
considered as a  candidate for describing the UV and IR physics of
the magnetic $SU(N_f-N_c)$ SYM with $N_f$ flavors of quarks.
Work related to issues discussed in section 2 appear also
in \cite{Dasgupta:2001um, Dasgupta:2001fg, Dasgupta:2001ac}.

We then study transitions in compact manifolds or singular spaces of
$G_2$ holonomy. Even
if we choose a particular construction of the latter, the local
geometry of the transitions we shall describe explicitly can then
occur in other $G_2$ manifolds. Precisely, we shall consider
orbifolds of the form (CY$\times
S^1)/\Z_2$. The $\Z_2$ acts again as an
inversion on $S^1$  and antiholomorphically on the CY,
 so that $J\to -J$ and $\Omega\to \bar \Omega$, where
$J$ and $\Omega$ are the \Ka form and holomorphic 3-form \cite{Joyce2}.
In general, the fixed point
set $\Sigma$ of the antiholomorphic involution on the CY is
then composed of disconnected special Lagrangian
3-cycles. Each componant in $\Sigma$
is promoted to an associative 3-cycle of $A_1$ singularities in
the 7-space. The first Betti number
$b_1$ of such an associative 3-cycle then counts the number
of chiral multiplets in the adjoint representation of an
$SU(2)$ gauge group. The model presented in Section 3.1, describes a
theory without massless adjoint fields. In the underlying CY,
disconnected 3-spheres approach each other till we reach a transition
where they intersect at a point. Then, they are replaced by the connected
sum $S^3 \# S^3$, which is topologically equivalent to a single
$S^3$. Transitions of this type might be related to the work of Joyce
\cite{Joyce:1999tz}. \footnote{In \cite{Kachru:2000vj}, disjoint homology
  3-spheres in different classes are connected to each other after a
  transition occurs. In our case, the 3-spheres are in the same
  class.}  Physically, in M-theory and a type IIA
orientifold limit, this amounts to the Higgsing $SO(4)\times SO(4)\to
SO(4)$ by a massless $(4,4)$ chiral multiplet. As discussed in Section
3.2, this Higgs branch {\em does not} exist classically in field
theory and is due to a dynamically generated superpotential. This
non-perturbative effect is described in the
brane construction by the classical bending of intersecting orthogonal NS
branes. In other phases, confinement of $SO(4)^2$ or $SO(4)$ also
take place, as in \cite{Atiyah:2000zz, Kaste:2001iq}.
{}From the brane point of view, these transitions are smooth, while
from the geometric point of view, they are argued to occur at finite
distance in moduli space, like standard conifold transitions between CY's.

In Section 4.1, the effects of a CY conifold transition \cite{Candelas:1990ug}
on M-theory
compactified on (CY$\times S^1)/\Z_2$ are considered. This
amounts to
$S^3$'s of $A_1$ singularities in M-theory that undergo flop transitions to
$\RP^3$'s with no singularity. In a type IIA orientifold limit, this
is described by a transition $S^3\to \RP^2$, with no RR
flux. It is interpreted as confinement together with a change of
branch in the scalar potential of neutral chiral multiplets, as in
\cite{Kaste:2001iq, Partouche:2001uq}. These
mixed effects are in contrast with the pure confining phenomenon
occuring in the non-compact cases based on $Spin(S^3)$ and
$T^*(S^3)$. Then, it is suggested that an orientifold of type IIA on a
compact CY with a 3-sphere of singularities might be dual to a type
IIA orientifold on the mirror CY.  The former describes  SYM in the
UV, while the latter could describe the IR. In Section 4.2, we consider
a situation where a non-Abelian Higgs mechanism takes place together
with a change of branch in the scalar potential. On one side of the
transition, this is described by D6 branes wrapped on two disconnected
$S^3$'s, each intersecting a third homology 3-sphere
between them at a point. At the
transition, this third 3-cycle undergoes a conifold transition to an
$S^2$, in the spirit of \cite{Greene:1996dh}.

Finally, in Section 5, we consider two very different situations where
3-cycles with non-trivial $b_1$ are involved. The models are still of
the form (CY$\times S^1)/\Z_2$. In Section 5.1, we focus on the case
where each $S^3$ occuring in Section 3.1 is replaced by
$S^2\times S^1$. This amounts to having an adjoint field
of the gauge group and hence a Coulomb branch. As an example, one of
the transitions involves a disconnected union $(S^2\cup S^2)\times
S^1$ that is becoming $(S^2\# S^2)\times S^1\cong S^2\times S^1$. In
general, such transitions concern the non-holomorphic $S^2$'s that are
part of 3-cycles and are expected to occur at
infinite distance in moduli space. In the brane picture, they correspond to
parallel NS branes (in the presence of RR background) that approach each other.
On the contrary, in Section 5.2, we
consider transitions at finite distance in moduli space where the
first Betti number $b_1$ of 3-cycles changes. Explicitely, we describe a
situation where four disconnected 3-spheres centered at the corners of
a square approach each other till they intersect at four points, giving
rise to a non-contractible loop passing through these points. In
total, the transition takes the form $\bigcup_{i=1}^{4} S^3 \to
\mbox{\large $\#$}_{i=1}^4 S^3 \cong S^2\times S^1$. Then, the radius
of the $S^1$ can
also decrease and we pass into a third phase: $S^2\times S^1\to S^3$,
with again a jump in $b_1$. In these cases, both the brane picture and
the field theory interpretation deserve to be studied further.

\section{Simple brane constructions, geometry and SYM}

\subsection{Type IIA branes versus type IIB geometry}
\label{IIB}

We first discuss systems which describe four dimensional SQCD at low energy.
Consider a brane configuration in the type IIA string theory
constructed out of
NS fivebranes and Dirichlet fourbranes (D4) (later, we shall also study
examples that include orientifold fourplanes (O4)) whose worldvolume
is stretched in the directions:
\begin{equation}
\begin{array}{lcl}
\mbox{NS}  &:& (x^0,x^1,x^2,x^3,x^4,x^5)\\
\mbox{NS'} &:& (x^0,x^1,x^2,x^3,x^8,x^9)\\
\mbox{D4/O4}   &:& (x^0,x^1,x^2,x^3,x^6)\, .
\label{0123}
\end{array}
\end{equation}
We separate the NS and NS' branes a distance $L_c$ in the direction $x^6$
and stretch $N_c$ D4 branes between them (see Figure 1(a));
we shall call these branes the
``color branes.'' The low energy theory on the
D4 branes is an $\N=1$, $SU(N_c)$ SYM in the $1+3$ dimensions common to
all the branes in Eq. (\ref{0123}). The gauge
multiplet corresponds to the low lying excitations of open strings stretched
between the D4 branes.
\begin{figure}[!h]
\centerline{\hbox{\psfig{figure=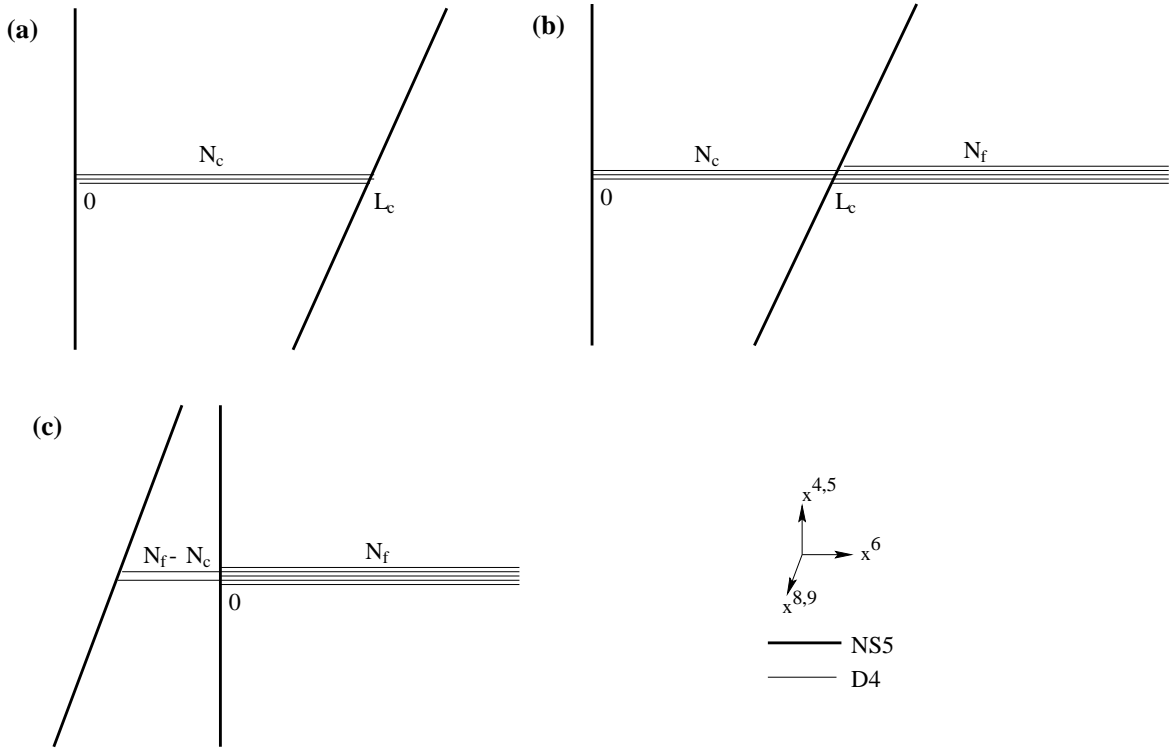,height=10cm}}}
\caption{{\bf Figure 1:}
{\footnotesize \em {\bf (a)} Realization of pure $SU(N_c)$ SYM
  with branes. {\bf (b)} Flavors of quarks are introduced by
  semi-infinite D4 branes. {\bf (c)} The magnetic dual description of {\bf
  (b)} is obtained by interchanging the positions of the Neveu-Schwarz
  branes.}}
\label{f1a}
\end{figure}
We will sometimes also add $N_f$ semi-infinite D4 branes stretched along the
direction $x^6$ on the other side of the NS' brane (see Figure 1(b));
we shall call these branes the ``flavor branes.''
The $SU(N_f)$ corresponding to open strings stretched
between the flavor branes is a global symmetry from the point of view
of the $SU(N_c)$ gauge theory on the color branes.
Open strings stretched between the color and flavor branes
correspond to $(N_c,N_f)$ hypermultiplets in $SU(N_c)\times SU(N_f)$,
and thus give rise to $N_f$ quark and anti-quark chiral multiplets
in the $SU(N_c)$ gauge theory.
While this can be guessed from the fact that in the vicinity
of the NS brane the system has an $\N=2$ supersymmetry, it cannot
be deduced in worldsheet perturbation theory due to the fact that
open strings confined to the vicinity of the NS brane are strongly
coupled~\footnote{For a separated stack of NS branes
this was verified by standard worldsheet techniques in \cite{egkrs}.}.

We may also consider a system where $x^6$ is compactified on a circle
of radius $R_6$. In this case we cannot have semi-infinite flavor branes,
but instead we can stretch $N_f$ D4 branes
of length $2\pi R_6-L_c$ along the left  and right side of NS and NS'
branes, respectively
(see Figure 2(a)). The low energy theory is now an $\N=1$,
$SU(N_c)\times SU(N_f)$ gauge theory with an $(N_c,N_f)$ hypermultiplet.
\begin{figure}[!h]
\centerline{\hbox{\psfig{figure=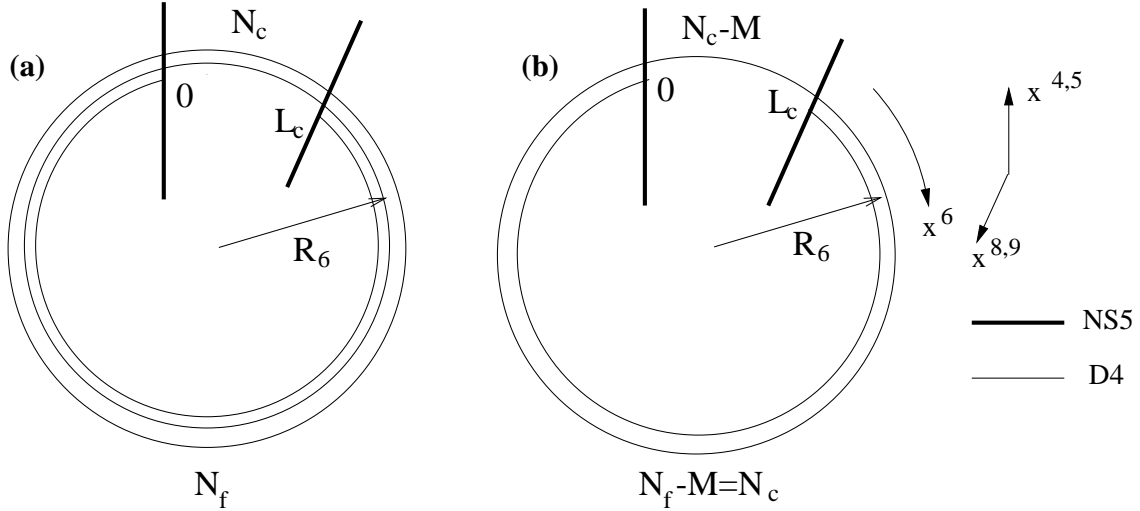,height=6.8cm}}}
\caption{{\bf Figure 2:} {\footnotesize \em {\bf (a)} Brane realization of
    $N=1$, $SU(N_c)\times SU(N_f)$ with an $(N_c,N_f)$
    hypermultiplet. {\bf (b)} The first step of a duality cascade.}}
\label{f2}
\end{figure}
A third system whose low lying theory is $\N=1$, $SU(N_c)\times SU(N_f)$
with a bi-fundamental hypermultiplet is described in Figure 3(a).
It consists of two NS branes and one NS' brane between them,
$N_c$ D4 branes of length
$L_c$ stretched between the first NS and the NS',
and $N_f$ D4 branes of length $L_f$
stretched between the NS' and the second NS brane.
In the limit where the second NS brane is sent to infinity we obtain
the system in Figure 1(b).
\begin{figure}[!h]
\centerline{\hbox{\psfig{figure=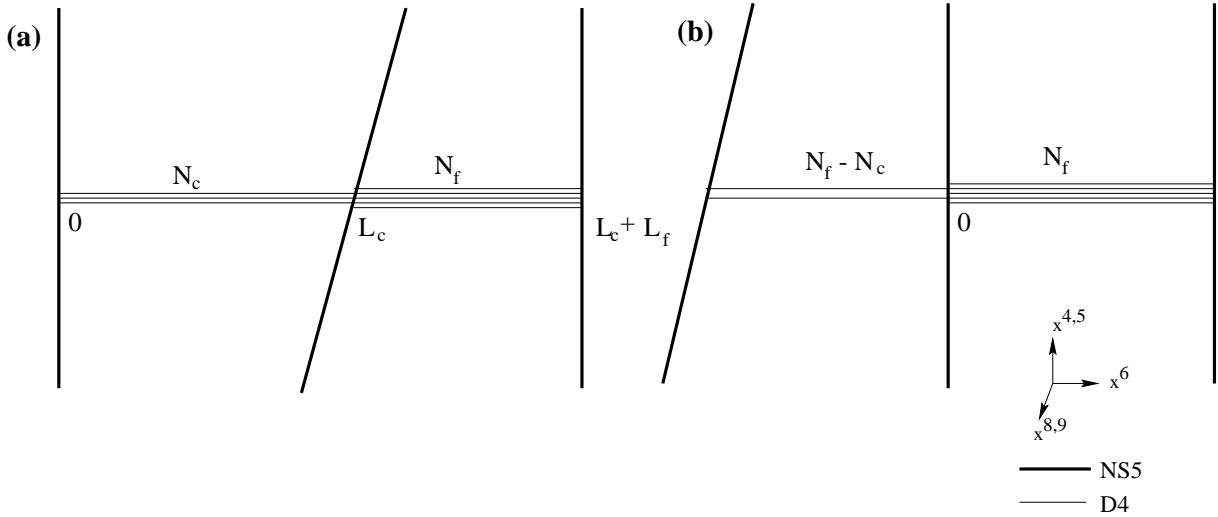,height=6.85cm}}}
\caption{{\bf Figure 3:}
{\footnotesize \em {\bf (a)} Brane realization of the ``electric''
  $SU(N_c)\times SU(N_f)$ SYM. {\bf (b)} Realization of the ``Magnetic''
dual $SU(N_f-N_c)\times SU(N_f)$ theory.}}
\label{f3}
\end{figure}

In all these systems we can add an O4 plane parallel to the D4 branes
(\ref{0123}). The sign of the RR charge of the orientifold is flipped
each time it is separated in two by a single Neveu-Schwarz
brane (see \cite{gk} for a review).
When the RR charge
of the orientifold is negative and there are $N_c$ D4 branes on top of
it~\footnote{Our convention is that $N_c$ is the {\it total} number
of D4 branes: ``Half'' D4 branes together with their mirror partners
under the orientifold reflection.}
the low energy theory has an $SO(N_c)$ gauge symmetry.
On the other hand, when the RR charge
of the orientifold is positive the gauge group is $Sp(N_c/2)$.

For the system of Figure 2(a), in the limit $R_6\to 0$, it is convenient to do
a T-duality $T_6$ in the direction $x^6$, bringing the system to a type IIB
description. In the limit where also
$L_c/R_6\to 0$, the NS and NS' system in Figure 2(a) can be regarded as
a Neveu-Schwarz fivebrane wrapped on the singular Riemann surface defined
by the  algebraic equation in $\Co^2$:
\begin{equation}
H(v,w)=vw=0~, \quad \mbox{where}\quad  v=x^4+ix^5~, \quad w=x^8+ix^9~.
\label{hvw}
\end{equation}
The $T_6$ duality turns the type IIA string theory
in the presence of such a fivebrane
into a type IIB theory on $\R^{1,3}\times \C$, where $\C$ is defined by
the algebraic equation in $\Co^4$:
\begin{equation}
F(v,w,z,z')=H(v,w)-zz'=vw-zz'=0~.
\label{fvw}
\end{equation}
This singular six dimensional space is the conifold
-- a cone with an $S^2\times S^3$ base. In the resolved conifold, there is
a single non-trivial cycle -- the blown up $S^2$. On the contrary,
the deformed conifold has only a non-trivial $S^3$
(this is discussed further in Section 2.2).

The relation between the two T-dual pictures is the following.
The distance $L_c/R_6$ in Figure 2(a) turns into a blow up parameter of
$S^2$.~\footnote{More precisely, there are two ``blow up'' modes of $S^2$:
A $\theta$ parameter due to a two index $B$-field on $S^2$,
and vol($S^2$).
The $\theta$ parameter is related to the separation $L_6=L_c$
of the fivebranes in the $x^6$ direction and hence to the YM coupling (see
below), while vol($S^2$) is related to a separation $L_7$ in $x^7$ and
hence to a FI D-term in the $\N=1$ $U(N_c)$ SYM (see \cite{gk}).}
On the other hand, the deformed conifold
\begin{equation}
F(v,w,z,z')=H(v,w)-zz'=vw-zz'=\mu
\label{fvwd}
\end{equation}
is T-dual to a Neveu-Schwarz fivebrane wrapped on the Riemann surface
\begin{equation}
H(v,w)=vw=\mu~.
\label{hvwd}
\end{equation}
The parameter $\mu$ in Eq. (\ref{fvwd}) is related to the radius
of $S^3$ in the deformed conifold.
In the fivebrane picture, when $\mu$ is turned on, the coincident
NS and NS' branes bend such that they do not pass through the origin
$v=w=0$ (see Figure 4). Thus, there is no gauge group.
\begin{figure}[!h]
\centerline{\hbox{\psfig{figure=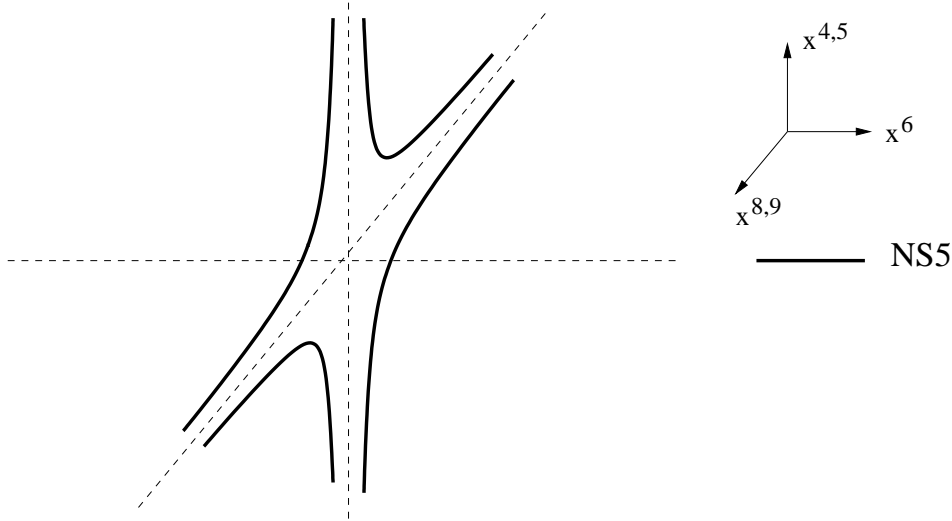,height=7cm}}}
\caption{{\bf Figure 4:}
{\footnotesize \em After two orthogonal Neveu-Schwarz branes
  intersect, they can bend. The gauge group has disappeared; it confines.}}
\label{f11}
\end{figure}

Next we discuss the T-dual description of the D4 branes in $\C$.
For simplicity, consider first the case $N_f=0$.
The color D4 branes in Figure 2(a) turn into $N_c$ D5 branes wrapped
on the $S^2$ cycle of the resolved conifold.
The limit $L_c/R_6\to 0$ is dual to the singular conifold where the $S^2$
cycle is degenerated and the D5 branes look like ``fractional'' D3 branes
at the tip of the conifold.
Since classically $L_c\sim 1/g_{YM}^2$, as we decrease $L_c$,
we increase the YM coupling $g_{YM}$.
Quantum mechanically, $g_{YM}$ is dimensionally transmuted into a dynamically
generated scale $\Lambda$; it runs towards strong coupling in the IR.
Hence, $L_c$ or its dual $S^2$ are dynamically degenerated.
It is claimed \cite{ks,Vafa:2000wi,Gopakumar:1999ki} that when
$S^2$ shrinks, the $S^3$
of the conifold dynamically
blows up to a size related to the QCD scale $\Lambda$ and that there are $N_c$
units of RR flux through $S^3$.
Hence, the geometrical conifold transition describes confinement of
$SU(N_c)$ SQCD. In the brane configuration of Figure 2(a) (and similarly for
the brane system of Figure 1(a)), this transition  corresponds to the
limit when
the NS and NS' branes intersect and bend away from the $v=w=0$ point,
as shown in Figure 4, in the presence of RR flux.

A small number of flavors $N_f<N_c$ in Figure 1(b) and 2(a)
does not change
the physics above: In type IIA, the intersection of the NS and NS' branes
still amounts to confinement and should again be dual to the conifold
transition $S^2\to 0\to S^3$ in type IIB.
However, when $N_f\geq N_c$ another physical transition is possible, namely
an $\N=1$ electric-magnetic duality.
Starting with the electric theory in Figure 1(b), this corresponds to
changing the position of the NS' brane from positive to negative coordinate
$x^6$. This gives rise
to a magnetic dual $\N=1$, $SU(N_f-N_c)$ SYM with $N_f$ flavors
(see Figure 1(c)).
Actually, when $3N_c>N_f\geq N_c$ this is expected to happen dynamically:
The system is driven towards strong coupling where the NS and NS' branes
intersect. Since physically on both sides of this transition we have a gauge
theory with matter and it is a question of convention to decide which one is
the electric or magnetic description~\footnote{The detailed structure
of the theories is missing for semi-infinite flavor D4 branes.},
we expect in the T-dual description
of type IIB a transition involving on each side 2-spheres. In other words,
the electric-magnetic duality should be realized as a geometric flop
transition $S^2\to 0\to \tilde S^2$ in type IIB, where $S^2$ and $\tilde S^2$
are the two possible small resolutions of a conifold~\footnote{Locally,
near the singularity, the two resolutions are isomorphic, but global
effects and/or the presence of D-branes can distinguish the two.} (see
the recent work \cite{Cachazo:2001sg}).

Similarly, in the system of Figure 2(a), when $N_c\geq N_f-N_c=M\geq 0$ the
dynamics
is expected to lead to the duality cascade \cite{ks}:
$SU(N_c)\times SU(N_c+M)\to SU(N_c-M)\times SU(N_c)\to
SU(N_c-2M)\times SU(N_c-M)\to \cdots \to SU(N_c-kM)\times SU(N_c+M-kM)$,
until $N_c-kM<M$ (the first step of the cascade is shown in Figure 2(b)).
After reaching the last step, the theory confines.
In the fivebranes description, this cascade is
dynamically  due to the bending of the
NS and NS' branes (see \cite{ks} for details). On the type IIB side,
this should correspond to the cascade of geometrical flops
and conifold transition:
$S^2\to 0\to \tilde S^2\to 0\to S^2\to 0\to\cdots\to S^2\to 0\to
S^3$. On duality cascades and Seiberg duality, see also
\cite{Dasgupta:2001fg, Dasgupta:2001ac}. 

In order to shed some light on the type IIB geometric description of
electric-magnetic duality, it is useful to recall the IIA branes/IIB geometry
duality relation for an $\N=2$ configuration. Concretely, we consider a system
of two parallel NS branes separated by a distance $L_c$
in the direction
$x^6$. Between them, $N_c$ D4 branes are stretched, whose coordinates in the
$v$-plane are $v_i$
$(i=1,...,N_c)$, describing the eigenvalues of the vacuum expectation
value (VEV)
of an adjoint field of the $SU(N_c)$ gauge theory in its Coulomb branch
(see Figure 5).
\begin{figure}[!h]
\centerline{\hbox{\psfig{figure=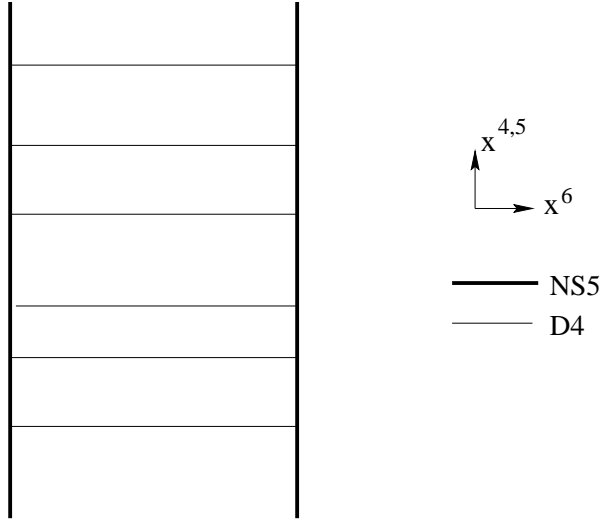,height=7cm}}}
\caption{{\bf Figure 5:}
{\footnotesize \em Brane realization of pure $\N=2$ $SU(N_c)$
  SYM in the Coulomb branch.
}}
\label{f4}
\end{figure}
After compactification of the $x^6$ direction, the function appearing in
Eq. (\ref{hvw}) is now $H(v,w)=w^2$ and the variable
$v$ is not constrained. The
algebraic equation (\ref{fvw}) in $\Co^4$ then becomes
\begin{equation}
F(v,w,z,z')=H(v,w)-zz'=w^2-zz'=0~,
\label{fvw2}
\end{equation}
which describes the ALE space $\Co^2/\Z_2$ times $\Co$, the local
geometry of a singular $K3\times T^2$ compactification of type IIB.
Now there is only one way to desingularize this space, namely
by blowing up an $S^2$ at the origin of the ALE space:
\begin{equation}
F(v,w,z,z')=H(v,w)-zz'=w^2-zz'=\mu~.
\label{fvw2b}
\end{equation}
This is T-dual in type IIA to a Neveu-Schwarz fivebrane on:
\begin{equation}
H(v,w)=w^2=\mu~,
\label{hvw2b}
\end{equation}
namely, two NS branes separated by a distance $L_c/R_6\sim\sqrt{\mu}$.
As in the $\N=1$ case,
$L_c/R_6$ in type IIA is mapped to a blow up parameter of the $S^2$ in
type IIB. Here, however, there is no physical (or geometrical) transition
occurring when the NS branes are coincident (or the $S^2$ vanishes).
Notice that the fact that the six dimensional geometry takes the form of
a product $\Co^2/\Z_2\times \Co$ implies that we have actually a blown up
$S^2$ at each point $v$ in $\Co$. Therefore, this $S^2$ has a one complex
dimensional moduli space and the $N_c$ wrapped D5 branes can independently
sit at any points $v_i$ $(i=1,...,N_c)$ in $\Co$, describing the Coulomb
branch of an $\N=2$ vector multiplet in the adjoint of $SU(N_c)$.

Let us treat now in the same spirit the brane system of Figure 3(a).
In order to
obtain a geometrical description of this configuration, we compactify the
direction $x^6$ on a circle of radius $R_6$ and perform a T-duality on it.
On the type IIB side, the CY geometry resulting from the presence of the
three type IIA NS branes is thus expected to contain two isolated
2-spheres with blow up parameters
$L_c/R_6$ and $L_f/R_6$ and intersecting at a point,
as depicted in Figure 6(a).
\begin{figure}[!h]
\centerline{\hbox{\psfig{figure=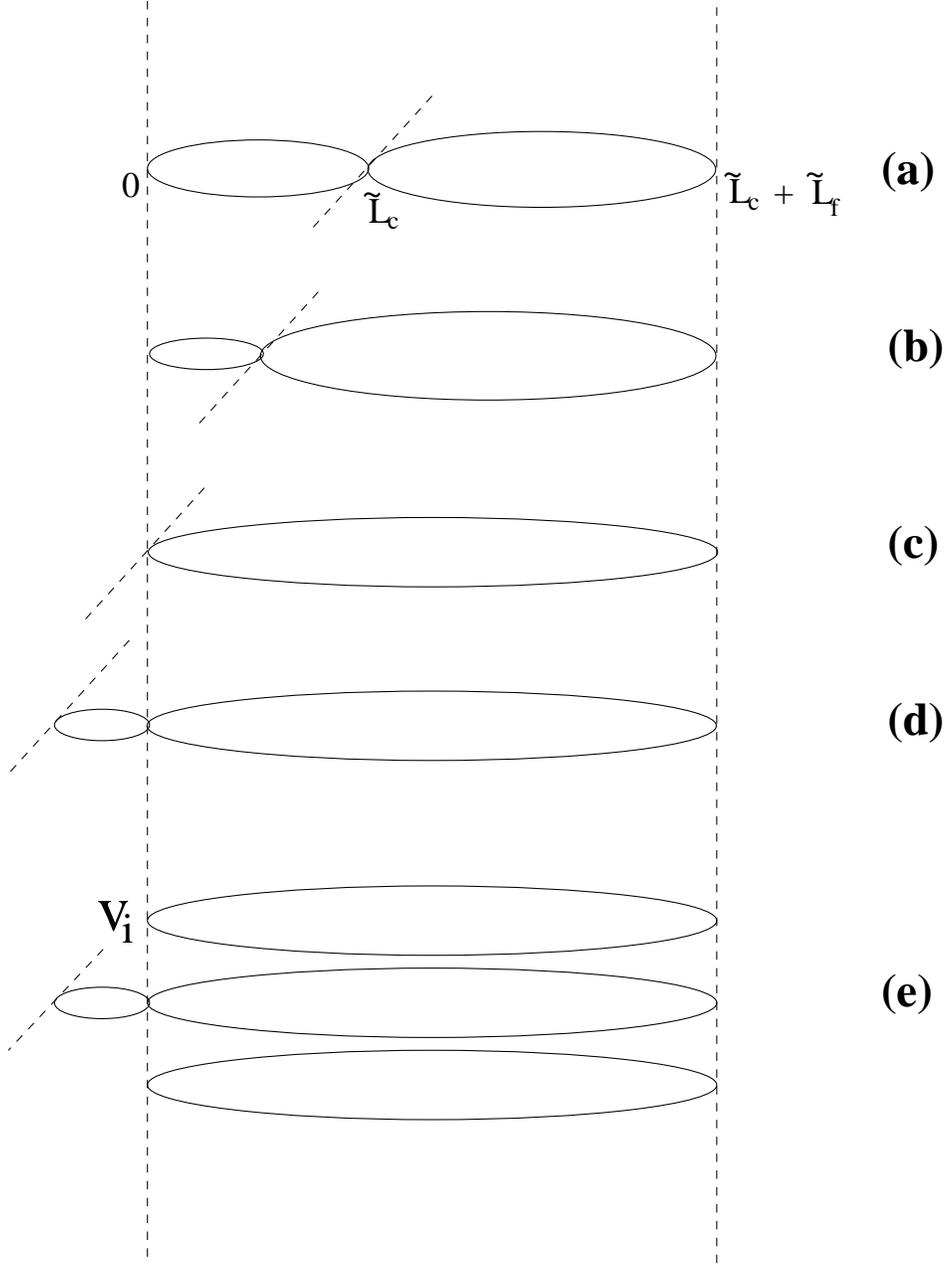,height=17cm}}}
\caption{ {\bf Figure 6:}
{\footnotesize \em {\bf (a)} Local type IIB geometry realizing
  $SU(N_c)\times SU(N_f)$ SYM, T-dual to Figure 3(a) indicated in
  dashes. Only the 2-cycles where D5 branes are wrapped are
  represented. {\bf (b,c,d)} Type IIB transition dual to the crossing
  of NS and NS' branes that describes Seiberg duality. {\bf (e)} One
  of the $\CP^1$'s has now a moduli space that parametrizes the
  ``meson'' VEV.} }
\label{f5}
\end{figure}
Clearly, this geometry is compatible with
the geometrical realization of the system of Figure 1(a) or 1(b) in the
limit $L_f/L_c\gg 1$. In addition, the
type IIA D4 branes give rise to $N_c$ ($N_f$) D5 branes wrapping the first
(second) of these $S^2$'s, generating an $SU(N_c)\times SU(N_f)$
gauge theory. The fact that the 2-spheres have to be
isolated is required by the fact that there is no adjoint field in any
of the gauge group factors.
Also, the 2-spheres have to intersect for the
$(N_c,N_f)$ hypermultiplet to be
massless. As a remark, the $\N=2$ version of the previous brane
configuration has already been considered in the literature
\cite{Karch:1998yv}.
It is obtained by replacing the middle NS' brane by a third parallel NS
brane~\footnote{This amounts to take $H(v,w)=w^3$ in Eq. (\ref{hvw}).}.
The dual type IIB geometric realization has been conjectured to consist
of a pair of 2-spheres intersecting at a point as before, each of them
living in a one complex dimensional family. As required, this geometry
describes the Coulomb branch of an $SU(N_c)\times SU(N_f)$  $\N=2$
gauge theory.

When $N_f\ge N_c$, let us describe what should be the geometrical picture
related to the magnetic dual brane configuration of Figure 3(b) obtained by
moving the NS' brane from positive to
negative values of its coordinate $x^6$. From the
brane point of view, we now have an $SU(N_f-N_c)\times SU(N_f)$
gauge theory with a hypermultiplet $(q,\tilde q)$ in the $(N_f-N_c,N_f)$
representation and a chiral field $M$ in the adjoint of $SU(N_f)$. The
$\N=2$ supersymmetry associated with the two parallel NS branes implies the
existence of a classical superpotential $\W=qM\tilde q$. After a
$T_6$ duality along the compact direction $x^6$, the type IIB configuration is
expected to give an isolated 2-sphere, which  we will denote by $\tilde S^2$,
with $N_f-N_c$ wrapped D5 branes on it and intersecting at one point a
second $S^2$ with $N_f$ D5 branes on it (see Figure 6(d)). However,
this second $S^2$
should live in a one complex dimensional family of 2-spheres in the CY, so that
the $N_f$ D5 branes can slide on any other
representative of the family. This would describe the
Coulomb branch of the $SU(N_f)$ gauge factor obtained by giving a
vacuum expectation value to the ``meson'' field $M$, as depicted in Figure
6(e). In total, Figure 6 illustrates a sequence of 2-spheres describing
a geometrical transition from an electric to a magnetic description.

To summarize, the singular type IIB geometry T-dual to an NS brane
intersecting an NS' brane at a point a finite distance away from another NS
brane should admit the three
different desingularizations described above. Two of
them are the small resolutions where a 2-sphere $S^2$ or $\tilde S^2$ is blown
up, while the third one consists in the appearance of a 3-sphere. In the next
section, we shall describe the mirror picture of these 3 phases.

\subsection{Mirror duality from type IIB to type IIA geometry}
\label{geoA}

Mirror symmetry relates generically a CY threefold in type IIB
to a different one in type IIA.
In particular, it maps the resolved conifold in type IIB
to the deformed conifold in type IIA, and vice versa.
Hence, the conifold transition of the
previous subsection,
\begin{equation}
\mbox{IIB}\quad :\qquad S^2\to 0\to S^3~,
\label{btrans}
\end{equation}
is mirror dual to:
\begin{equation}
\mbox{IIA}\quad :\qquad S^3\to 0\to S^2~.
\label{atrans}
\end{equation}
Recall that both  transitions in Eqs. (\ref{btrans}, \ref{atrans})
are T-dual to the transition when NS and NS' branes approach, intersect
and then bend.

To be concrete, let us describe explicitly the type IIA geometry
on the deformed conifold $T^*(S^3)$, defined in Eq. (\ref{fvwd});
here we write it in the form:
\begin{equation}
z_1^2+ z_2^2+ z_3^2+ z_4^2= \mu~,
\label{con}
\end{equation}
where $z_{i}$ $(i=1,2,3,4)$ are complex coordinates and $\mu$
can be chosen to be real and positive without loss of generality.
The base $S^3$ is identified with the fixed point set of the
antiholomorphic involution
$z_i\to\bar z_i$. Denoting by $x_i$ and $y_i$ the real and imaginary parts
of $z_i$, it is  automatically a special Lagrangian
submanifold, whose equation is:
\begin{equation}
y_i=0 \quad , \quad x_1^2+ x_2^2+ x_3^2+ x_4^2= \mu~,
\label{s3mu}
\end{equation}
showing that its volume is related to $\mu$.

Let us concentrate first on the
realization in this context of the pure $SU(N_c)$ SYM theory.
We saw in the previous section how the D4 branes of Figures 1(a)
turn into D5 branes wrapped on $S^2$ in type IIB. The mirror picture in type
IIA thus involves $N_c$ D6 branes  wrapping the $S^3$ of Eq. (\ref{s3mu}).
The massless spectrum contains a $U(N_c)$ gauge group.

At $\mu=0$, the deformed conifold becomes a cone whose apex at $z_i=0$ can be
blown up to an $S^2$. The resolved conifold obtained this way can  be written
as
\begin{equation}
\left\{
\begin{array}{l}
z_1^2+ z_2^2+ z_3^2+ z_4^2 = 0\\
(z_3-iz_4)\xi_1 + (z_1-iz_2)\xi_2 = 0~,
\end{array}
\right.
\label{res1}
\end{equation}
where $\xi_{1,2}$ are projective coordinates of $S^2\cong \CP^1$.
In this form, it is clear that any non-singular point on the conifold is
lifted to a single point on the
resolved conifold, while at the singular point $z_i=0$ on the conifold,
$\xi_{1,2}$ are not fixed and parametrize a full $\CP^1$.
Alternatively, the resolved conifold can be rewritten as
\begin{equation}
\left\{
\begin{array}{l}
(z_1+iz_2)\xi_1 - (z_3+iz_4)\xi_2 = 0\\
(z_3-iz_4)\xi_1 + (z_1-iz_2)\xi_2 = 0~,
\end{array}
\right.
\label{asres}
\end{equation}
where it takes the explicit form of an $O(-1)+ O(-1)$ bundle over
$\CP^1$.~\footnote{In fact,
since $\xi_1$ and $\xi_2$ must not vanish simultaneously, the determinant
of the coefficients of $\xi_{1,2}$ in Eq. (\ref{asres})
must vanish identically.
This determinant is $\sum_i z_i^2$, hence
one can replace the first equation
in (\ref{asres}) by $\sum_i z_i^2=0$.}
Now, the transition (\ref{atrans}) takes
the system of $N_c$ D6 branes on $S^3$ to an $S^2$ with $N_c$ units of RR flux
\cite{Vafa:2000wi}.

Similarly, we can consider the situation in Figure 1(a) with an O4 plane
added on top of the $N_c$ D4 branes. In type IIB, it
amounts to performing an orientifold projection that fixes the $S^2$ so that
there is an O5 plane wrapped on $S^2$.
In the mirror picture of type IIA, we must have an O6 plane wrapped on $S^3$.
Thus, one considers the orientifold projection on the deformed conifold
\cite{Sinha:2000ap}:
\begin{equation}
z_1^2+ z_2^2+ z_3^2+ z_4^2= \mu~, \quad
\mbox{where}\quad (z_i;\bar z_i)\equiv (\bar z_i;z_i)~,
\label{con/orient}
\end{equation}
combined with the exchange of the left and right movers on the worldsheet.
Adding $N_c$ D6 branes on top of the O6 plane gives rise to an $SO(N_c)$ or
$Sp(N_c/2)$ gauge theory.
To identify what the orientifold projection on the deformed
conifold becomes after the transition to the resolved conifold, one extends
the orientation reversal action to the variables $\xi_{1,2}$. Actually, there
is a unique way to do it and the IIA background becomes
\begin{equation}
\left\{
\begin{array}{l}
(z_1+iz_2)\xi_1 - (z_3+iz_4)\xi_2 = 0\\
(z_3-iz_4)\xi_1 + (z_1-iz_2)\xi_2 = 0
\end{array}~, \quad \mbox{where} \quad (z_i,\xi_1,\xi_2;c.c.)\equiv
(\bar z_i, -\bar \xi_2, \bar \xi_1; c.c.)~.
\right.
\label{res/orient}
\end{equation}
Note that a fixed point of the involution would have
$\xi_1=\xi_2=0$ which is forbidden. As expected, the involution is therefore
freely acting since the $S^3$ has disappeared. Thus, there are neither O6
planes nor
D6 branes in this phase. However, there is still a 2-cycle in this background
since the base $S^2$ of the
resolved conifold is now replaced by $\RP^2$. By conservation
of RR charge, there are also ${N_c\over 2} \mp 2$ units of flux on $\RP^2$,
where $\mp$ is the sign of the O6 plane.  Physically,
the $SO(N_c)$ or $Sp(N_c/2)$ group confines \cite{Sinha:2000ap}.

\vspace{.3cm}

\noindent {\em \large Introducing matter in the type IIA geometry}
\vspace{.2cm}

\noindent
In type IIB, the implementation of matter and flavor symmetry was done by
considering two 2-cycles intersecting at a point with D5 branes wrapped on
them. Thus, in type IIA we are looking for another 3-cycle that
intersects $S^3$. In \cite{Ooguri:2000bv}, such a geometry was
considered~\footnote{The aim of \cite{Ooguri:2000bv} was to provide
  some quantitative check of the  pure Yang-Mills duality conjecture
  of \cite{Vafa:2000wi,Gopakumar:1999ki}.}.
In the deformed conifold of Eq. (\ref{con}), an $S^3$ was identified
as the fixed point set of the antiholomorphic involution $z_i\to\bar z_i$.
Another SLAG is fixed by the involution
$z_{1,2}\to \bar z_{1,2}$, $z_{3,4}\to -\bar z_{3,4}$. It is given by
\begin{equation}
y_{1,2}=x_{3,4}=0\quad , \quad x_1^2+ x_2^2= \mu+ y_3^2+ y_4^2~.
\label{s2s1}
\end{equation}
In this equation, $y_{3,4}$ are arbitrary and parametrize a complex plane
$\Co$. Also, the modulus of $x_1+ix_2$ is fixed but not its phase. Altogether,
Eq. (\ref{s2s1}) defines a 3-cycle of topology $\Co\times S^1$.
Note that the intersection between this cycle and the $S^3$ of Eq. (\ref{s3mu})
satisfies
\begin{equation}
x_{3,4}=y_i=0\quad , \quad x_1^2 +x_2^2=\mu~,
\label{s1A}
\end{equation}
which is an $S^1$.
If we wrap
$N_f-N_c$ D6 branes on the $S^3$ and $N_f$ D6 branes on $\Co\times S^1$,
we thus obtain an $SU(N_f-N_c)$ gauge theory with $N_f$
hypermultiplets of quarks $(q,\tilde q)$ in the fundamental
representation -- a ``magnetic'' theory. Note that for a geometry
where $\Co\times S^1$ would
be replaced by $S^2\times S^1$ at finite volume, the $SU(N_f)$ flavor symmetry
would be gauged. Also, there would be a  chiral field $M$ in the adjoint
representation of $SU(N_f)$ coupled to the quarks via a classical
superpotential $\W=qM\tilde q$. However, in our case $M\equiv 0$
since the $N_f$ hypermultiplets are massless.

On the resolved conifold, we already saw that the  $S^3$ fixed by
$z_i\to \bar z_i$ in $T^*(S^3)$ is replaced by an $S^2$. To see what $\Co\times
S^1$ becomes after the conifold transition, one extends to
the resolved conifold the antiholomorphic involution that was fixing
it:
$z_{1,2}\to \bar z_{1,2}$, $z_{3,4}\to -\bar z_{3,4}$, $\xi_1\to\bar\xi_2$,
$\xi_2\to\bar\xi_1$. The new fixed point set satisfies
\begin{equation}
y_{1,2}=x_{3,4}=0\quad , \quad \lambda\equiv \xi_1/\xi_2= 1/\bar \lambda\quad
,\quad
\left\{
\begin{array}{l}
x_1^2+x_2^2=y_3^2+y_4^2\\
(iy_3+y_4)\lambda + (x_1-ix_2) = 0
\end{array}
\right.~,
\label{res}
\end{equation}
where $\lambda$ thus parametrizes an $S^1$.
For any non-vanishing $y_3+iy_4$, \ie parametrizing $\Co^*$,
$\lambda$ is uniquely determined and the phase of $x_1+ix_2$
parametrizes an $S^1$.
On the contrary, at the origin $y_3+iy_4=0=x_1+ix_2$, it is the phase
$\lambda$ that is arbitrary. Altogether, the topology of this 3-cycle is
thus again $\Co\times S^1$ and it intersects the $S^2$ ($z_i=0$, $\xi_1/\xi_2$
arbitrary) along its equator ($z_i=0$, $\xi_1/\xi_2=\lambda$ arbitrary).
Therefore, the $N_f$ D6 branes wrapped on $\Co\times S^1$
remain present on both sides of the transition. However, the $N_f-N_c$
D6 branes wrapped on $S^3$ in the deformed conifold have disappeared but
imply the presence of $N_f-N_c$ units of RR flux on $S^2$ in
the resolved conifold. Thus, in this geometrical phase,
the open string sector does not contain massless gauge bosons and quarks
any more. This is expected in the IR of the gauge theory we started
with on $T^*(S^3)$,
due to confinement, where a mass gap is generated~\footnote{Actually, this
is expected when $N_f>3N_c$; in the window
${3\over 2}N_c<N_f<3N_c$ one expects an interacting conformal field theory
of quarks and gluons \cite{Seiberg:1995pq},
while for $N_f\leq {3\over 2}N_c$ the
magnetic theory is IR free. It is not clear to us how to distinguish
the various cases from geometry.}.

Since the above duality is an extention of the conjecture of
 \cite{Vafa:2000wi} where quarks are included, it  would be very interesting
to give quantitative checks of it, for example  in the spirit of
 \cite{Acharya:2001hq}.

The generalization to $SO(N_f-N_c+4)\times Sp(N_f/2)$ ($N_f$ even) and
 $Sp((N_f-N_c-4)/2)\times SO(N_f)$ groups is
 straightforward. One has to consider the orientifold projection on
 $T^*(S^3)$ given in Eq. (\ref{con/orient}) with $N_f-N_c\pm 4$ D6
 branes on top of the O6 plane wrapped on $S^3$. Also, there are $N_f$ flavor
 branes wrapped on $\Co/\Z_2\times S^1$, where $\Z_2$ acts as
 $y_3+iy_4\to -y_3-iy_4$. The singular points on this 3-cycle are precisely the
 $S^1$ of the intersection with $S^3$. After the conifold transition, the
 geometry is given in Eq. (\ref{res/orient}). The O6 plane and
 $N_f-N_c\pm 4$
 D6 branes on $S^3$  are replaced by $\RP^2$ with ${N_f-N_c
 \over 2}$ units of RR flux, while the $N_f$ flavor branes remain.

In Section 2.1 we saw that when matter in the fundamental representation
is introduced, both from the brane point of view of Figures 1-3
and the type IIB geometry, one can also describe electric-magnetic duality.
For completeness, we next review
how this can also be done in type IIA geometry.
In fact, there exists an alternative local construction
described in \cite{Ooguri:1997ih}, where two phases
are naturally related to each other by  Seiberg duality, when expected.

Consider the type IIA compactification
on a CY, whose local geometry is described by
\begin{equation}
\left\{
\begin{array}{l}
V^2+{V'}^2=(Z-a_1)(Z-a_2)\\
W^2+{W'}^2=Z-b~,
\label{ef}
\end{array}
\right.
\end{equation}
where $V$, $V'$, $W$, $W'$, $Z$ are complex variables and $a_1$, $a_2$, $b$
are complex parameters. This manifold contains two non-trivial
circles $S^1_V$, $S^1_W$ parametrized by the rotation angles of two
$SO(2,\R)$ matrices acting on the column vectors of components
$V,V'$ and $W,W'$, respectively. When $Z$ equals $a_1$ or $a_2$, $S^1_V$
degenerates. Similarly, when $Z$ equals $b$, $S^1_W$ degenerates.
Thus, by considering the segment $[a_1,a_2]$ in the $Z$-plane and
$S^1_V,\, S^1_W$ as a fiber on it, one obtains a 3-cycle of topology
$S^2\times S^1$. Similarly, it can be seen that two other 3-cycles of
topology $S^3$ can be associated to
the segments $[a_1,b]$ and $[b,a_2]$. We shall
denote such a 3-cycle by its base segment in the $Z$-plane.
Let us wrap $N_c$ D6 branes on $[a_1,b]$ and $N_f$ D6 branes on
$[b,a_2]$. Requiring an unbroken  $\N=1$ supersymmetry in space-time implies
that $a_1$, $b$ and $a_2$ are aligned in the $Z$-plane so that we can
choose them to be along the $\Re e(Z)$-axis. If we start with a
configuration $a_1<b<a_2$, we thus have branes wrapped on two $S^3$'s.
Notice that at the intersection $Z=b$ of their base segments, $S^1_W$ vanishes
but $S^1_V$ is of
finite size. Hence the two $S^3$'s intersect along a circle.
Thus, the brane system generates an $\N=1$ $SU(N_c)\times SU(N_f)$
gauge theory in four dimensions with a massless hypermultiplet $(Q,\tilde Q)$
in the $(N_c,N_f)$ representation.

The previous CY can now be deformed \cite{Ooguri:1997ih} by changing the
complex structure $b$
such that $b<a_1<a_2$. If $N_f>N_c$, the wrapped branes will combine such
that we end up with $N_f-N_c$ D6 branes wrapped on the 3-sphere $[b,a_1]$ and
$N_f$ others wrapped on the 3-cycle $[a_1,a_2]$. Note that the latter is of
topology $S^2\times S^1$ and intersects the 3-sphere along the finite size
circle $S^1_V$ that sits at $Z=a_1$. As a result, this system describes
an $\N=1$ $SU(N_f-N_c)\times SU(N_f)$ gauge theory with a hypermultiplet
$(q,\tilde q)$ in the $(N_f-N_c, N_f)$ representation coupled to a chiral
field $M$ in the adjoint of $SU(N_f)$, with a classical superpotential
$\W= qM\tilde q$. Thus, the above geometric transition realizes
the same electric-magnetic duality we encountered in the brane and type IIB
pictures of Figures 3 and 6.
It can be summarized schematically by
the sequence
\begin{equation}
\mbox{IIA}\quad :\qquad (S^3,S^3)\to (0,\mbox{singular 3-cycle})\to
(\tilde S^3, S^2\times S^1)~,
\end{equation}
where
on each side of the transition the 3-cycles intersect along an $S^1$.
Notice that the first of these $S^3$'s can actually be deformed into
$\tilde S^3$ by considering a path $(b-a_1)\to -(b-a_1)$ in complex
structure moduli where the 3-sphere never vanishes.
Such a path corresponds in the brane
picture of Figures 1 and 3 to a motion in the plane $(x^6,x^7)$, hence
turning on also a FI D-term in the low energy SYM \cite{egk}.
This complex structure deformation in type IIA
\begin{equation}
\mbox{IIA}\quad :\qquad S^3\to 0\to \tilde S^3~,
\end{equation}
is mirror dual to
the 2-sphere flop in the type IIB geometry:
\begin{equation}
\mbox{IIB}\quad :\qquad S^2\to 0\to \tilde S^2~.
\end{equation}
The generalization of the type IIA geometric
description of Seiberg duality for $SO$ and $Sp$ groups is realized by
implementing an orientifold projection \cite{Ooguri:1997ih}.

The link from these electric and magnetic pictures to the brane configuration
of Figure 3 is found
as follows \cite{Ooguri:1997ih}.
Each of the equations in (\ref{ef}) takes the form of an elliptic
fibration over the $Z$-plane with a monodromy transformation around each
point $Z=a_{1,2}$ or $Z=b$, respectively. After two T-dualities
on the circles $S^1_V$ and $S^1_W$ to another type IIA description,
this is translated into the presence of three Neveu-Schwarz fivebranes.
Two of them are parallel and can be identified by convention with NS branes
(see Eq. (\ref{0123})), while the third one can be identified with
an NS' brane.
Also, the transverse coordinate $x^6+ix^7$ can be identified with $Z$ in
Eq. (\ref{ef}). Therefore, the two NS branes sit at $x^6=a_1$ and $x^6=a_2$,
while the NS' brane sits at $x^6=b$.
Since the 3-cycles in the geometric approach are $S^1_V\times S^1_W$
fibrations over segments in the $Z$-plane, the D6 branes wrapped on them have
Dirichlet boundary conditions in the directions dual to the circles on which
we T-dualize. As a result, they give rise to D4 branes stretched between the
NS and NS' branes along the direction $x^6$, as in Figure 3.

To conclude this section,
we would like to make the link between the geometric
descriptions of type IIA and type IIB in Section 2.1. The brane
picture of Section 2.1 is related to the type IIB geometry
by a T-duality $T_6$.
It is also related to the type IIA geometry
via two T-dualities that act on the
phases of the complex planes $x^4+ix^5$ and  $x^8+ix^9$. As a
result, the IIB and IIA geometries are related into each other by three
T-dualities, as expected for mirror descriptions of CY threefolds
\cite{Strominger:1996it}.

\subsection{Type IIA geometry versus M-theory geometry}

Lifting flat parallel type IIA D6 branes to M-theory gives a
purely geometrical compactification on a Taub-NUT space
\cite{Townsend:1995kk, Sen:1997kz}.
Some cases of D6 branes in curved backgrounds have also been considered
\cite{Acharya:2000gb, Atiyah:2000zz, Gomis:2001vk, Edelstein:2001pu,
  Aganagic:2001nx, Aganagic:2001jm, Brandhuber:2001yi,
  Hernandez:2001bh, Gomis:2001vg}.
In general, a type IIA compactification on a manifold of reduced holonomy
with D6 branes wrapped on cycles is lifted in M-theory to a purely
gravitational background of different reduced holonomy
\cite{Gomis:2001vk, Edelstein:2001pu}. More precisely,
if one considers the type IIA string on a manifold $\M_d$,
the M-theory compact
space will have locally the form $\M_d\times S^1$. However, the $S^1$ is
non-trivially fibered over $\M_d$ so that its radius vanishes
identically on the submanifolds on which D6 branes are wrapped. Once
we are in M-theory, one can ask whether the transitions
considered in type IIA with branes would have some simple
geometrical interpretations.

We first review the situation for the geometrical transition
associated to confinement of pure $\N=1$ SYM theory
\cite{Acharya:2000gb, Atiyah:2000zz, Atiyah:2001qf}.
Consider M-theory compactified on $Spin(S^3)$, the spin bundle over
$S^3$ \cite{Gibbons:1990er}. The metric of this manifold is known,
while topologically it looks like the space
\begin{equation}
|u_1|^2+|u_2|^2-|u_3|^2-|u_4|^2=V~,
\label{spin}
\end{equation}
where $u_i$ $(i=1,...,4)$ are complex variables and $V$ is a real
parameter~\footnote{In fact, Eq. (\ref{spin}) is just a model for
topological properties that are relevant for our purpose. In particular,
the metric derived from this equation is {\em not} of $G_2$ holonomy
and {\em not} even Ricci flat.}.
$Spin(S^3)$ admits an $SU(2)^3$ isometry group in which
various $U(1)$ subgroups can be chosen to perform a Kaluza-Klein (KK)
reduction to a weak coupling type IIA description.
An example of such a $U(1)$ acts in the model of Eq. (\ref{spin}) as
\begin{equation}
U(1)~~:\quad (u_3,u_4)\to (e^{i\alpha} u_3,e^{i\alpha}u_4)~.
\label{U}
\end{equation}
For the massless spectrum in
M-theory to contain some non-Abelian gauge group, one can  consider
orbifolds of the previous manifold.
For instance, we can choose the discrete subgroup of the $U(1)$
isometry generated by
\begin{equation}
\sigma~~:\quad (u_3,u_4)\to (e^{2i\pi/N_c} u_3, e^{2i\pi/N_c}u_4)~,
\end{equation}
and focus on  $Spin(S^3)/\Z_{N_c}$.

Let us consider first the case $V>0$. The fixed point set of $\sigma$ and the
$U(1)$ action satisfies
\begin{equation}
u_3=u_4=0\quad , \quad |u_1|^2+|u_2|^2=V~,
\label{s3M}
\end{equation}
which is the $S^3$ base of an $\R^4/\Z_{N_c}$ fibration.
This describes a pure $\N=1$ $SU(N_c)$ gauge theory.
The dimensional reduction along $S^1\cong U(1)$
gives a type IIA string theory compactified on a CY threefold
with a non-trivial $S^3$ cycle, $T^*(S^3)$ of Eq. (\ref{con}), under the
identification $\mu\equiv V$, together with $N_c$ D6 branes
wrapped on $S^3$ \cite{Atiyah:2000zz}.
As explained in Section 2.2, two T-dualities translate this system
to the type IIA brane configuration of Figure 1(a).

When $V<0$, the orbifold is freely acting, as can be seen from the fixed
point set (\ref{s3M}) which is now empty. Therefore, there is no trace anymore
of the gauge group, a fact that has been interpreted  in \cite{Atiyah:2000zz}
as
confinement of $SU(N_c)$. Geometrically, there is an $S^3$ flop transition
at $V=0$, where the 3-sphere (\ref{s3M}) present for $V>0$ is replaced by a
Lens space $S^3/\sigma$
\begin{equation}
u_1=u_2=0\quad , \quad -|u_3|^2-|u_4|^2=V~, ~~ \mbox{where} ~~ (u_3,u_4)\equiv
e^{2i\pi/N_c}(u_3,u_4)~,
\label{lens}
\end{equation}
for $V<0$. Schematically, the transition in M-theory takes the form:
\begin{equation}
\mbox{M-theory}\quad : \qquad S^3\to 0\to S^3/\sigma~.
\label{Mtrans}
\end{equation}
The Lens space can be seen as a Hopf fibration over $S^2$,
whose fiber is precisely $U(1)$. Upon KK reduction along
$S^1\cong U(1)$, it gives rise to
the $S^2$ of the resolved conifold with $N_c$ units of RR flux through it.
This is again consistent with the interpretation of confinement,
as reviewed in Section 2.2.

Similarly, we can consider singularities in M-theory in the $D$-series.
For $N_c\geq 8$ and even, one thus defines $Spin(S^3)/\tau$,
where $\tau$ has two generators \cite{Sinha:2000ap}
\begin{equation}
\tau~~:\quad (u_3,u_4)\to (e^{2i\pi/(N_c-4)} u_3,
e^{2i\pi/(N_c-4)}u_4)\quad\mbox{and}\quad (u_3,u_4)\to (\bar u_4,-\bar u_3)~.
\label{tau}
\end{equation}
This definition of $\tau$ is equivalent to the dihedral
group $D_{N_c/2}$, which implies that physically there is an
$SO(N_c)$ gauge group for $V>0$ that confines for $V<0$. When $V>0$, the
dimensional reduction along $S^1\cong U(1)$ gives
$N_c/2$ D6 branes, an O6 plane and $N_c/2$ mirror branes
wrapped on the $S^3$ of $T^*(S^3)$ in type IIA. This is precisely the
orientifold model of Eq. (\ref{con/orient}). When $V<0$, the first
generator
of $\tau$ implies that the reduction of the Lens space $S^3/\tau$
gives an $S^2$ with $N_c-4$ units of RR flux. Identifying the complex
coordinate of this $S^2$ with $-u_3/u_4$, the second generator of
$\tau$ implies the modding action
$-u_3/u_4\to \bar u_4/\bar u_3$.
This background is precisely the orientifold of the resolved
conifold in Eq. (\ref{res/orient}) under the identification
$\xi_1=-u_3$, $\xi_2=u_4$.

Actually, the classical type
IIA description on the deformed conifold for $\mu\equiv V\gg0$ provides a good
approximation of the physics only in the UV, where the non-Abelian gauge
group is weakly coupled.
When $\mu\equiv V$ is still positive but decreases, the SYM
coupling increases and the naive classical description becomes
less accurate. When we pass to negative values of $V$ in M-theory, the
situation gets even worse.
In general, in the type IIA reduction it is necessary to compute
worldsheet instanton contributions in order to describe
strong coupling effects in the corresponding gauge theory,
like confinement.
As an example, when $Spin(S^3)/\sigma$ is reduced along $S^1\cong
U(1)$, the type IIA worldsheet instantons are computed by closed
string topological amplitudes.
In \cite{Aganagic:2001jm},
$Spin(S^3)/\sigma$ is instead reduced along another
$S^1\cong U(1)''$ (we shall define later in Eq. (\ref{U''})). For $V>0$, the
type IIA background is in this case $\Co\times \Co^2/\Z_{N_c}$, thus
describing an explicit $SU(N_c)$ gauge
symmetry in space-time~\footnote{In addition, there is a D6 brane
wrapped on a SLAG of topology $\Co/\Z_{N_c}\times S^1$.}. This
classical background is accurate to describe the SYM physics in the
UV. However, for $V<0$, the classical type IIA background is still
$\Co\times \Co^2/\Z_{N_c}$,~\footnote{With a D6 brane
wrapped on a SLAG of topology $\Co \times S^1$.} which cannot be
trusted for describing accurately the $SU(N_c)$ gauge theory in the IR.
This is due to the presence of large instanton corrections arising
from open string worldsheets with disk topology. To compute
them, one can map the system in this
phase to a dual one, namely a type IIB string theory on the
mirror CY with a D5 brane wrapped on a curve.

We have reviewed that for even $N_c\geq 8$ the lift to M-theory
of a type IIA orientifold on $T^*(S^3)$ with an O6 plane and a total of
$N_c$ D6 branes wrapped on $S^3$ generates an $SO(N_c)$ gauge
group~\footnote{Actually, the M-theory
backgrounds based on orbifolds of $Spin(S^3)$ are the lifts of type
IIA string theories in the infinite string coupling limit. For finite string
coupling, one has to replace $Spin(S^3)$ by a $G_2$ holonomy
manifold constructed in \cite{Brandhuber:2001yi}.}.
In the following we shall discuss the case $N_c=4$ that will help us for
treating models in later sections.
In the notation of Eq. (\ref{con})
and defining $x^{10}$ as a coordinate along a circle $S^1$ of radius
$R_{10}$, let us consider
\begin{equation}
{T^*(S^3)\times S^1 \over w\I}~,
\quad\mbox{where} \quad w:~ z_i\to \bar z_i\quad \mbox{and}\quad  \I:~
x^{10}\to-x^{10}~.
\label{so4}
\end{equation}
For $\mu>0$, this orbifold is topologically $S^3\times
(\R^3\times S^1)/\Z_2$, where $\Z_2$ acts as an inversion on each
coordinate of $\R^3$ and $x^{10}$. Therefore, there are two copies of
$S^3$ of $A_1$ singularities, one at $x^{10}=0$ and the other at
$x^{10}=\pi R_{10}$, generating an $SU(2)^2$ gauge symmetry.
The metric of the double cover is simply
\begin{equation}
ds^2=ds^2_{CY}+ {R_{10}^2\over l_{\mbox{\scriptsize p}}^2} \left(
dx^{10} \right)^2~,
\label{metric}
\end{equation}
where $ds^2_{\mbox{\scriptsize CY}}$ is the CY metric and
$l_{\mbox{\scriptsize p}}$ is the eleven-dimensional Planck
mass. Thus, identifying $R_{10}$ with the type IIA coupling and taking
the limit $R_{10}\to 0$, the geometric background becomes $T^*(S^3)/w$,
\ie an orientifold of type IIA with an O6 plane wrapped on
$S^3$. Since these remarks apply as well for compact CY's, where
the total RR charge must cancel, there must be two D6 branes (and their
mirrors) on top of the O6 plane. Altogether, the branes and
orientifold generate an $SO(4)\cong SU(2)^2$ gauge group as in
M-theory \cite{Kachru:2001je}.

At $\mu=0$, $T^*(S^3)$ used in the construction of the orbifold has
become a cone and we first choose to desingularize it by blowing up
an $S^2$. Therefore, from the 7-dimensional point of view, we are
now considering M-theory on
\begin{equation}
{\mbox{resolved conifold}\times S^1 \over w'\I}~,
\label{so4res}
\end{equation}
where, in the notations of Eq. (\ref{asres}),  the involution $w'\I$
acts as
\begin{equation}
w':~ \left\{
\begin{array}{l}
z_i\to \bar z_i\\
\xi_1\to -\bar\xi_2~,~~ \xi_2\to \bar\xi_1
\end{array}\right.
\qquad \mbox{and}\qquad  \I:~
x^{10}\to-x^{10}~.
\label{toto1}
\end{equation}
Since $w'$ is freely acting, there is no obvious trace of the non-Abelian gauge
group in M-theory, a fact that we are again going to interpret as
confinement. Actually, the two $S^3$'s of $A_1$ singularities
present in the space (\ref{so4}) for $\mu>0$ have become  $(S^2\times
S^1)/w'\I\cong \RP^3$ in the manifold
(\ref{so4res}).~\footnote{Since $(S^2\times
S^1)/w'\I$ is closed with  $\pi_1\left( (S^2\times
S^1)/w'\I\right) =\Z_2$, $(S^2\times
S^1)/w'\I\cong \RP^3$ by Thurston elliptization conjecture.} As explained
near Eq. (\ref{Mtrans}), such flop transitions $S^3\to 0 \to \RP^3$
characterize confinement of each $SU(2)$ factor.
Taking the limit $R_{10}\to 0$ in Eq. (\ref{metric}) sends us
again to an orientifold of type IIA on $(\mbox{resolved conifold})/w'$,
where there are no fixed points \ie no orientifold plane. By conservation of
charge, there are no D6 branes and thus no gauge group, as in
M-theory. Actually, this background is precisely the one given in
Eq. (\ref{res/orient}) that has been shown to describe the confining
phase of $SO(4)$: An orientifold of the resolved conifold with no RR
flux on $\RP^2$ since ${N_c\over 2}-2=0$. Thus the reduction $R_{10}\to 0$ of
the M-theory backgrounds (\ref{so4}) and (\ref{so4res}) is similar to
the reduction along $S^1\cong U(1)$ of $Spin(S^3)/\tau$ for the case
of an $SO(4)$ gauge theory.

However, when $\mu=0$ in Eq. (\ref{so4}), we can alternatively desingularize
the orbifold by passing into the
phase $\mu<0$ of $(T^*(S^3)\times S^1)/w\I$.~\footnote{Due to the
  action of the antiholomorphic involution $w$ on $T^*(S^3)$, we have now
  to distinguish the cases $\mu>0$ and $\mu<0$.} In that case, $w$ is
freely acting and
there is no obvious sign of the non-Abelian gauge group present for
$\mu>0$. Again, there are  flop transitions $S^3\to 0 \to S^3/w\cong \RP^3$ in
M-theory~\footnote{$S^3$ for $\mu>0$ and $S^3/w$ for $\mu<0$ are
  parametrized by $\Re e(z_i)$ and $\Im m(z_i)$ in Eq. (\ref{con}),
respectively.} and this third phase should also describe confinement
of $SU(2)^2$.~\footnote{As
will be seen in Section 4, when one considers the similar
  phase diagram where an arbitrary CY replaces $T^*(S^3)$ in the
  orbifold (\ref{so4}), the two phases where $SO(4)$ confines are distinguished
  by the number of neutral chiral multiplets.}
Sending  $R_{10}\to 0$ in Eq. (\ref{metric}), one obtains
a description in terms of
an orientifold of type IIA on $T^*(S^3)/w$, with no fixed points \ie
no orientifold plane. As before, conservation of RR
charge implies that there are no D6 branes and thus no gauge group, as
in M-theory. This orientifold background is very different from the one
considered in the previous paragraph that
describes accurately confinement of $SO(4)$ once closed string
worldsheet instanton corrections are taken into account. Locally, when
passing from $\mu>0$ to $\mu<0$ in the orientifold of type IIA on
$T^*(S^3)/w$, a transition $S^3\to 0\to S^3/w\cong \RP^3$ occurs
instead of $S^3\to 0 \to\RP^2$ we had before. In the phase $\mu<0$, one
then has to compute corrections, eventually by considering a dual
description relevant for calculations.

\vspace{.3cm}

\noindent {\em \large A new duality in type IIA}

\vspace{.2cm}

\noindent
The duality of Eq. (\ref{atrans}),
where branes on $S^3$ and flux on $S^2$ are interchanged,
is more easily understood when lifted to M-theory \cite{Atiyah:2000zz}.
Starting again from M-theory on $Spin(S^3)/\sigma$, we are going
to consider now a KK reduction along a different M-theory
circle and obtain a new duality conjecture in type IIA.

Let us define another subgroup of isometry that acts as
\begin{equation}
U(1)'~~:\quad (u_1,u_2)\to (e^{i\beta} u_1,e^{i\beta}  u_2)~,
\label{U'}
\end{equation}
whose fixed point set is given in Eq. (\ref{lens}).
For $V<0$, this set is a Lens space $S^3/\sigma$ and actually the
whole 7-manifold satisfies
$Spin(S^3)/\sigma=Spin(S^3/\sigma)\cong \R^4\times S^3/\sigma$, where
$U(1)'$ acts only on $\R^4$.
Thus, upon KK reduction along $S^1\cong U(1)'$, the type IIA geometry
becomes $T^*(S^3/\sigma)\cong \R^3\times S^3/\sigma$,
where a D6 brane is wrapped on the base $S^3/\sigma$.
An equation for $S^3/\sigma$ is (\ref{s3mu}), where $\mu$ is now identified
with $-V>0$ and a $\Z_{N_c}$ modding action on $x_i$ is understood.
The manifold $T^*(S^3/\sigma)$ is thus the deformed conifold of
Eq. (\ref{con}) on which the modding action on $\Re e(z_i)$ is lifted
to $z_i$ by holomorphicity:
\begin{equation}
z_1^2+ z_2^2+ z_3^2+ z_4^2= -V~,\quad \mbox{where}\quad
\left\{ \begin{array}{l}
\pamatrix{z_1 \cr z_2}\equiv
\pamatrix{\cos(\omega) &
\mbox{\footnotesize -\!}\sin(\omega) \cr
\sin(\omega) &
\phantom{\mbox{\footnotesize -\!}}\cos(\omega)}\pamatrix{z_1 \cr z_2} \\
\pamatrix{z_3 \cr z_4}\equiv
\pamatrix{\cos(\omega) &
\mbox{\footnotesize -\!}\sin(\omega) \cr
\sin(\omega) &
\phantom{\mbox{\footnotesize -\!}}\cos(\omega)}\pamatrix{z_3 \cr z_4}
\end{array}\right.~,
\label{conlens}
\end{equation}
with $\omega=2\pi/N_c$.
Clearly, we thus have
$T^*(S^3/\sigma)=T^*(S^3)/\sigma$ where we use
the same symbol $\sigma$ to refer to an orbifold
  action on the whole manifold or restricted to $S^3$.
To make contact with the notation of Eq. (\ref{spin}), the relations between
the coordinates of the Lens spaces in $Spin(S^3)/\sigma$ and
$T^*(S^3/\sigma)$ are $u_3=x_1+ix_2$, $u_4=x_3+ix_4$.
In this phase, the massless spectrum in M-theory
and type IIA contains a chiral field whose scalar
is $\Upsilon\equiv C+i\mu/N_c$, where $C$ is the expectation
value of the 3-form on $S^3/\sigma$ whose volume is
$-V\equiv \mu$. In type IIA, there is also an $\N=1$ $U(1)$ vector
multiplet on the worldvolume of the brane.

For $V>0$, $Spin(S^3)/\sigma \cong \R^4/\Z_{N_c}\times S^3$, where
the 3-sphere is defined in Eq. (\ref{s3M}). Since on this $S^3$, $U(1)'$
acts freely, the 3-sphere is a Hopf
fibration over an $S^2$, whose fiber is $S^1\cong U(1)'$.
Therefore, $S^3$ is reduced in type IIA to an $S^2$ with
one unit of
RR two-form flux on it, while the $\R^4/\Z_{N_c}$ fibration gives rise
to an $A_{N_c-1}$ singularity over the $S^2$. Physically, the flux on
$S^2$ is interpreted as a magnetic FI term
\cite{Antoniadis:1996vb,Partouche:1997yp} that breaks $\N=2\to \N=1$
in string theory \cite{Taylor:2000ii} (see also \cite{Kiritsis:1997ca,
  Mayr:2001hh} for supersymmetry breaking in string theory).
To be more precise concerning the geometry,
since the KK reduction of $Spin(S^3)$ along
$S^1\cong U(1)'$ is the resolved conifold, the reduction of
$Spin(S^3)/\sigma$ gives in type IIA the resolved
conifold in which the $\R^4$ fiber over $S^2$ is modded by $\Z_{N_c}$.
Since in Eq. (\ref{asres}) $(\xi_1,\xi_2)$ parametrizes
$\CP^1\cong S^2$, while the $z_i$'s constrained by the
equations parametrize $\R^4$, the type IIA background takes the form:
\begin{equation}
\left\{
\begin{array}{l}
(z_1+iz_2)\xi_1 - (z_3+iz_4)\xi_2 = 0\\
(z_3-iz_4)\xi_1 + (z_1-iz_2)\xi_2 = 0
\end{array}
\right.
~,\quad \mbox{where}\quad
\left\{
\begin{array}{l}
z_1\pm iz_2\equiv e^{\pm i\omega}(z_1\pm iz_2)\\
z_3\pm iz_4\equiv e^{\pm i\omega}(z_3\pm iz_4)
\end{array}\right. ~.
\label{resmod}
\end{equation}
In this phase, the massless spectrum in M-theory and type IIA contains a
non-perturbative $SU(N_c)$ gauge group. The theta angle and gauge
coupling are given in type IIA by the complexified \Ka modulus of $S^2$:
$T\equiv B+i \,\mbox{vol}(S^2)$,
where $B$ is the NS-NS
2-form flux on $S^2$. In M-theory variables, $T\equiv C+i V$,
where $C$ is the flux of the 3-form on $S^3$ whose volume is $V$.
Note that in type IIA, the dimensional reduction of the RR 3-form
on $S^2$ gives rise to
an additional perturbative $\N=1$ vector multiplet. Combined with the
non-Abelian factor,
the string theory gauge group is thus $U(N_c)$.

In M-theory, the two phases $V<0$
and $V>0$ are dual to each other in the sense that one passes  from
one to the other by a change of energy scale from the IR $(V<0)$ to the UV
$(V>0)$ of the corresponding gauge theory.
Therefore, the type IIA descriptions to which they descend upon
dimensional reduction should also be dual in that sense. Note that this is
consistent with the fact that the two backgrounds of Eqs. (\ref{conlens})
and (\ref{resmod}) are related by a conifold transition since
the modding actions on the $z_i$'s are equivalent.
Hence, we are led to conjecture:

\vspace{.2cm}

\noindent {\em In type IIA,  a $\CP^1$ of $A_{N_c-1}$ singularity with one unit
of RR flux through it is dual to a Lens space $S^3/\Z_{N_c}$ with one D6-brane
wrapped on it. They describe the UV and IR physics of a pure $\N=1$ $SU(N_c)$
gauge theory, respectively.}

\vspace{.2cm}

In fact, the $SU(N_c)$ gauge group explicit in the UV confines in the
IR, while the diagonal $U(1)$ remains spectator through the
transition. As in the large $N$ duality conjecture of \cite{Vafa:2000wi},
$\Upsilon$ should be interpreted physically
as the lowest component of the gaugino condensate superfield,
$g_s \mbox{Tr} W^\alpha W_\alpha=\Upsilon+\cdots$, where $W^\alpha$ is the
supersymmetric field strength of the $SU(N_c)$ factor.
In particular, it would be very
interesting to rederive in the context of the above duality conjecture
the relation between $T$ and $\Upsilon$:
\begin{equation}
(e^\Upsilon-1)^N=e^{-T}~.
\end{equation}

The generalization of the duality conjecture to the $SO(N_c)$ groups
(for even $N_c\ge 8$) is straightforward. Consider M-theory on
$Spin(S^3)/\tau$. For $V<0$, the reduction along $S^1\cong U(1)'$
gives the manifold $T^*(S^3/\tau)$, where $S^3/\tau$ is the Lens space
associated to the dihedral group, and a D6 brane is wrapped on
$S^3/\tau$. The action on $u_{3,4}$ of the two generators of $\tau$ in
Eq. (\ref{tau}) are translated to an action on $x_i$ in
Eq. (\ref{s3mu}) due to the identification $u_3=x_1+ix_2$,
$u_4=x_4+ix_5$. Then, holomorphicity implies that
$T^*(S^3/\tau)=T^*(S^3)/\tau$ in this phase. The
defining equation of this manifold  is given
in Eq. (\ref{conlens}), where $\omega=2\pi/(N_c-4)$, and a second
identification
\begin{equation}
(z_1,z_2,z_3,z_4)\equiv (z_3,-z_4,-z_1,z_2)
\label{ide}
\end{equation}
is understood.
For $V>0$, the reduction gives an $S^2$ with one unit of RR flux and a
$D_{N_c/2}$ singular fibration over it. Globally, this space is given
in Eq. (\ref{resmod}) with the additional identification (\ref{ide}).
Finally,  the previous considerations can also be
applied to the $E_{6,7,8}$ groups.

\vspace{.3cm}

\noindent {\em \large Introducing flavor in the M-theory geometry?}
\vspace{.2cm}

\noindent
We have seen how two phases associated to the UV and IR physics in
a pure SYM theory realized in type IIA can be lifted to M-theory. In
Section 2.2, confinement in presence of massless quarks multiplets
have also been considered in type IIA string theory. Our aim is now to propose
an M-theory description that could realize such a SYM theory coupled to
quarks.

With this in mind, we first consider
another isometry subgroup considered in
\cite{Aganagic:2001jm,Atiyah:2001qf}, $U(1)''$, that corresponds in
the model of Eq. (\ref{spin}) to the
action
\begin{equation}
U(1)''~~:\quad (u_2,u_3)\to (e^{i\gamma} u_2, e^{i\gamma}u_3) ~.
\label{U''}
\end{equation}
If we consider the
discrete subgroup of $U(1)''$ generated by
\begin{equation}
\rho~~:\quad (u_2,u_3)\to (e^{2i\pi/N_f} u_2, e^{2i\pi/N_f}u_3)~,
\end{equation}
we can construct the orbifold $Spin(S^3)/\Z_{N_f}$. The fixed point set of
$\rho$ and $U(1)''$ is
\begin{equation}
u_2=u_3=0\quad , \quad |u_1|^2-|u_4|^2=V~,
\label{s2s1M}
\end{equation}
whose topology is $\Co\times S^1$ and is the base of
an $\R^4/\Z_{N_f}$ fibration~\footnote{For $V>0$, $\Co\times S^1$ is
  parametrized by
$u_4$ and the phase of $u_1$, while for $V<0$, the roles of $u_4$ and $u_1$
is reversed.}.
If $\Co$ was replaced by a finite volume $\CP^1$,
since $b_1(\CP^1\times S^1)=1$, this would
generate an $SU(N_f)$ gauge theory with one chiral field in the adjoint
representation. However, since we deal with the infinite volume base
$\Co\times S^1$,
these fields are frozen and we are left
with an $SU(N_f)$ global symmetry.
{}From the results of \cite{Aganagic:2001jm,Atiyah:2001qf}, the
dimensional reduction of
$Spin(S^3)/\rho$ along $S^1\cong U(1)''$ gives a type IIA string
theory on the space $\Co^3$, where $N_f$ coincident D6 branes are
wrapped on a SLAG of topology $\Co\times S^1$, thus producing an
$SU(N_f)$ global symmetry in type IIA.

{}For integers $N_f>N_c$, let us denote by $\tilde \sigma$ the generator
similar to $\sigma$, where $N_c$ is replaced by $N_f-N_c$:
\begin{equation}
\tilde \sigma~~:\quad (u_3,u_4)\to (e^{2i\pi/(N_f-N_c)} u_3,
e^{2i\pi/(N_f-N_c)}u_4) ~.
\end{equation}
Since $\tilde \sigma$ and $\rho$ commute, the group they generate is
of finite order. We shall denote it by $\{\tilde \sigma,\rho\}$.
We now consider the space $Spin(S^3)/\{\tilde \sigma,\rho\}$ as a
good candidate for describing a ``magnetic'' $\N=1$ $SU(N_f-N_c)$ gauge theory
with $N_f$ flavors of  quarks.

For $V>0$, the singular point
set of $\tilde \sigma$ is subject to a quotient by $\rho$
implying an identification $u_2\equiv e^{2i\pi/N_f}u_2$ in Eq. (\ref{s3M}).
As a result, the 3-space $S^3/\Z_{N_f}$ has singular points:
\begin{equation}
u_2=u_3=u_4=0\quad ,\quad |u_1|^2=V~,
\label{s1M}
\end{equation}
which is an $S^1$. Similarly, the singular point set of $\rho$ is subject
to a quotient by $\tilde \sigma$, implying an identification
$u_4\equiv e^{2i\pi/(N_f-N_c)}u_4$ in Eq. (\ref{s2s1M}). Therefore,
the resulting fixed point set $\Co/\Z_{N_f-N_c}\times S^1$ is singular along
the $S^1$ given in (\ref{s1M}). Notice
that the group element $\tilde \sigma \rho$ does not introduce new singular
points since its fixed points are also
given by (\ref{s1M}). To summerize, the orbifold points of
$Spin(S^3)/\{\tilde \sigma,\rho\}$ consist of two singular 3-spaces
$S^3/\Z_{N_f}$ and $\Co/\Z_{N_f-N_c}\times S^1$ glued together
precisely along their $S^1$ of singularities,
as depicted in Figure 7.
\begin{figure}[!h]
\begin{picture}(100,280)
\centerline{\hbox{\psfig{figure=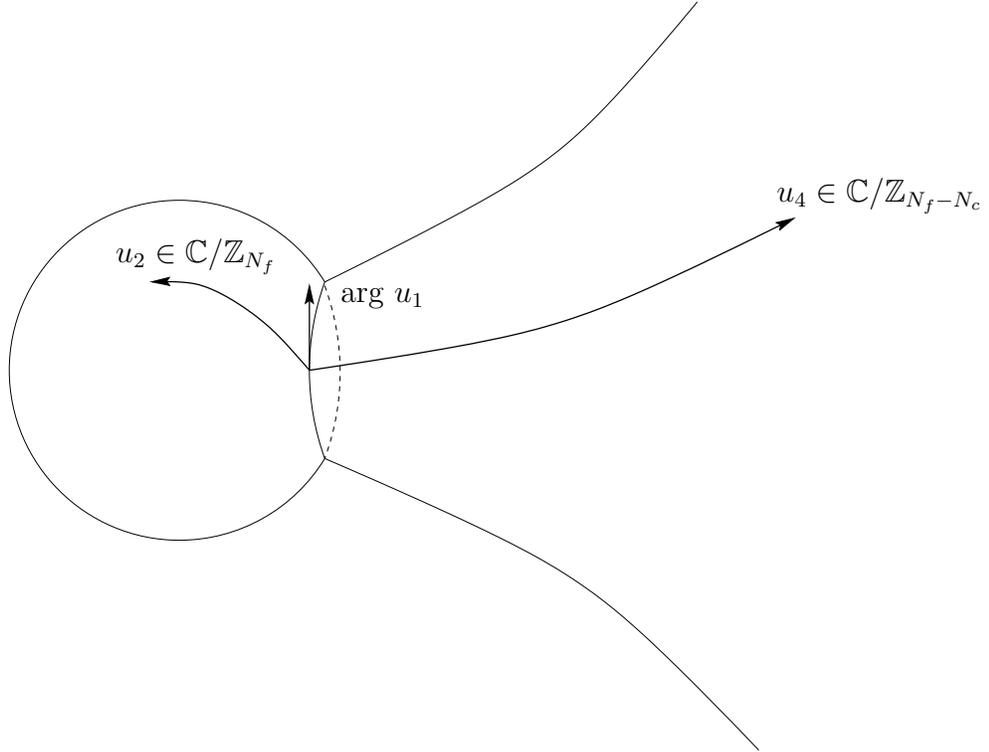,height=10cm}}}
\put(-74,208){$u_4\in \Co/\Z_{N_f-N_c}$}\put(-324,185){$u_2\in
  \Co/\Z_{N_f}$}\put(-239,170){arg $u_1$}
\end{picture}
\caption{{\bf Figure 7:}
{\footnotesize \em An $S^3/\Z_{N_f}$ of $A_{N_f-N_c-1}$
  singularities and a $\Co/\Z_{N_f-N_c}\times S^1$ of $A_{N_f-1}$
  singularities. The two bases intersect along an $S^1$.}}
\label{f6}
\end{figure}

{}From an M-theory point of view, it is
not yet known how to determine the massless spectrum of this background.
Actually, there should be gauged $SU(N_f-N_c)$
and global $SU(N_f)$ symmetries arising from the $\R^4/\Z_{N_f-N_c}$
and $\R^4/\Z_{N_f}$ fibrations over the bases $S^3/\Z_{N_f}$ and
$\Co/\Z_{N_f-N_c}\times S^1$, respectively. In addition, there could be
massless matter localized at the intersection $S^1$ of the bases,
in the bifundamental representation of $SU(N_f-N_c)\times SU(N_f)$,
that could be interpreted as $N_f$ flavors of massless quarks
$(q,\tilde q)$.
In that case, this geometry would be
a good candidate for describing the lift to M-theory of the
``magnetic'' theory described in Section \ref{geoA}: In type IIA on $T^*(S^3)$,
$N_f-N_c$ and
$N_f$ D6 branes are wrapped on $S^3$ and $\Co\times S^1$ that are
intersecting along $S^1$.  However, we have not been
able to find a KK reduction of M-theory on $Spin(S^3)/\{\tilde \sigma,\rho\}$
that would send us to this type IIA picture.

For $V<0$, on one hand
$\tilde \sigma$ and $\tilde \sigma\rho$ are freely acting.
On the over hand, the fixed
point set of $\rho$ is given by Eq. (\ref{s2s1M}) with the
identification $u_4\equiv e^{2i\pi/(N_f-N_c)}u_4$, giving rise to a topology
$\Co\times S^1/Z_{N_f-N_c}\cong \Co\times S^1$. Hence, we are left with an
$SU(N_f)$ global symmetry in M-theory associated with the $\R^4/\Z_{N_f}$
fibration over $\Co\times S^1$. Again, this could be interpreted as the
IR physics of the confining $SU(N_f-N_c)$ gauge theory with matter.
Upon KK reduction, this would
be consistent with the type IIA configuration described after Eq. (\ref{res}):
$N_f$ D6 branes wrapped on $\Co\times S^1$ in the resolved conifold
with $N_f-N_c$ units of RR flux through the $S^2$.

As a final remark, M-theory on $Spin(S^3)/\{\tilde \sigma,\rho\}$ can be mapped
to various type IIA models that could shed some light on the physics
described by this geometry. As an example, for $V>0$, the KK reduction
on $S^1\cong U(1)''$ results in a type IIA string theory on a
background $\Co^2/\Z_{N_f-N_c}\times \Co$. In addition, there are
$N_f$ coincident D6 branes wrapped on a SLAG $\Co/\Z_{N_f-N_c}\times S^1$. This
situation is treated in \cite{Aganagic:2001jm} for a single wrapped D-brane.

\section{Connecting/disconnecting and vanishing 3-spheres}
\label{without}

In this section we consider a compact M-theory background where various
transitions take place such as the connection of two disjoint 3-spheres of
singularities. We first describe the M-theory construction and
determine the massless spectrum in each phase. The models we shall
consider are similar to those studied in \cite{Kaste:2001iq}.
Then we shall give
quantitative and physical interpretations of the transitions by mapping
them to dual descriptions involving some local geometries or brane
configurations considered in the previous sections.

\subsection{The M-theory geometrical setup}
\label{model1}

We start by considering  a  two-parameter sub-set of CY's $\C_1$ in the
family  $\CP^4_{11222}[8]$. The
Hodge numbers of the threefolds are $h_{11}=2$ and $h_{12}=86$. The
defining polynomial
\begin{equation}
p_1 \equiv z_6^4(z_1^8 + z_2^8-2\phi z_1^4z_2^4)+(z_3^2-\psi z_2^4z_6^2)^2
+z_4^4+z_5^4=0\; ,
\label{p1}
\end{equation}
is written in terms of two complex parameters $\psi$, $\phi$ and the
projective coordinates subject to two scaling
actions $\Cs$, whose weights are given in the following table:
\begin{equation}
\begin{tabular}{c|cccccc}
 & $z_1$ & $z_2$& $z_3$&$ z_4$& $z_5$&$ z_6$ \\ \hline
$\Cs_1$& $0$&$0$&$1$&$1$&$1$&$1$ \\
$\Cs_2$& $1$&$1$&$0$&$0$&$0$&$-2$
\end{tabular} \label{scalings}
\end{equation}
The presence of the variable $z_6$ together with the second scaling action
is due to the blow up of the $\Z_2$ singularity sitting at $z_1=z_2=0$
in the ambient $\CP^4_{11222}$: In
the resulting toric space, there are  excluded  sets
\begin{equation}
(z_1,z_2)\neq (0,0) \quad \mbox{and}\quad
(z_3,z_4,z_5,z_6)\neq(0,0,0,0)\; .
\label{forbid}
\end{equation}
Notice that we have chosen to work in a slice of the complex structure moduli
space by fixing to zero most of the coefficients of the monomials allowed
by the
$\Cs$ actions.

A CY in the family $\C_1$ is singular when the equation
(\ref{p1}) is non-transverse, \ie\ when $p_1=0$, $dp_1=0$.
This happens only in charts where $z_2$ and $z_6$ do not vanish so that we can
rescale them to 1. Then, the singularities occur
\begin{equation}
\begin{array}{lllll}
\mbox{for any $\psi$,}&  \mbox{at }\phi=+1 &~:~ & (i^k,1,\pm
\sqrt{\psi},0,0,1)& ,(k=0,...,3)~,\\
& \mbox{or }\phi=-1 &~:~ & (i^k e^{i\pi/4},1,\pm \sqrt{\psi},0,0,1)
&,(k=0,...,3)~,\\
\mbox{and for any $\phi$,}& \mbox{at }\psi=\pm \sqrt{\phi^2-1} &~:~ &
(\phi^{1/4}i^k,1,0,0,0,1)& ,(k=0,...,3)~.
\label{nodes}
\end{array}
\end{equation}
It happens that the determinant of second
derivatives $\det(\partial_A\partial_Bp_1)$ $(A,B=1,3,4,5)$ has two
vanishing eigenvalues at these points. Therefore, these isolated
singularities are not nodal points. We shall come back to this remark later.
Since the maps
$(\phi, z_1)  \to (-\phi, e ^{ i\pi/4} z_1)$ and  $(\psi, z_6)
\to (-\psi,i z_6)$ leave $p_1$
invariant, the threefolds associated to $(\pm \phi, \pm \psi)$ are one and
only one. Thus, from the CY point of view, we could consider
complex parameters $\psi$ and $\phi$ such that $\Re e(\phi),\Re
e(\psi)  \geq 0$ only.

However, we are interested in $\N=1$ compactifications of M-theory on
orbifolds of the form similar to Eq. (\ref{so4}):
\begin{equation}
\G_1={\C_1\times S^1 \over w\I}\; ,
\label{g1}
\end{equation}
where the involution $w\I$ acts simultaneously on the CY and $S^1$ as:
\begin{equation}
w:~ z_i\to \bar{z}_i\quad (i=1,...,6)~~, \qquad \I :~x^{10}\to
-x^{10}\; . \label{s}
\end{equation}
Actually, for $w$ to be a
symmetry of $\C_1$, we restrict now $\phi$ and $\psi$ to real
values. From the point of view of $\G_1$, since $z_1 \to e ^{
  i\pi/4} z_1$ and  $z_6\to iz_6$ do not commute with $w$, $\phi$ and
$-\phi$ as well as $\psi$ and $-\psi$ are no longer equivalent.

We now determine the fixed point set of the orbifold. Since these
points satisfy $(z_i,\bar z_i,x^{10})\equiv
(\bar z_i,z_i,-x^{10})$, they are described by two copies of the
special Lagrangian 3-cycle $\Sigma$ in $\C_1$ fixed by $w$. One copy
sits at $x^{10}=0$, while the second sits at $x^{10}=\pi
R_{10}$.~\footnote{In general, $\Sigma$ itself is a union of disconnected
3-cycles. When the first Betti number $b_1$ of one of these
components is non-vanishing, it is conjectured \cite{Joyce2} that it
is possible to blow up the singularities lying on it (if some other
condition is also satisfied). In that case,
the holonomy of the resulting space is $G_2$. When $b_1$ of each
component vanishes, none of them can be desingularized and the holonomy is
the semi-product of $SU(3)$ and $\Z_2$. However, the
four-dimensional physics of M-theory on these spaces is in both cases
$\N=1$ and we shall refer to them  as $G_2$ orbifolds.}
Defining $x_i$ and $y_i$ to be the real and imaginary parts of $z_i$,
$\Sigma$ is given by $y_i=0$ and Eq. (\ref{p1}) for real
unknowns $x_i$:
\begin{equation}
x_6^4(x_1^8+x_2^8-2\phi x_1^4x_2^4)+(x_3^2-\psi x_2^4x_6^2)^2+x_4^4+x_5^4=0
 \; .
\label{p1r}
\end{equation}
Note that $x_6$ cannot vanish in this equation, since otherwise it would imply
$x_{3,4,5}=0$ as well, which is forbidden (see
Eq. (\ref{forbid})). Thus, we can rescale $x_6$
to 1. Similarly, $x_2$ can always be rescaled to 1 since $x_2=0$
would also imply $x_{1}=0$.
In Eq. (\ref{p1r}), the remaining unknowns are not projective anymore
and we can solve for $x_1^4$:
\begin{equation}
x_1^4 =\phi \pm\sqrt{\phi^2-\left[1+(x_3^2-\psi)^2+x_4^4+x_5^4\right]}\; .
\label{def3}
\end{equation}
Clearly, for $\phi<1$, there is no real solution for $x_1^4$ and $\Sigma$
is therefore empty. For $\phi=1$, there are solutions
for $x_3^2=\psi$, $x_4=x_5=0$, $x_1^4=1$, when $\psi\geq 0$ and $\Sigma$
consists of four points
\begin{equation}
\Sigma=\{(\pm 1,1,\pm \sqrt{\psi},0,0,1)\}\, .
\end{equation}
At these points, the CY is also singular, as can be seen from
Eq. (\ref{nodes}).

The situation for $\phi>1$ is more delicate. By defining
the variables
\begin{equation}
X_1= x_1^4-\phi \;\; ,\quad X_3=x_3^2-\psi\;\; , \quad X_j=x_j^2 \;
\mbox{sign}(x_j)\; , \quad
(j=4,5)\; ,
\label{u}
\end{equation}
the equation for $\Sigma$ turns out to be
\begin{equation}
X_1^2+X_3^2+X_4^2+X_5^2=\phi^2-1\; , \label{Sph}
\end{equation}
which describes an $S^3$ of radius $\sqrt{\phi^2-1}$.
However, whereas $X_{4,5}$ and $x_{4,5}$ are in one-to-one
correspondence, the map from $x_1$ to $X_1$ is two-to-one as can be seen
from the relation
$-\sqrt{\phi^2-1}\leq x_1^4-\phi\leq\sqrt{\phi^2-1}$ that gives two
disconnected sets of solutions
\begin{equation}
\begin{array}{lll}
&&0<(\phi-\sqrt{\phi^2-1})^{1/4}
\leq x_1\leq (\phi+\sqrt{\phi^2-1})^{1/4}\; \\ &\mbox{ and }&
\phantom{0}-(\phi+\sqrt{\phi^2-1})^{1/4}
\leq x_1\leq -(\phi-\sqrt{\phi^2-1})^{1/4}<0\; .
\end{array}
\label{xx}
\end{equation}
Thus, the total fixed point set $\Sigma$ consists of the disjoint union of
two isomorphic sets $\Sigma_+$ and $\Sigma_-$, where
$x_1$ is always strictly positive in the former and strictly negative in the
latter. We shall concentrate now on $\Sigma_+$, keeping in mind
that one obtains $\Sigma_-$ from $\Sigma_+$ by changing $x_1\to -x_1$.
Finally, we have to translate the fixed point coordinate $X_3$ into
$x_3$.  From the inequalities
$-\sqrt{\phi^2-1}\leq x_3^2-\psi\leq\sqrt{\phi^2-1}$, the following
discussion arises, which is illustrated by Figure 8 that represents the
projection $x_4=x_5=0$ of $\Sigma_+$ in the plane $(x^1,x^3)$:
\begin{figure}[!h]
\centerline{\hbox{\psfig{figure=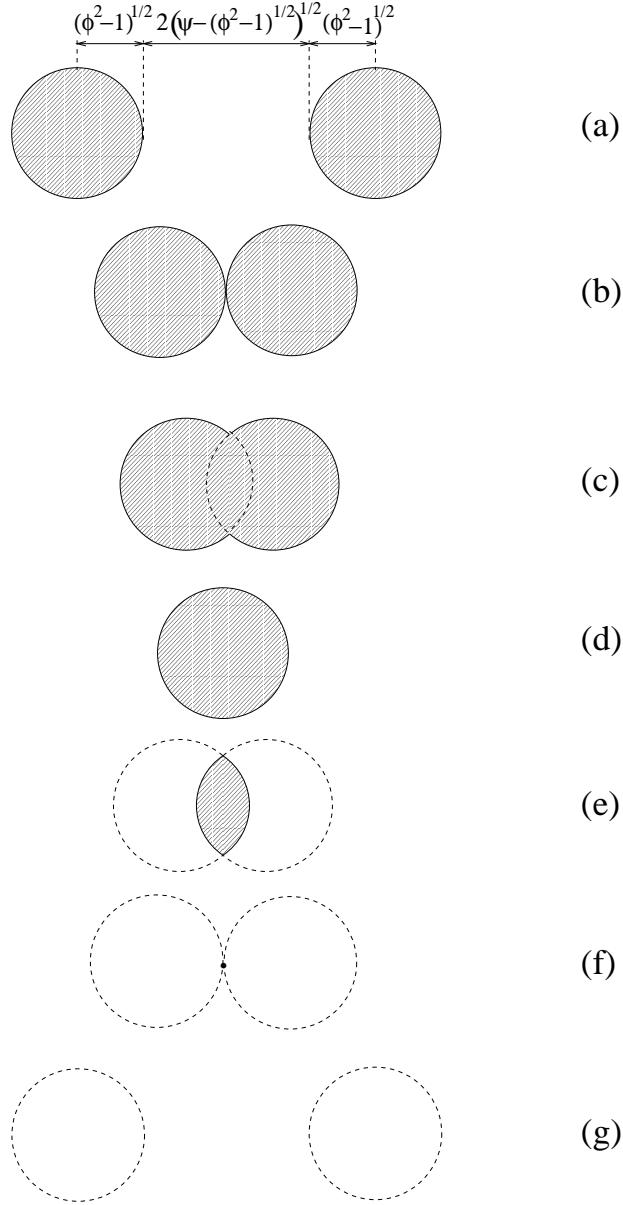,height=16cm}}}
\caption{{\bf Figure 8:}
{\footnotesize \em {\bf (a)}  $\Sigma_+$ is composed of two
  3-spheres that apprach each other. {\bf (b)}
  They intersect at a singular point of the manifold. {\bf (c,d,e)}
They are connected. Topologically, it is equivalent to a single
$S^3$. {\bf (f)} The $S^3$ shrinks to a singular point of the
manifold. {\bf (g)} $\Sigma_+$ is empty.}}
\label{f8}
\end{figure}
\vskip .2in

$\bullet$ For $\psi > \sqrt{\phi^2-1}$: There exist two disjoint sets
of solutions for $x_3$
\begin{equation}
\begin{array}{rll}
&&0<(\psi-\sqrt{\phi^2-1})^{1/2}
\leq x_3\leq (\psi+\sqrt{\phi^2-1})^{1/2}\; \\ &\mbox{ or }&
\phantom{0}-(\psi+\sqrt{\phi^2-1})^{1/2}
\leq x_3\leq -(\psi-\sqrt{\phi^2-1})^{1/2}<0\; ,
\end{array}
\label{xxx}
\end{equation}
so that $\Sigma_+$ is composed of two disconnected 3-spheres (see
Figure 8(a)):
\begin{equation}
\mbox{for $\psi > \sqrt{\phi^2-1}$} \quad : \qquad \Sigma_+ = S^3 \cup S^3\; .
\end{equation}
\vskip .2in

$\bullet$ For $\psi = \sqrt{\phi^2-1}$: The two previous $S^3$'s intersect at
one point $(\phi^{1/4}, 1,0,0,0,1)$, so that  $\Sigma_+$ is singular at
this point (see Figure 8(b)).
Note that the CY is also singular at the same point (see Eq. (\ref{nodes})).
\vskip .2in

$\bullet$ For $-\sqrt{\phi^2-1}<\psi < \sqrt{\phi^2-1}$: In this
phase, the set
of solutions for $x_3$ takes the form
\begin{equation}
\begin{array}{rll}
&&0\leq x_3\leq (\psi+\sqrt{\phi^2-1})^{1/2}\; \\ &\mbox{ or }&
\phantom{0}-(\psi+\sqrt{\phi^2-1})^{1/2}
\leq x_3\leq 0\; ,
\end{array}
\label{xxxx}
\end{equation}
so that
\begin{equation}
\mbox{for $-\sqrt{\phi^2-1}<\psi < \sqrt{\phi^2-1}$} \quad :
\qquad \Sigma_+ = S^3 \; ,
\end{equation}
where this $S^3$ is actually the connected sum of the two 3-spheres we
had before (see Figures 8(c,d,e)).
\vskip .2in

$\bullet$ For $\psi = -\sqrt{\phi^2-1}$: The size of the previous
$S^3$ has vanished and we have a singular 3-cycle (see Figure 8(f)) :
\begin{equation}
\mbox{for $\psi = -\sqrt{\phi^2-1}$} \quad : \qquad \Sigma_+ =
\{(\phi^{1/4},1,0,0,0,1)\}\; ,
\end{equation}
which again corresponds to a singular point in the CY (see Eq.
(\ref{nodes})).
\vskip .2in

$\bullet$ Finally, for $\psi < -\sqrt{\phi^2-1}$: There is no solution for
$x_3$ and we have (see Figure 8(g)):
\begin{equation}
\mbox{for $\psi < -\sqrt{\phi^2-1}$} \quad : \qquad \Sigma_+ = \emptyset\; .
\end{equation}
\vskip .2in

To summarize,  at fixed $\phi>1$ and according to the values of $\psi$,
the fixed point set $\Sigma_+$ is:
\begin{equation}
\begin{array}{lll}
\psi>\sqrt{\phi^2-1},~~~ &
\Sigma_+=S^3\cup S^3 &\mbox{(disconnected union)}\; , \\
\psi=\sqrt{\phi^2-1},~~~ &
\Sigma_+=S^3\cup S^3 &\mbox{(intersecting at one singular point)}\; , \\
|\psi|<\sqrt{\phi^2-1},~~~ & \Sigma_+=S^3\#S^3\cong S^3 &
\mbox{(connected sum of the
  $S^3$'s)}\; , \\
\psi=-\sqrt{\phi^2-1},~~~ &
\Sigma_+=\{(\phi^{1/4},1,0,0,0,1)\}&\mbox{(one singular point)}\; ,\\
\psi<-\sqrt{\phi^2-1},~~~ & \Sigma_+= \emptyset \; &\mbox{(no fixed point)}\; .
\end{array} \label{ss1}
\end{equation}
This determines three phases in the $(\phi,\psi)$ plane drawn in
Figure 9, where
the topology of $\Sigma_+$ is represented. We now describe the massless
spectrum in each phase.
\begin{figure}[!h]
\centerline{\hbox{\psfig{figure=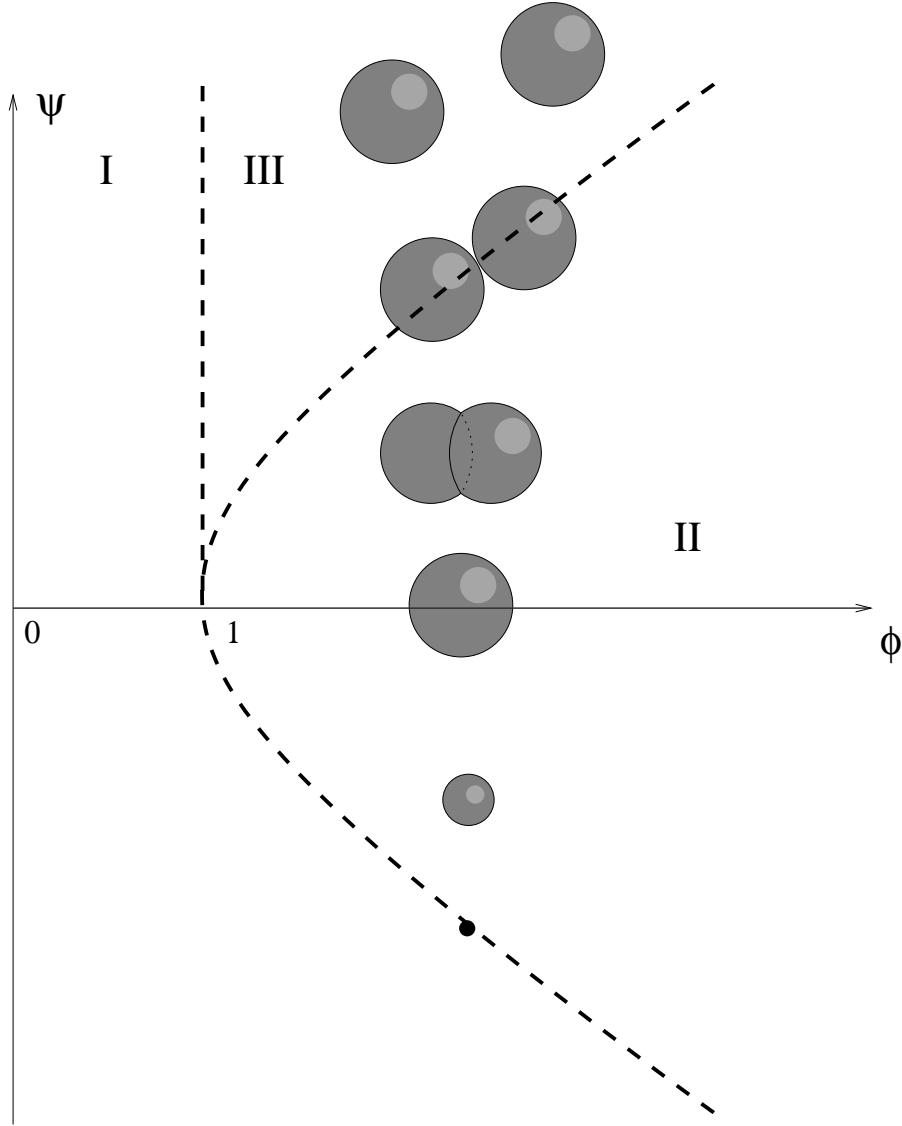,height=15cm}}}
\caption{{\bf Figure 9:} {\footnotesize \em Phase diagram of the model based on
  $\G_1$. $\Sigma_+$ is represented for various values of
  $(\phi,\psi)$. When non-empty, it is composed of 3-spheres. The
  transitions are physical and should occur at finite distance in
  moduli space.}}
\label{f99}
\end{figure}

\vspace{.3cm}

\noindent {\em \large -- Spectrum in phase I:
$\phi<1$ or $\phi\geq 1,\, \psi<-\sqrt{\phi^2-1}$}
\vspace{.2cm}

\noindent
In this phase, the compact
space is smooth and the four-dimensional massless spectrum is obtained
by dimensional reduction of the eleven-dimensional supergravity
multiplet \cite{Papadopoulos:1995da}. Besides the
$\N=1$ gravity multiplet, the resulting spectrum
contains $b_2$ vector multiplets and $b_3$
neutral chiral multiplets~\footnote{These multiplets are in fact linear
  multiplets due to a Peccei-Quinn symmetry remanent of the gauge invariance $C
\to C+d\Lambda_2$ of the eleven dimensional supergravity 3-form.},
where $b_{2}$ and $b_{3}$ are the Betti
numbers of $\G_1$.
{}From our construction of $\G_1$, $b_{2,3}$ can be expressed in terms of
$h_{11}$ and $h_{12}$, the Hodge numbers of $\C_1$. Actually, a
2-cycle on the orbifold arises from a 2-cycle in $\C_1$ which is even under
$w$. Let us define the number of even (odd) 2-cycles in $\C_1$ to be
$h_{11}^+$ ($h_{11}^-$). On the other hand,
the product of $S^1$ by any of the $h_{11}^-$ odd 2-cycles of $\C_1$
gives rise to a 3-cycle on $\G_1$. In
addition, the 3-cycles even under $w$ in $\C_1$ remain in the
orbifold. Noticing that there are as many even as odd 3-cycles in
$\C_1$, one obtains
\begin{equation}
b_2=h_{11}^+ \quad , \quad b_3= \frac{h_{30}+h_{03}}{2} +
\frac{h_{21}+h_{12}}{2}  +h_{11}^-=1+h_{12}+h_{11}^-\;.
\label{b23}
\end{equation}
In the present case, there are $h_{11}=2$ cohomology classes on the
CY. The first one being the pullback on $\C_1$ of
the \Ka form of $\CP^1_{11222}$, it is odd under $w$. The second
one is Poincar{\'e} dual of the holomorphic blow up  $\CP^1$ at
$(z_1,z_2)=(0,0)$, which is also odd under $w$.
As a result, we have
$h_{11}^+=0$ and the Betti numbers of $\G_1$ are
\begin{equation}
b_2=0 \qquad \mbox{ and } \qquad b_3=1+86+2=89\; .
\label{q}
\end{equation}
Thus, we have
\begin{equation}
\mbox{89 chiral multiplets } \qquad \mbox{and} \qquad
\mbox{no gauge group .}
\end{equation}

\vspace{.3cm}

\noindent {\em \large -- Spectrum in phase II:  $\phi>1, \,
  |\psi|<\sqrt{\phi^2-1}$}
\vspace{.2cm}

\noindent
In this phase, the massless spectrum still contains the states arising
from the reduction from eleven to four dimensions of the supergravity
multiplet.  This gives
the $\N=1$ gravity multiplet together with 89 chiral
multiplets.
In addition \cite{Acharya:2000gb,Kaste:2001iq}, there are states localized on
the fixed point set
$\left(\Sigma_+\cup \Sigma_-\right) \times \{0,\pi R_{10}\}$, which is composed
of four disconnected copies of $S^3\#S^3\cong S^3$.
At any point of one of these 3-spheres,
the geometry looks like $\R^4/\Z_2\times S^3$, where $S^3$ is
parametrized by the $x_i$'s, while $\R^4/\Z_2$ accounts for the
$y_i$'s and $x^{10}$. Thus, there is an $SU(2)$ gauge group arising
from M-theory on $\R^4/\Z_2$ further compactified to four dimensions on
$S^3$.
Since $b_1(S^3)=0$, this results in an  $\N=1$ vector multiplet of $SU(2)$
with no adjoint matter for each $S^3$. Including the spectrum arising
from the four disjoint 3-spheres in $\Sigma\times \{0,\pi R_{10}\}$, we obtain
\begin{equation}
\mbox{1 vector multiplet of }SU(2)^4 \qquad \mbox{and} \qquad
\mbox{89 neutral chiral multiplets .}
\end{equation}
Note that since $b_1(S^3)=0$, it is not possible to desingularize the $A_1$
singularities to obtain a smooth $G_2$-holonomy manifold.

\vspace{.3cm}

\noindent {\em \large -- Spectrum in phase III:  $\phi>1, \,
  \psi>\sqrt{\phi^2-1}$}
\vspace{.2cm}

\noindent
This case is treated as the previous one. The only difference is
that each $S^3$ in phase II is replaced by
a disjoint union $S^3\cup S^3$ in phase III. Therefore the massless
spectrum is
 \begin{equation}
\mbox{1 vector multiplet of }SU(2)^8 \qquad \mbox{and} \qquad
\mbox{89 neutral chiral multiplets .}
\end{equation}

\subsection{Brane and field theory interpretation of the geometrical phases}

We have seen that there are three geometrical  phases in Figure 9. Two
transitions, II$\to$I and  III$\to$I, correspond to passing from
the UV to the IR of the SYM theories that confine. This was
explained in a non-compact model in \cite{Atiyah:2000zz,Sinha:2000ap},
reviewed in Section 2.3, and generalized to compact $G_2$ manifolds of the form
$(CY\times S^1)/w\I$ in \cite{Kaste:2001iq}. We now have to understand
the third transition, III$\to$II. We do not know how to determine the
physics it describes from a pure M-theoretic point of view, due to
additional massless matter that may occur at the transition.
Therefore, we shall map the local geometries involved in each
transition in M-theory to local configurations of branes in type IIA which
are similar to those considered in Section 2.1.

We first reduce the M-theory background in phase II to a type IIA
orientifold on $\C_1/w$ by identifying the size $R_{10}$ of $S^1$
with the string coupling. This is similar to what we did for the space
of Eq. (\ref{so4}). In $\G_1$, the set $\Sigma_+\times\{0,\pi
R_{10}\}$ is composed of two 3-spheres that descend in type IIA to a
single $S^3$ on which two D6 branes on top of an orientifold sixplane O6
(as well as two mirror branes)
are wrapped \cite{Kachru:2001je}.~\footnote{We only consider the
  vicinity of the component $\Sigma_+$ of $\Sigma$; the discussion for
  the second one $\Sigma_-$ is independent and similar.}
Two  T-dualities of the form described
in Section 2.2 take this, locally,
to the type IIA brane configuration in Figure 10:
NS and NS' branes (see Eq. (\ref{0123}))
are separated along an O4 plane in the direction
$x^6$.
\begin{figure}[!h]
\centerline{\hbox{\psfig{figure=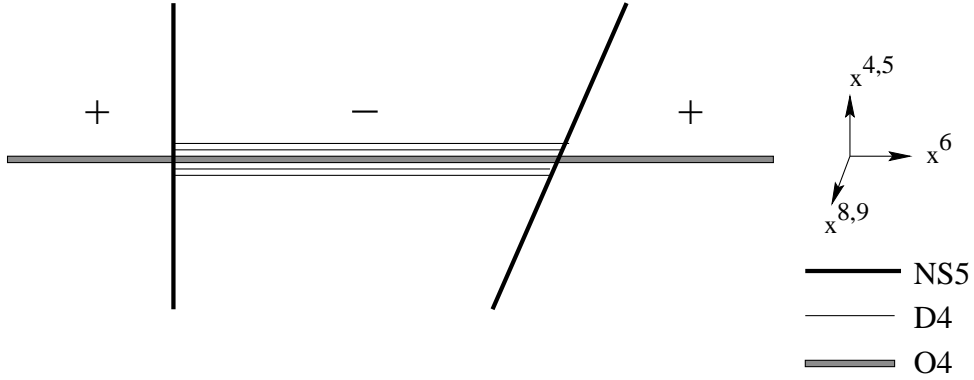,height=5cm}}}
\caption{{\bf Figure 10: }
{\footnotesize \em Brane realization of $SO(4)$ SYM.}}
\label{f9}
\end{figure}
The orientifold has a negative RR charge between the
fivebranes and positive charge on the other sides
(see \cite{gk} for a review).
Two D4 branes and their mirror images are stretched between the
NS and NS' branes. The low energy gauge theory is indeed
a four dimensional $\N=1$ SYM with gauge group $SO(4)\cong SU(2)^2$,
as in the M-theory description.

On the line  $\psi=-\sqrt{\phi^2-1}$ in Figure 9, the NS and NS'
branes intersect at a point as in Figure 11(d).
After the transition from phase II to phase I, the gauge group $SO(4)$
has disappeared. As explained in Section 2.1, in the brane picture,
this is obtained dynamically by the bending of the
fivebranes away from the orientifold, as shown in Figure 11(e).
In M-theory, as in the non-compact models considered in Section 2.3,
the transition should correspond for any of the four 3-spheres of
$(\Sigma_+\cup \Sigma_-)\times \{0,\pi R_{10}\}$ to a flop $S^3\to
0\to \RP^3$.
Since this transition in M-theory occurs by varying
$\psi$, which is the real part of a complex structure modulus of the
underlying CY $\C_1$, $\RP^3$ is the lift to the orbifold $\G_1$ of the
3-cycle of topology $S^3/w\cong \RP^3$ in $\C_1/w$ that exists for
$\psi<-\sqrt{\phi^2-1}$ and vanishes at the transition. As an example,
for the particular transition $(\psi= 0,\phi>1)\to (\psi= 0,\phi<1)$
along the $\psi\equiv 0$ axis of Figure 9, the ``flopped'' 3-sphere of
$\C_1$ can be
seen as the fixed point locus of an antiholomorphic involution of the
form $z_1\to \bar z_2$,  $z_2\to \bar z_1$, $z_j\to \bar z_j$
$(j=3,4,5,6)$.~\footnote{Up to phases appearing in the definition of
  the involution \cite{Kaste:2001iq}.}
At any other point of the transition
II$\to$I, the ``flopped'' 3-sphere in $\C_1$ is a generic
special Lagrangian 3-cycle and cannot be seen as the fixed point locus of an
antiholomorphic involution that acts globally on the CY.
Finally, we postpone to Section 4 the discussion of the transition
II$\to$I from the point of view of the type IIA orientifold on $\C_1/w$.

Next, we consider the transition III$\to$II. We start by reducing the
M-theory geometry in phase III of Figure 9 to a type IIA orientifold
description. Again focusing only on
$\Sigma_+\times \{0,\pi R_{10}\}$ in $\G_1$, for
$\psi\gg\sqrt{\phi^2-1}$, locally, we should get two systems in ten
dimensions, each of which being T-dual to the brane configuration in Figure 10.
Since the two systems are connected at the transition
$\psi=\sqrt{\phi^2-1}$ (see Figure 9), their dual brane
configuration should be located on the same orientifold, as is shown
in Figure 11(a).
\begin{figure}[!h]
\centerline{\hbox{\psfig{figure=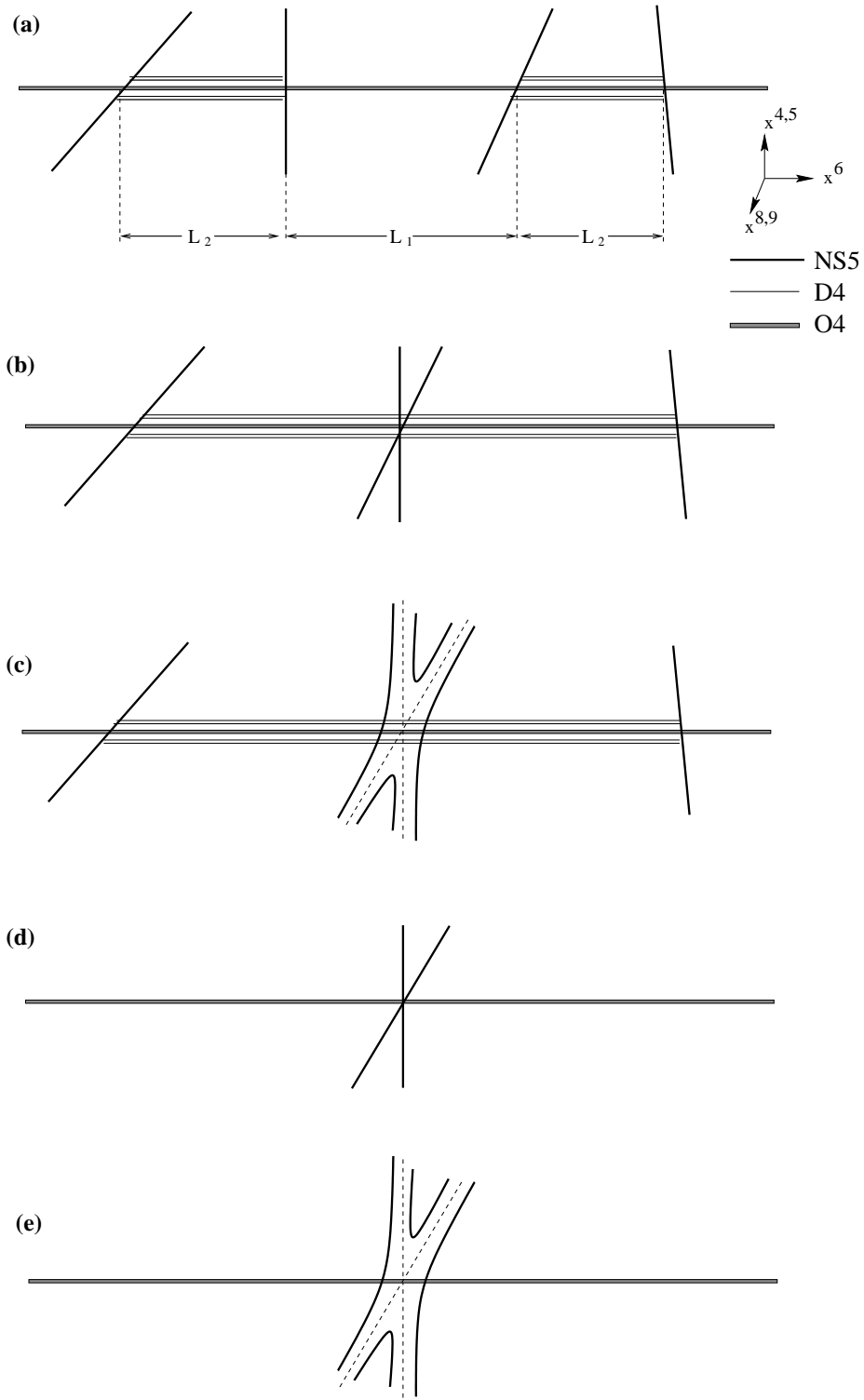,height=20cm}}}
\caption{{\bf Figure 11: } {\footnotesize \em {\bf (a)} Two $SO(4)$
    systems of branes approach each other. {\bf (b)} The middle branes
  intersect. {\bf (c)} Then, they bend. {\bf (d)} The left over
  Neveu-Schwarz branes approach each other till they intersect. {\bf (e)}
  Finally, they bend.} }
\label{f10}
\end{figure}

As $\psi$ approaches $\sqrt{\phi^2-1}$, these
two systems approach each other. More precisely, the quantity
$\psi-\sqrt{\phi^2-1}$ is associated with the distance $L_1$ between the
two systems in Figure 11(a),
while $\sqrt{\phi^2-1}$ corresponds to the distance $L_2$ between
the NS and NS' in each of the two systems (see also Figure 8(a) for
the geometric description).
The relative orientation between the various Neveu-Schwarz fivebranes
varies as we change $\psi$ and $\phi$.
In the limit $\psi\to\sqrt{\phi^2-1}$,
however, the two middle fivebranes are
not parallel~\footnote{This is due to
the existence of a 3-sphere ``between the two systems''
which shrinks to $0$ size in the limit
$\psi\to\sqrt{\phi^2-1}$. In fact, there are other $S^3$'s that
collapse simultaneously to the same point implying that the resulting
singularity is more severe than a standard nodal point (we recall that
the determinant of second derivative vanishes at this point). All
these $S^3$'s are related under the map $(\psi, z_6)  \to (-\psi, i z_6)$
to the 3-spheres present in phase I that vanish at
$\psi=-\sqrt{\phi^2-1}$.}.
Moreover, the other two fivebranes are not mutually parallel with any
of the other fivebranes, as shown in Figure 11(b).

The low energy gauge theory is an $\N=1$ $SO(4)\times SO(4)$ SYM
with a $Q\in (4,4)$ chiral multiplet.
In phase III \ie Figure 11(a), $Q$ is massive since it corresponds
to the open strings stretched from the D4 branes of one system
to those of the other. However, since the $G_2$ orbifold in M-theory is
compact, the mass parameter of $Q$ is actually a field $M$ whose VEV,
classically, is
a flat direction. This is described in field theory by a tree level
superpotential
\begin{equation}
\W_{\mbox{\scriptsize tree}}=MX~,
\label{wtree}
\end{equation}
where
\begin{equation}
X= Q^2
\label{xqq}
\end{equation}
is the gauge singlet built out of $Q\in (4,4)$.~\footnote{More precisely,
there is also
a massive adjoint chiral multiplet $(\Phi_L,\Phi_R)$
of $SO(4)\times SO(4)$ coupled to $Q$, which upon integrating it out
may generate a non-trivial superpotential for $X$
(see \cite{Giveon:1998sn} for the details in the $SU(N_L)\times SU(N_R)$ case);
it is plausible that, due to the symmetric nature of the M-theory background
which we consider, such terms will vanish in the corresponding SYM
(the analog of the case $N_L=N_R$ and $\mu_L=-\mu_R$ in \cite{Giveon:1998sn}).}
Crossing the line $\psi=\sqrt{\phi^2-1}$ from phase III to phase II
in Figure 9, there is a breaking of $SO(4)\times SO(4)$ to
a single $SO(4)$. In the brane picture, this is realized
dynamically by the bending of the middle
fivebranes away from the orientifold, as shown in Figure 11(c).
In the low energy SYM, this quantum effect is taking us away from the
origin of the classical moduli space by giving a VEV to
the bifundamental $Q$, thus going to the Higgs branch where
$SO(4)\times SO(4)$ is broken to the diagonal $SO(4)$. Since this
branch does not exist in the perturbative superpotential of
Eq. (\ref{wtree}), the Higgs mechanism that occurs is a
non-perturbative effect.
Next we shall see that it is possible to describe it in gauge theory.

A dynamical superpotential for $X$ is
generated~\footnote{This can be shown, for instance, by ``integrating in''
\cite{Intriligator:1994uk} $Q$ to pure $\N=1$ $SO(4)\times SO(4)$ SYM
with $\W_{\mbox{\tiny down},\pm}\propto (\Lambda_{1,\mbox{\tiny
down}}^3\pm \Lambda_{2,\mbox{\tiny down}}^3)$.} and has two
physically distinct phases described by:
\begin{equation}
\W_{\mbox{\scriptsize dyn},\pm}(X)={X^2\over\Lambda_{\pm}}~,\qquad
\Lambda_{\pm}={\Lambda_1\pm \Lambda_2\over 2}~,
\label{wdx}
\end{equation}
where $\Lambda_1$ and $\Lambda_2$ are the ``QCD scales''
of the $SO(4)\times SO(4)$, respectively.
In our case, the M-theory geometry is such that the gauge couplings of
the two $SO(4)$ factors are equal:
\begin{equation}
\Lambda_1=\Lambda_2\equiv\Lambda~,
\label{lll}
\end{equation}
hence~\footnote{In the case $\Lambda_-=0$ there is no dynamical superpotential
for $X$: $\W_{\mbox{\tiny dyn},-}(X,\Lambda_-=0)=0$, since
$\W_{\mbox{\tiny down},-}=0$.
Alternatively, this can be understood as the limit $\Lambda_-\to 0$
in Eq. (\ref{wdx}) which sets the constraint $X=0$.}
\begin{equation}
\Lambda_+=\Lambda~, \qquad \Lambda_-=0~.
\label{lllo}
\end{equation}
To obtain the phase structure which we see in the M-theory
geometrical picture we assume that there is also a dynamically
generated term for $M$:~\footnote{This is the {\it only} term in
$M$ and $\Lambda$ which is consistent with the symmetries; it is
similar to the quantum potential generated
for the ``meson'' in a ``magnetic'' $\N=1$ theory \cite{Seiberg:1995pq}.
Also, the $1/4$ normalization of this term is required for the total
superpotential to admit a vacuum.}
\begin{equation}
\W_{\mbox{\scriptsize dyn},\pm}(M)={\Lambda_{\pm}M^2\over 4}~.
\label{wdm}
\end{equation}
Altogether, the quantum superpotential we
expect
is~\footnote{Note that $\W$ is self-dual under the strong-weak coupling
duality ${\Lambda\over 2\mu}\leftrightarrow {2\mu\over\Lambda}$
together with ${X\over\mu^2}\leftrightarrow {M\over\mu}$, where $\mu$
is some mass scale.}:
\begin{equation}
\W_{\pm}(X,M)={X^2\over\Lambda_{\pm}}+MX+{\Lambda_{\pm}M^2\over 4}~.
\label{qw}
\end{equation}
Let us see now that $\W_\pm$ have the required properties.

Varying $\W_{\pm}$ with respect to $X$ and $M$
give rise to the same equation of motion:
\begin{equation}
X={-{1\over 2}\Lambda_{\pm}M~.}
\label{seom}
\end{equation}
Hence the quantum moduli space is composed of two branches $\M_{\pm}$
which intersect at $M=X=0$.
In the case (\ref{lllo}), the branch $\M_+$ consists of:
\begin{equation}
\M_+\quad :\qquad X=-{1\over 2}\Lambda_+ M~.
\label{mplus}
\end{equation}
This is the Higgs branch associated to the geometrical phase II
in Figure 9.
On the other hand, the branch $\M_-$ consists of:
\begin{equation}
\M_-\quad :\qquad X=0\quad , \quad  M\quad {\rm anything}~.
\label{mminus}
\end{equation}
We associate this branch with phase III.

Finally, the geometrical phase transition from III to I in Figure 9 is
similar to the transition from II to I. In the dual brane picture, phase
III is described by the system of Figure 11(a) and the transition
consists in approaching, intersecting and then bending the NS and NS'
of each system of branes separately but simultaneously.

To summerize, a phase diagram with the transitions
in the brane configurations is shown in Figure 12.
\begin{figure}[!h]
\centerline{\hbox{\psfig{figure=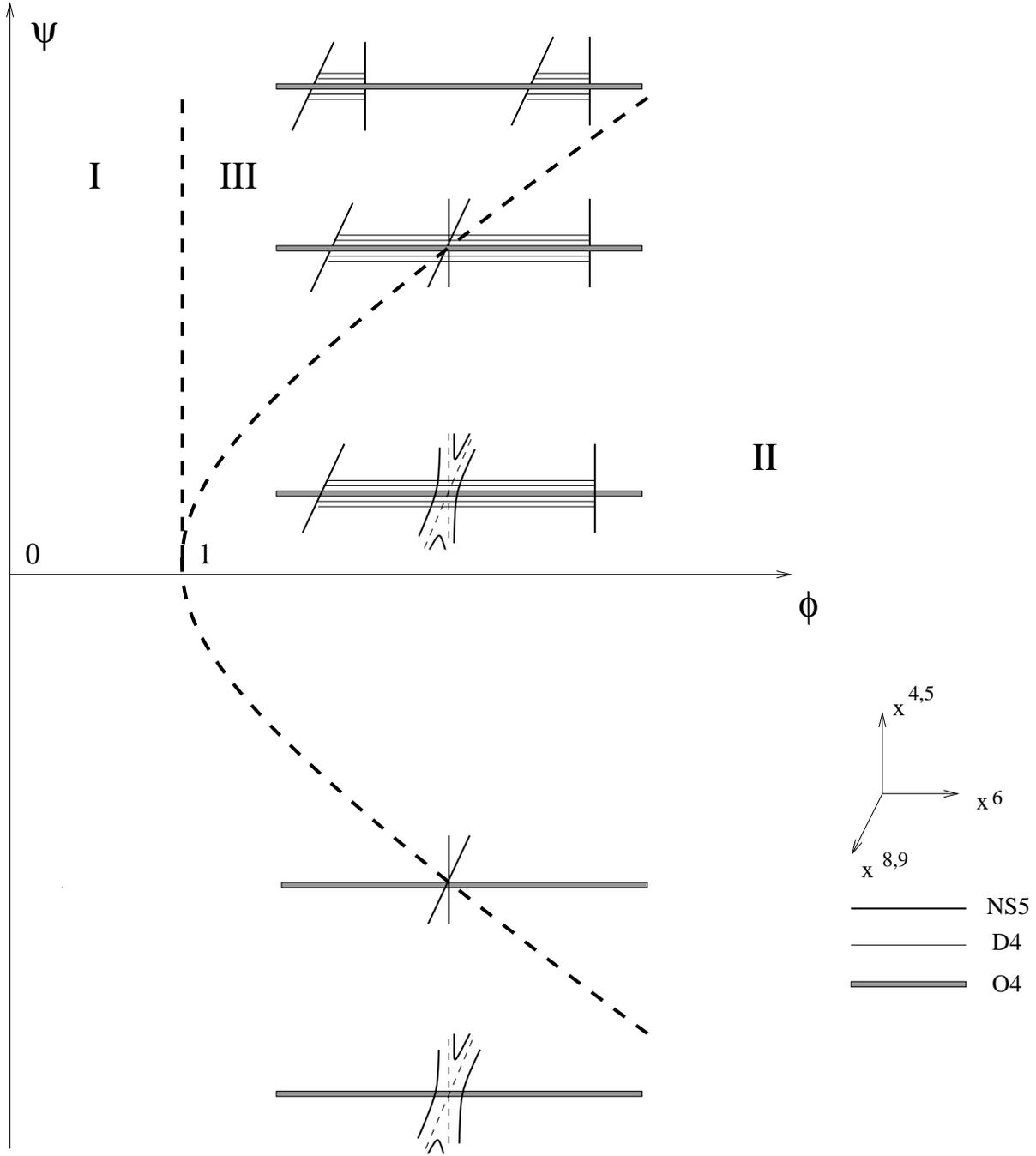,height=18cm}}}
\caption{{\bf Figure 12:}
{\footnotesize \em Phase diagram of the model based on
  $\G_1$. The brane system dual to $\Sigma_+$ is represented for
  various values of $(\phi,\psi)$.} }
\label{f12}
\end{figure}
All these transitions are smooth in the brane picture and should occur at
finite distance in moduli space in the M-theory geometry. In Figure
12, they are represented by dashed lines.

\section{Type II orientifolds, confinement and Higgs mechanism}


Three examples of phase transitions II$\to$I, III$\to$I and
III$\to$II have been considered in Section 3. In each case, from the
M-theory point of view, 3-spheres experience flop transitions, while from
the brane point of view, this is
realized by the bending of Neveu-Schwarz fivebranes.
The transitions were interpreted in field theory as confinement (or
quantum Higgs effect) of $SO(4)$ gauge groups. To clarify some points
from the type IIA orientifold descriptions, let us relate these
effects to the transitions that occur in the local model $(T^*(S^3)\times
S^1)/w\I$ defined in Eq. (\ref{so4}).

In the non-compact $G_2$ orbifold $(T^*(S^3)\times S^1)/w\I$, we saw
in Section 2.3 that one
can perform a conifold transition on the underlying CY threefold
$T^*(S^3)$. Thus, one passes into the moduli space of
$\left( (\mbox{resolved conifold})\times S^1\right)/w'\I$ that is
related to confinement. In Section 4.1, we shall consider the
similar transition in the compact case. We shall see that in general the
change in the Hodge numbers of the underlying CY has important physical
consequences. In addition to confinement, a change of branch in the
scalar potential of chiral multiplets takes place.

However, we note that these CY conifold transitions do not
occur in the specific model we studied in
Section 3. This is due to the fact that for $\C_1$, at the local
singularities of Eq. (\ref{nodes}), the determinant of second
derivatives vanishes.
Instead, the transitions for the compact model of Section 3 correspond
in the local model $(T^*(S^3)\times S^1)/w\I$ to passing from
positive values of $\mu$ defined in Eq. (\ref{con}) to negative values
of $\mu$. As explained in Section 2.3, when $\mu<0$, quantum
corrections in the classical type IIA orientifold obtained by sending
$R_{10}\to 0$ have to be taken into account in order to describe
accurately confinement. As an example, one can consider a dual
description to compute these corrections.
At the end of Section 4.1, we shall comment on the situation where the
non-compact CY threefold $T^*(S^3)$ is replaced by a compact one.

Finally, the case of a non-Abelian Higgs mechanism together with a
change of branch in the scalar potential will be considered in
Section 4.2. This will be treated in the spirit of \cite{Greene:1996dh}.

\subsection{Confinement, scalar potential and conifold transitions}

To describe a conifold transition in the underlying CY of a $G_2$
orbifold, we consider the second model treated in \cite{Kaste:2001iq},
\begin{equation}
\G_2=(\C_2\times S^1)/w\I\; ,
\end{equation}
where $\C_2$ is another sub-family of CY threefolds
in $\CP^4_{11222}[8]$ defined by
\begin{equation}
p_2 \equiv z_6^4(z_1^8 + z_2^8-2\phi z_1^4z_2^4)+(z_3^2-\phi z_6^2
z_2^4)^2+(z_4^2-\phi z_6^2 z_2^4)^2+(z_5^2-\phi z_6^2 z_2^4)^2=0~.
\label{def33}
\end{equation}
The antiholomorphic involution $w$ acts again as $z_i\to \bar
z_i$, while $\phi$ is a real parameter. At the particular values
$\phi=\pm 1$, $\C_2$ becomes a conifold,
whose nodal points~\footnote{The determinant of second derivatives does
not vanish at these singularities.} are
\begin{equation}
\begin{array}{llll}
\mbox{for }\phi=+1 &~:~ & (i^k,1,\pm1,\pm1,\pm1,1)& ,(k=0,...,3)~,\\
\mbox{for }\phi=-1 &~:~ & (i^k e^{i\pi/4},1,\pm1,\pm1,\pm1,1) &,(k=0,...,3)\; ,
\label{nodes22}
\end{array}
\end{equation}
where the $\pm$ signs are all independent.
The phase diagram  of the
fixed point set $\Sigma\times \{0,\pi R_{10}\}$ of this model is then
obtained as in Section 3.1 \cite{Kaste:2001iq}:
\begin{equation}
\begin{array}{lll}
\phi>1, ~~~& \Sigma= \bigcup_{n=1}^{16} S^3 & \mbox{(disconnected and of radii
  }\sqrt{\phi^2-1})\; , \\
\phi=1,~~~ & \Sigma=\{(\pm 1,1,\pm 1,\pm 1,\pm 1,1)\}&\mbox{(\ie 16
  points)}\; ,\\
\phi<1,~~~ & \Sigma={ \emptyset} &\mbox{(no fixed
  points)}\; .
\end{array}
\end{equation}
{}From Eq. (\ref{nodes22}), notice that at $\phi=1$, the 16 vanishing
3-spheres coincide with singular points of $\C_2$, where we shall
consider a conifold transition.

In M-theory, as in Section 3.1, the 3-spheres of $A_1$ singularities
for $\phi>1$
give rise to an $SU(2)^{32}$ gauge group. In addition, from
Eq. (\ref{b23}), there are $\N=1$ vector and chiral multiplets in
M-theory. Sending $R_{10}\to 0$, one
obtains a type IIA orientifold on $\C_2/w$, with two D6 branes
(and their mirror partners) and an O6
plane wrapped on each of the 16 $S^3$'s of $\Sigma$. There is thus an
$SO(4)^{16}$ gauge group as in M-theory.  Also, it is shown in
\cite{Vafa:1996gm} that the massless
spectrum arising from invariant forms on an orientifold of CY in type
IIA is consisting of~\footnote{Actually, only freely acting cases were
  considered in \cite{Vafa:1996gm} but their arguments apply to
  singular orientifolds as well.}
\begin{equation}
h_{11}^+ ~~ \mbox{vector multiplets$\quad$ and$\quad$}   1+h_{12}+h_{11}^-~~
\mbox{neutral chiral multiplets~.}
\label{specIIA}
\end{equation}
Hence, the spectra  in the type IIA orientifold and in M-theory
coincide. As in the case based on $\C_1$, we have $h_{11}^+=0$ and we
have in total
\begin{equation}
\mbox{1 vector multiplet of }SO(4)^{16} \qquad \mbox{and} \qquad
\mbox{89 neutral chiral multiplets .}
\end{equation}

At $\phi=1$, we  now choose to blow up an $S^2$ at each node of the
conifold $\C_2$. The CY threefold $\C_2'$ obtained this way can be written as
\begin{equation}
\left\{
\begin{array}{ll}
\phantom{-}\left[ z_6^2(z_1^4- z_2^4)+i(z_3^2- z_6^2 z_2^4)\right]\xi_1 +
\left[ (z_4^2-z_6^2 z_2^4)-i(z_5^2-z_6^2
  z_2^4)\right]\xi_2=0 \\
-\left[ (z_4^2-z_6^2 z_2^4)+i(z_5^2- z_6^2
  z_2^4)\right]\xi_1 + \left[ z_6^2(z_1^4- z_2^4)-i(z_3^2- z_6^2
  z_2^4)\right]\xi_2 =0
\end{array}\right.~,
\label{c2'}
\end{equation}
where $\xi_{1,2}$ are projective coordinates parametrizing $\CP^1\cong
S^2$ as in Eq. (\ref{asres}). The Hodge numbers of $\C_2'$ are
$h_{11}'=3$ and $h_{12}'=55$. In M-theory, we have moved into the
moduli space of
\begin{equation}
\G_2'={\C_2'\times S^1 \over w'\I}~,
\label{so4res'}
\end{equation}
where the involution takes the form \cite{Kaste:2001iq}
\begin{equation}
w':~\left\{
\begin{array}{l}
z_i\to \bar z_i\\
\xi_1\to -\bar\xi_2~,~~ \xi_2\to \bar\xi_1
\end{array}\right.
\qquad \mbox{and}\qquad  \I:~
x^{10}\to-x^{10}~.
\label{toto2}
\end{equation}
The orbifold is freely acting, hence the spectrum is given by the
Betti numbers $b_2'$ and $b_3'$ of $\G_2'$. From Eq. (\ref{b23}),
since $h_{11}^{'-}=0$,~\footnote{Two of the $(1,1)$ forms were already
odd on $\C_2/w$ and the third one dual to the $\CP^1$ parametrized
by $\xi_1/\xi_2$ is also odd under $w'$.}
we have
\begin{equation}
\mbox{59 chiral multiplets } \qquad \mbox{and} \qquad
\mbox{no gauge group .}
\end{equation}
Similarly, in the
orientifold description on $\C_2'/w'$, there are no fixed points and
thus no orientifold sixplane and D6 branes. From Eq. (\ref{specIIA}),
the spectrum is also consisting of 59 chiral multiplets.

Locally, the transition from
$\G_2$ (for $\phi>1$) to $\G_2'$ amounts to a flop $S^3\to
(\CP^1\times S^1)/w\I\cong \RP^3$ for each $SU(2)$ factor. In the IIA
orientifold limit $R_{10}\to 0$, it is translated into a conifold
transition $S^3\to \RP^2$
with no RR flux for each $SO(4)$ factor, as was the case between the
local models of Eqs. (\ref{con/orient}) and (\ref{res/orient}).
This describes confinement for each $SU(2)^2\cong SO(4)$ factor from
both points of view. However, in the compact case, due to the change
in the Hodge numbers of the underlying CY, the number
of neutral chiral multiplets also varies in the transition. This was
discussed in detail in \cite{Kaste:2001iq,Partouche:2001uq}
and interpreted as a change of
branch in the scalar potential. Actually, passing from $\C_2'/w'$ to
$\C_2/w$ in type IIA orientifolds, the number of flat directions in
the potential changes at the conifold point due to the appearance of
additional massless states. These states are black hole chiral
multiplets associated with membranes wrapped on the vanishing $\RP^2$'s,
as in the $\N=2$ case \cite{Strominger:1995cz, Greene:1995hu}.

Let us return now to the model based on $\G_1$.
{}From  M-theory, brane and field theory points of view, the
transitions of Section 3 concerned the non-Abelian gauge groups only,
since both the number of neutral chiral multiplets (89 in our example)
and $U(1)$ vector multiplets (0 in our example) were
constant~\footnote{Actually,
one chiral multiplet parametrizes the path followed in
moduli space, while the other chiral and Abelian vector multiplets are
relegated to the role of spectators.}.
In the description in terms of type IIA  orientifold on $\C_1/w$, let
us focus as an example on the transition II$\to$I. As discussed in Section
3.2, locally this should amount to a flop $S^3\to 0 \to S^3/w\cong \RP^3$,
as it is the case for the non-compact model $T^*(S^3)/w$, when passing
from $\mu>0$ to $\mu<0$. To describe quantitatively confinement in
these cases,
quantum corrections have to be taken into account. To compute them, it
should a priori be possible to map the system in the phase with D6 branes
and an O6 plane on $S^3$ to a type IIB description on the mirror CY
with D5 branes and an O5 plane on an isolated curve. This would be a
strategy in
the spirit of \cite{Hori:2000ck,Aganagic:2000gs,Aganagic:2001nx},
applied to SLAG's of various topologies and in presence of orientifold planes.
Another strategy was also proposed in \cite{Mayr:2001xk} and relates
threefolds with D-branes to fourfolds.
However, we reviewed in Section 2.2 that two D6 branes
(with their mirror partners)
and an O6 plane on $S^3$ is dual to an $\RP^2$ with no RR flux in an
orientifold of type IIA \cite{Sinha:2000ap}. One can then ask if in the
compact case there would be a similar closed string dual description
in the sense of \cite{Vafa:2000wi,Sinha:2000ap}, involving $\RP^2$
and describing the same massless neutral chiral and
Abelian vector multiplets?

Thus, we are looking for a type IIA orientifold on $\tilde
\C_1/\tilde w$, where $\tilde \C_1$ is a CY admitting  an
antiholomorphic involution $\tilde w$. Let $\tilde h_{11}^\pm$ be
the number of even and odd 2-forms on $\tilde \C_1$ under $\tilde w$,
and $\tilde h_{12}$ the number of 3-forms. From
Eq. (\ref{specIIA}), in order to have an equal number of $U(1)$ vector
multiplets and neutral chiral multiplets on $\tilde \C_1/\tilde w$ and
$\C_1/w$, we need $\tilde h_{11}^+=h_{11}^+$ and
$1+\tilde h_{12} +\tilde h_{11}^-=1+ h_{12} +h_{11}^-$. Taking the sum
of these equations, we thus have
\begin{equation}
\tilde h_{11}^+=h_{11}^+ \qquad \mbox{and}\qquad
\tilde h_{11} +\tilde h_{12}=h_{11} +h_{12} ~.
\label{condim}
\end{equation}
The second of these equations is suggestive. Actually,
if one supposes that there could be a general rule for relating a
generic CY orientifold to a dual CY orientifold, the latter could
hardly involve anything but the
original manifold or its mirror. Now, since the dual type IIA orientifold
should locally involve $\RP^2$ instead of $S^3$, we are led to
ask whether an orientifold of type IIA on a CY threefold could be dual to an
orientifold of type IIA on the mirror CY. In the mirror CY, there
should be isolated $\CP^1$'s mirror to the 3-spheres on which branes
and orientifold planes are wrapped~\footnote{In general, a SLAG is
  mirror to an holomorphic cycle. Here, we consider the case where the
  holomorphic cycle is an isolated 2-sphere.}. Since $\tilde w$ is required to
act on these $\CP^1$'s so that they become $\RP^2$'s, as in
Eqs. (\ref{toto1}) and (\ref{toto2}), $w'$ is freely acting. Hence,
there are no  O5 planes and, due to the vanishing of the RR charge, no
D5 branes. As expected to describe confinement, the non-Abelian gauge
group of the original orientifold does not show up in the present one.

Also, we note that for the mirror $\widetilde{\mbox{CY}}$ of a
threefold, the orientifold of type IIA on
$\widetilde{\mbox{CY}} / \tilde w$
is obtained in the limit $R_{10}\to 0$ of M-theory on $\tilde \G=
( \widetilde{\mbox{CY}} \times S^1 ) / \tilde w\I$.
M-theory on the two backgrounds $\G={\left( \mbox{CY}\times S^1\right)
  /  w\I}$ and $\tilde \G$ has an
equal number of Abelian vector and neutral chiral massless multiplets.
In addition,
locally, the $S^3$'s of $A_1$ singularities in $\G$ are replaced by
$(\CP^1\times S^1)/\tilde w\I\cong \RP^3$ in $\tilde \G$, as was the
case at the beginning of this section. Thus, $\G$ and $\tilde \G$
could also be dual.

Finally, one can consider more general type IIA orientifolds that
involve an even number $N_c$ of D6 branes on top of an O6$_\pm$
wrapped on $S^3$ in a CY. Then, this configuration might be related to
an orientifold of type IIA on the mirror CY with ${N_c\over 2}\pm 2$
units of RR flux on $\RP^2$. Similarily, the mirror statement in type IIB
would be equivalent.

\subsection{Non-Abelian Higgsing, scalar potential and conifold transitions}

We have seen that for orientifolds of compact CY's, confinement
can be described
by a conifold transition. We would like to see now that in a different
set up of branes and orientifolds, a conifold transition can also
describe a non-Abelian Higgs mechanism. In all these cases, these effects
are combined with a change of branch in the scalar potential.

In \cite{Greene:1996dh}, the Higgsing of $\N=2$
$U(1)$ vector multiplets in  type II compactifications on
CY manifolds in terms of confinement of magnetic flux was
studied. In that work, 3-spheres that vanish at some
conifold locus in complex structure moduli space were considered.  In
general, these 3-spheres are not independent in homology. Instead,
there classes satisfy linear combinations that vanish. Due to these linear
relations, a 3-cycle that meets such an $S^3$ must meet at
least another one.

Let us consider now the case illustrated in Figure 13(a).
\begin{figure}[!h]
\centerline{\hbox{\psfig{figure=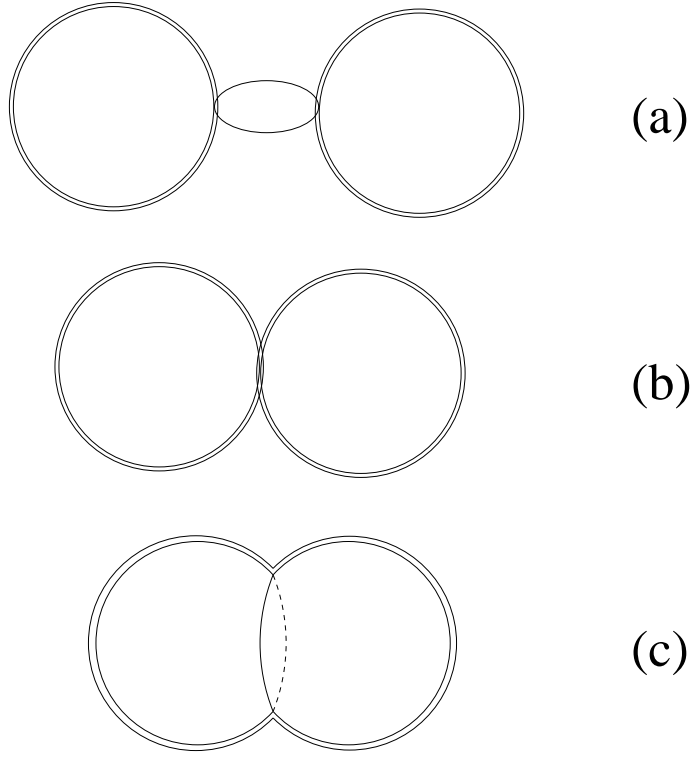,height=10cm}}}
\caption{{\bf Figure 13: }
{\footnotesize \em A conifold transition describing the Higgs
  mechanism $SO(N_c)\times SO(N_c)\to SO(N_c)$.}}
\label{f14}
\end{figure}
There
are two $S^3$'s intersecting a third 3-cycle at a point. An
equal even number $N_c$ of D6 branes on top of an O6$_-$ plane are wrapped on
the 3-spheres. The non-Abelian gauge group is then
$SO(N_c)\times SO(N_c)$. Since the $S^3$'s are fixed by the
orientifold projection, the 3-cycle that intersect them is non-orientable. When
it is $\RP^3$, we can go to a locus in complex structure moduli space where it
vanishes, as in Figure 13(b). Then, blowing up the singularity to a 2-cycle,
the 3-spheres have become 3-chains connected along their boundary, a
blow up 2-sphere, as can be seen in Figure 13(c). The brane system
then
consists of $N_c$ D6 branes and an O6$_-$ plane wrapped on the single
$S^3$, which is the connected sum of the two original
3-spheres. Therefore, the gauge group has become the
diagonal $SO(N_c)$ and the whole transition is a non-Abelian Higgs
mechanism. Note that simultaneously, as in the case of confinement
treated previously, the number of massless neutral chiral
multiplets has also changed due to the variation of Hodge numbers of the
CY through the conifold transition.

To conclude, we signal that an example of the above conifold
transition is realized in the model based on
$\C_3/w$, where $\C_3$ is a CY threefold living in a sub-family of
$\CP^4_{11222}[8]$. The defining polynomial of $\C_3$ is
\begin{equation}
p_3 \equiv z_6^4(z_1^8 + z_2^8-2\phi z_1^4z_2^4)+(z_3^2-\psi_1 z_6^2
z_2^4)^2+(z_4^2-\psi_2 z_6^2 z_2^4)^2+(z_5^2-\psi_3 z_6^2 z_2^4)^2=0~,
\label{def333}
\end{equation}
where $\phi$ and $\psi_{1,2,3}$ are real, while
$w$ acts as usually as $z_i\to\bar z_i$. The relevant transition occurs
at $\psi_2,\psi_3>\psi_1=\sqrt{\phi^2-1}$.

\section{Transitions involving 3-cycles with $b_1>0$}

Up to now, we have considered transitions in M-theory that involve
3-cycles with vanishing first Betti number only. We would like now
to consider in the same spirit 3-cycles with
non-trivial $b_1$. In Section 5.1, we shall deal with transitions
where $b_1=1$ on both sides of the transitions. On the contrary,
we shall focus in Section 5.2 on an example where
$b_1$ of the 3-cycles changes at the transition.
In the first (second) case, the transitions are expected
to be at infinite (finite) distance in moduli space.

\subsection{Replacing $S^3$'s by $S^2\times S^1$'s}

Let us consider models involving 3-cycles of topology
$S^2\times S^1$. In particular, we would like to see what the
transitions of 3-spheres in Section 3 become when each $S^3$ is
replaced by an $S^2\times S^1$. Following \cite{Kaste:2001iq}, one can
start from the model based on $\C_1$ and take  an orbifold
$(z_4,z_5)\to (-z_4,-z_5)$ on it that is blown up. The CY manifold
$\C_4$ obtained this way has Hodge numbers $h_{11}=3$ and $h_{12}=55$
and is defined by a polynomial
\begin{equation}
p_4 \equiv z_6^4(z_1^8 + z_2^8-2\phi z_1^4z_2^4)+(z_3^2-\psi z_2^4z_6^2)^2
+z_7^2(z_4^4+z_5^4)=0\; ,
\label{def4}
\end{equation}
for complex $\phi$ and $\psi$, where there are three $\Cs$ scaling
actions, whose  weights  are
\begin{equation}
\begin{tabular}{c|ccccccc}
 & $z_1$ & $z_2$& $z_3$&$ z_4$& $z_5$&$ z_6$&$ z_7$ \\ \hline
$\Cs_1$& $0$&$0$&$1$&$1$&$1$&$1$&$0$ \\
$\Cs_2$& $1$&$1$&$0$&$0$&$0$&$-2$&$0$ \\
$\Cs_3$& $0$&$0$&$0$&$1$&$1$&$0$&$-2$
\end{tabular}
\label{sc}
\end{equation}
The forbidden set in the ambient space is now
\begin{equation}
(z_1,z_2)\neq (0,0) \quad \mbox{,}\quad
(z_4,z_5)\neq(0,0)\quad \mbox{and}\quad
(z_3,z_6)\neq(0,0)\; .
\label{forbid2}
\end{equation}
There are singular points
that occur in charts where $z_2$ and $z_6$ do not vanish, so that they
can be rescaled to one.  These singularities arise for specific values
of $\phi$ and $\psi$:
\begin{equation}
\begin{array}{lllll}
\mbox{for any $\psi$,}&  \mbox{at }\phi=+1 &~:~ & (i^k,1,\pm
\sqrt{\psi},z_4,z_5,1,0)& ,(k=0,...,3)~,\\
& \mbox{or }\phi=-1 &~:~ & (i^k e^{i\pi/4},1,\pm \sqrt{\psi},z_4,z_5,1,0)
&,(k=0,...,3)~,\\
\mbox{and for any $\phi$,}& \mbox{at }\psi=\pm \sqrt{\phi^2-1} &~:~ &
(\phi^{1/4}i^k,1,0,z_4,z_5,1,0)& ,(k=0,...,3)~,
\label{cirsing}
\end{array}
\end{equation}
where the projective coordinates $(z_4,z_5)\equiv (\lambda
z_4,\lambda z_5)$ $(\lambda \in \Cs_3)$ parametrize a $\CP^1$.

We now restrict  $\phi$, $\psi$ to real values in order to
consider the $G_2$ orbifold
\begin{equation}
\G_4={\C_4\times S^1\over w\I}~,
\end{equation}
where the involution is
\begin{equation}
w:~ z_i\to \bar{z}_i\quad (i=1,...,7)~~, \qquad \I:~
\quad x^{10}\to
-x^{10}\; . \label{s2}
\end{equation}
Looking for the fixed points of this involution, Eq. (\ref{def3}) is
modified to
\begin{equation}
x_1^4 =\phi
\pm\sqrt{\phi^2-\left[ 1+(x_3^2-\psi)^2+x_7^2(x_4^4+x_5^4)\right] }\; ,
\label{def4fp}
\end{equation}
where the real projective coordinates $(x_4,x_5)\equiv(l x_4,l x_5)$
$(l \in
\R^*$) parametrize a circle. Choosing a point $(x_4,x_5)$ on this
circle, Eq. (\ref{def4fp}) can be written in terms of the variables $X_{1,3}$
defined in Eq. (\ref{u}) and $X_7=x_7\sqrt{x_4^4+x_5^4}$:
\begin{equation}
X_1^2+X_3^2+X_7^2=\phi^2-1~,
\label{op}
\end{equation}
which is an $S^2$ of radius $\sqrt{\phi^2-1}$ when $\phi\ge 1$.
Since the rest of the
discussion is identical to what was done after  Eq. (\ref{Sph}),
we know that each $S^3$ of $A_1$ singularities of radius
$\sqrt{\phi^2-1}$ in $\G_1$ is replaced in $\G_4$ by a cycle $S^2\times S^1$
of $A_1$ singularities where the radius of $S^2$ is
$\sqrt{\phi^2-1}$. Similarly, two $S^3$'s a
distance $2\left[ \psi-\sqrt{\phi^2-1} \right]^{1/2}>0$
away in $\G_1$ become two $S^2$'s a
distance $2\left[ \psi-\sqrt{\phi^2-1} \right]^{1/2}>0$
away in $\G_4$, altogether multiplied by $S^1$.
As a result, for $\phi<1$, the fixed point
set $\Sigma\times \{0,\pi R_{10}\}$ is empty, while for $\phi\ge 1$,
$\Sigma$ is composed of two identical copies $\Sigma_+$ and $\Sigma_-$
of topology summarized as follows:
\begin{equation}
\begin{array}{lll}
\psi>\sqrt{\phi^2-1},~~~ &
\Sigma_+=(S^2\times S^1)\cup( S^2\times S^1) &\mbox{(disconnected
  union)}\; , \\
\psi=\sqrt{\phi^2-1},~~~ &
\Sigma_+=(S^2\cup S^2)\times S^1
 &\mbox{(with a singular $S^1$ of intersection)}\; , \\
|\psi|<\sqrt{\phi^2-1},~~~ & \Sigma_+=(S^2 \# S^2)\times S^1 \cong
  S^2\times S^1 &\mbox{(with a connected
  sum of $S^2$'s)}\; , \\
\psi=-\sqrt{\phi^2-1},~~~ &
\Sigma_+=\{(\phi^{1/4},1,0,x_4,x_5,1,0)\}&\mbox{(a singular $S^1$)}\; ,\\
\psi<-\sqrt{\phi^2-1},~~~ & \Sigma_+= \emptyset \; &\mbox{(no fixed point)}\; .
\end{array} \label{ss2}
\end{equation}
In figure 14, the fixed point set $\Sigma_+\times \{0\}$ in $\G_4$ is
represented for various values of the parameters. We describe now the
M-theory spectrum in each phase.
\begin{figure}[!h]
\centerline{\hbox{\psfig{figure=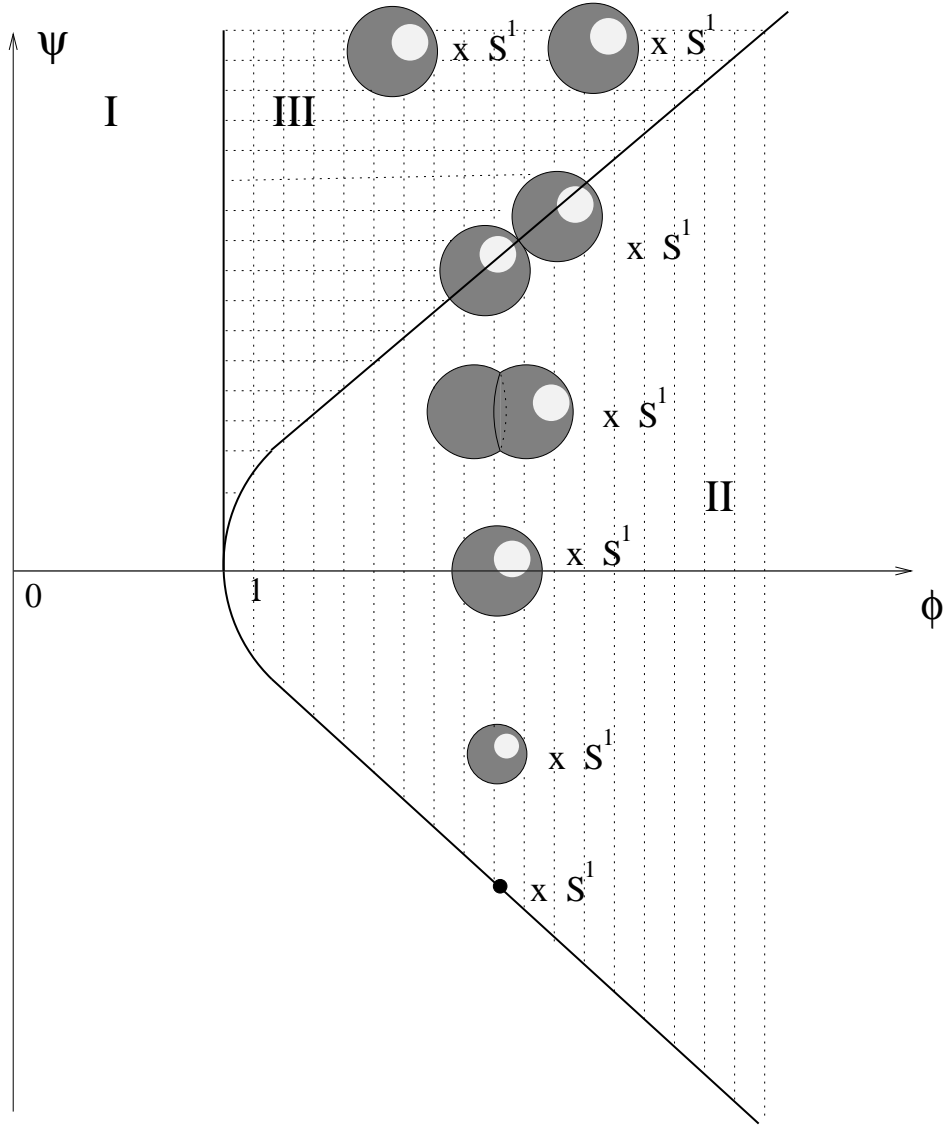,height=15cm}}}
\caption{{\bf Figure 14:}
{\footnotesize \em Phase diagram of the model based on
  $\G_4$. $\Sigma_+$ is represented for various values of
  $(\phi,\psi)$. When non-empty, it is composed of 2-spheres times
  circles. The transitions are not expected to be physical and
  should occur at infinite distance in moduli space. Hatches indicate
  that we sit in the Coulomb branch parametrized by axes orthogonal to
  the figure.}}
\label{f20}
\end{figure}

\vspace{.3cm}

\noindent {\em \large -- Spectrum in phase I:
$\phi<1$ or $\phi\geq 1,\, \psi<-\sqrt{\phi^2-1}$}
\vspace{.2cm}

\noindent
In this case, since the involution is freely acting, the massless spectrum is
given  by Eq. (\ref{b23}). Since $d(z_4/z_5)\wedge d(\bar
z_4/\bar z_5)$ is odd under $w$, the blow up $\CP^1$ at
$(z_4,z_5)=(0,0)$ in $\C_4$ is also odd. Thus $h_{11}^+=0$ and we have
\begin{equation}
\mbox{59 chiral multiplets } \qquad \mbox{and} \qquad
\mbox{no gauge group .}
\end{equation}

\vspace{.3cm}

\noindent {\em \large -- Spectrum in phase II:  $\phi>1, \,
  |\psi|<\sqrt{\phi^2-1}$}
\vspace{.2cm}

\noindent
We still have the previous spectrum associated to the invariant forms on the
orbifold. In addition, for each $S^2\times S^1$, if
we do not desingularize the $A_1$ singularity, there is an  $SU(2)$ gauge group
with $b_1(S^2\times S^1)=1$ chiral multiplet in the adjoint
representation \cite{Acharya:2000gb, Kaste:2001iq}.
Because $b_1>0$, blowing up the $A_1$ singularity is expected to be
possible~\footnote{under some additional circumstances
  \cite{Joyce2}.}. This amounts to
going into the Coulomb branch of $SU(2)$ by giving a VEV to the adjoint matter
field. In this branch, $SU(2)$ is broken to $U(1)$ and the matter in
the adjoint of $SU(2)$ gives rise to a single massless neutral chiral
multiplet.
In phase II, we have at most four singular $S^2\times S^1$'s that can be
desingularized independently. Therefore, the total spectrum takes one on the
following forms:
\begin{eqnarray}
&\mbox{an $SU(2)^{4-k}\times U(1)^k$ gauge group$ ~,~~~(k=0,...,4)\,
  ,$} &\nonumber  \\
&\mbox{with 1 adjoint chiral multiplet of $SU(2)^{4-k}$} & \label{spII}\\
&\mbox{and $59+k$ neutral chiral multiplets .}&\nonumber
\end{eqnarray}
In Figure 14, the hatches in phase II signal that as soon as we sit in
the Coulomb branch of an $SU(2)$ gauge group, we are actually at a
point in moduli
space with a non-zero coordinate in a direction orthogonal to the
$(\phi,\psi)$ plane.

\vspace{.3cm}

\noindent {\em \large -- Spectrum in phase III:  $\phi>1, \,
  \psi>\sqrt{\phi^2-1}$}
\vspace{.2cm}

\noindent
The only difference between phase III and phase II is that there
are eight instead of four $S^2\times S^1$'s of $A_1$ singularities that
can be blown up. As a result, the spectrum is now
\begin{eqnarray}
&\mbox{an $SU(2)^{8-k}\times U(1)^k$ gauge group$ ~,~~~(k=0,...,8)\,
  ,$} &\nonumber  \\
&\mbox{with 1 adjoint chiral multiplet of $SU(2)^{8-k}$} & \label{spp}\\
&\mbox{and $59+k$ neutral chiral multiplets .}&\nonumber
\end{eqnarray}
As before, phase III is hatched in Figure 14 to indicate that at fixed
$\phi$ and $\psi$, we can still move in the Coulomb branches.

\vspace{.3cm}

\noindent {\em \large Brane interpretation}
\vspace{.2cm}

\noindent
To understand what is happening at the transitions, we would like to
translate them to the language of brane constructions in type IIA.

Let us start by reducing M-theory on $\G_4$ in phase II, when we sit
at the origin of the Coulomb branches (\ie $k=0$ in Eq. (\ref{spII})). Sending
$R_{10}$ to zero, the set $\Sigma_+\times\{0,\pi R_{10}\}$ of $A_1$
singularities gives rise to a single copy of $S^2\times S^1$ with an
O6 plane coincident with two D6 branes (and their mirror partners)
wrapped on it~\footnote{We again only consider the
component $\Sigma_+$ of $\Sigma$; the discussion for the second
one $\Sigma_-$ is independent and similar.}.
In type IIA orientifold language, this system generates
an $SO(4)\cong SU(2)^2$ gauge group with $b_1(S^2\times S^1)=1$ adjoint
field. Since this $S^2\times S^1$ lives in a one complex
dimensional moduli space of SLAG's, the two D6 branes can slide on any
other 3-cycle in this family. This describes the Coulomb
branches $SO(4)\to SU(2)\times U(1)$ or  $SO(4)\to U(1)^2$
corresponding to $k\neq 0$ in Eq. (\ref{spII}).

After two T-dualities, the local configuration is described by the
type IIA system of Figure 15: Two parallel NS branes orthogonal to an
O4 plane together with two D4 branes sitting at $x^4+ix^5=v_{1,2}$ and
their mirror images at $-v_{1,2}$ (see (\ref{0123}) for conventions).
\begin{figure}[!h]
\centerline{\hbox{\psfig{figure=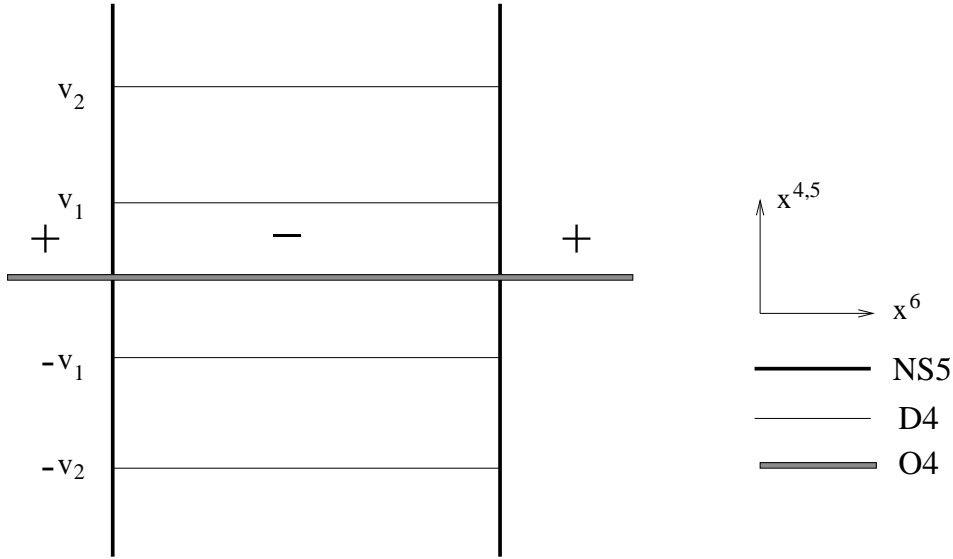,height=7.5cm}}}
\caption{{\bf Figure 15: }
{\footnotesize \em Brane realization of $\N=2$ $SO(4)$ SYM in
  the Coulomb branch.
}}
\label{f16}
\end{figure}
This describes an $\N=2$ $SO(4)$ vector multiplet in the Coulomb
branch of the theory. However, the usual classical gauge coupling
$\tau_{\mbox{\scriptsize cl}}$ is
promoted in our case to a full dynamical $\N=1$ chiral field whose VEV
is a flat direction. Geometrically, its scalar component is associated with
$\psi+\sqrt{\phi^2-1}$, the inverse gauge coupling,
complexified by the M-theory 3-form, the theta angle. In phase
II, $\psi+\sqrt{\phi^2-1}>0$ and quantum corrections should imply
that we remain in the Coulomb branches parametrized by $v_{1}$
($v_{2}$) for the first (second) $SU(2)$ factor \cite{Seiberg:1994rs}.

If $\psi$ decreases, when $\psi+\sqrt{\phi^2-1}=0$ in Figure 14, the
two NS branes are coincident and we are at infinite classical  gauge
coupling (see Figure 16(d)).
\begin{figure}[!h]
\centerline{\hbox{\psfig{figure=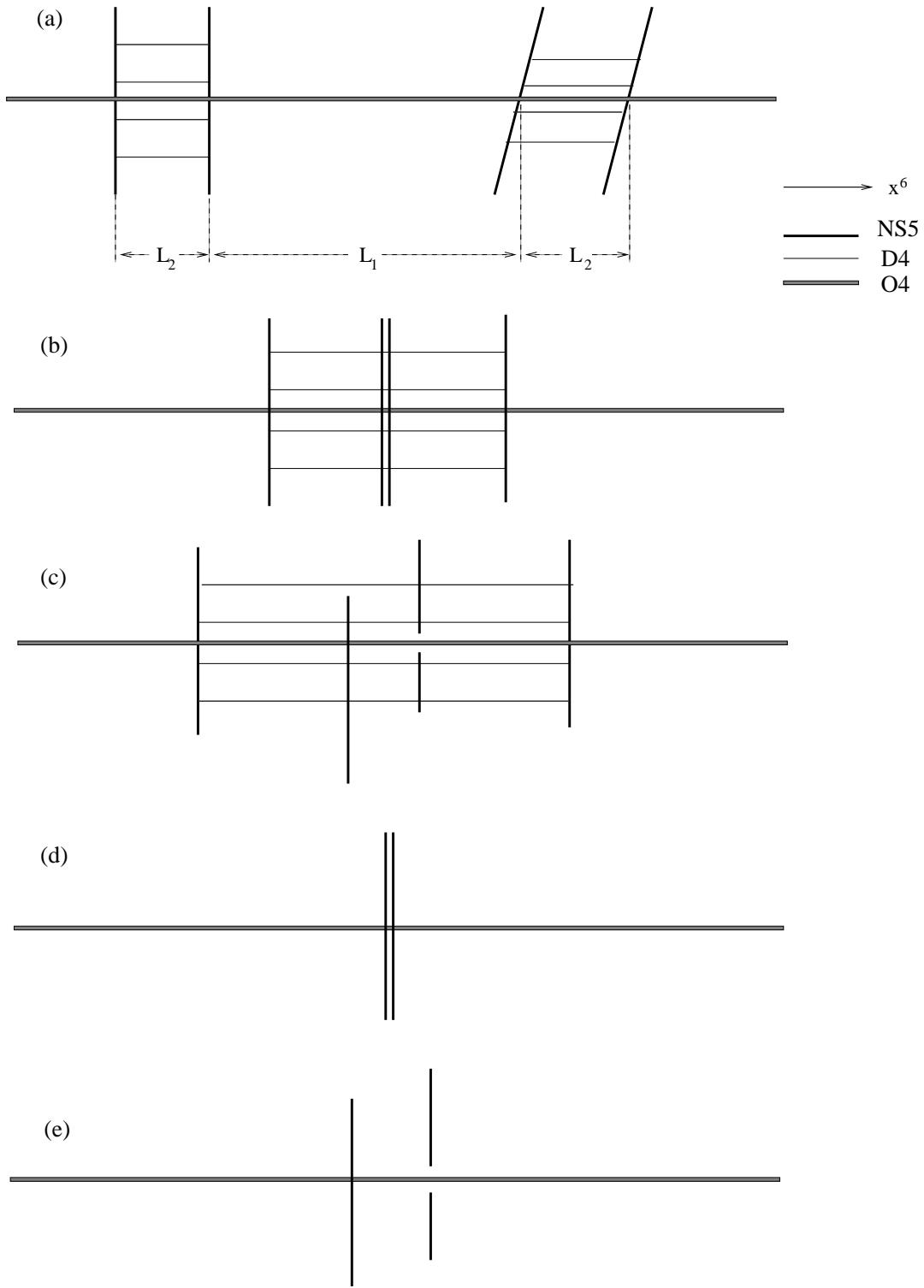,height=20cm}}}
\caption{{\bf Figure 16: }
{\footnotesize \em {\bf (a)} Two $SO(4)$ systems of branes with adjoint
  fields approach each other. {\bf (b)} The middle branes coincide. {\bf (c)}
  Then, they separate. {\bf (d)} The left over NS branes approach
  each other till they coincide. {\bf (e)} Finally, they separate.}}
\label{f17}
\end{figure}
For
$\psi+\sqrt{\phi^2-1}<0$ we have passed into phase I, where the gauge theory
and adjoint matter have disappeared. In the brane picture, this phase is
realized by separating the two NS branes in the directions
$x^{7,8,9}$, while keeping them parallel so that they are mirror to
each other with respect to the O4 plane,
as indicated on Figure 16(e). Thus, phases II and I are
described by parallel NS branes separated in different
directions. It amounts to sending a classical parameter to infinity, the gauge
coupling, and then passing into a different theory. Also, as we reviewed
in Section 2.2, systems of parallel NS branes are mapped in type
IIB geometry to ALE spaces such as Eq. (\ref{fvw2}), and can be
desingularized in only one way given in Eq. (\ref{fvw2b}).
We thus expect the transition II$\to$I to occur at infinite distance in moduli
space, not corresponding to a physical transition such as a Higgs
mechanism. In other words, when we are in phase II, quantum
effects forbid us to leave the Coulomb branches $v_{1,2}\neq 0$ and we
are an infinite distance away from the boundary domain wall
$\psi=-\sqrt{\phi^2-1}$ where the moduli space dimension jumps.
Notice that in \cite{Kaste:2001iq}, such an infinite distance away
transition was associated to an $S^2\times S^1$ flop transition, where
only $S^2$ vanishes.

Let us consider now the transition III$\to$II. In phase III of Figure 14,
  the brane picture is consisting of two sets of branes similar to
Figure 15 along a common O4 plane. The two systems of parallel
Neveu-Schwarz fivebranes are non-mutually parallel, as shown in
Figure 16(a). When $\psi$ approaches $\sqrt{\phi^2-1}$, the distance
$L_1$ between these systems decreases. Also, we expect their relative
angle to decrease, so that at the transition
$\psi=\sqrt{\phi^2-1}$ the two central NS branes are
parallel and collide. This has to be the case due to the fact that the
  geometrical transitions  $\psi\stackrel{>}{\to} \sqrt{\phi^2-1}$
and $\psi\stackrel{<}{\to}- \sqrt{\phi^2-1}$ are related into each other by
the map $(\psi, z_6)  \to (-\psi, i z_6)$ in the defining polynomial
Eq. (\ref{def4}). From the geometric point of view carracterized by
  the Neveu-Schwarz branes only, they are therefore identical.
At the origin of the Coulomb branch
(\ie $k=0$ in Eq. (\ref{spp})), the low energy theory is
an $\N =1$ $SO(4)\times SO(4)$ SYM with a massless chiral
multiplet in the adjoint and a massive $Q\in (4,4)$ chiral
multiplet~\footnote{The transition can also be considered when we sit
  in the Coulomb branch, as actually is described in Figure 16.}.
When $L_1=0$, the two central NS branes are coincident (see Figure
16(b)) and we pass into phase II by separating them in the directions
$x^{7,8,9}$ (see Figure 16(c)). As before, a brane transition that
consists of approaching, colliding and separating two parallel NS
branes is expected to be unphysical. Hence, the
line $\psi=\sqrt{\phi^2-1}$  should  be at infinite distance in
moduli space.

Finally, the transition III$\to$I in Figure 14 is similar to the
transition II$\to$I. To summarize, the model based on $\G_4$ should
give rise to three distinct components of moduli space. Therefore, we
have separated the different domains in  Figure 14 by continuous lines.

\subsection{3-cycles transitions with non-constant $b_1$}

In this Section, our aim is to describe other types of transitions
where the first Betti number of the involved 3-cycles change. As will
be precised at the end of this section, they should occur at finite
distance in moduli space.

We start again with a
two-parameter sub-family of manifolds $\C_5$ within $\CP^4_{11222}[8]$,
whose defining polynomial is
\begin{equation}
p_5 \equiv z_6^4(z_1^8 + z_2^8-2\phi z_1^4z_2^4)+(z_3^2-\psi z_6^2
z_2^4)^2+(z_4^2-\psi z_6^2 z_2^4)^2+z_5^4=0~,
\label{defp5}
\end{equation}
where $\phi$ and $\psi$ are complex. This family has members
where singular points occur~\footnote{\label{foo}The determinant of second
  derivatives at these singularities has one vanishing eigenvalue.}:
\begin{equation}
\begin{array}{lllll}
\mbox{for any $\psi$,}&  \mbox{at }\phi=+1 &~:~ &
(i^k,1,\pm \sqrt{\psi},\pm \sqrt{\psi},0,1)& ,(k=0,...,3)\;,\\
& \mbox{or }\phi=-1 &~:~ &
(i^k e^{i\pi/4},1,\pm \sqrt{\psi},\pm \sqrt{\psi},0,1)
&,(k=0,...,3)\;,\\
\mbox{and for any $\phi$,}& \mbox{at }\psi=\pm \sqrt{\phi^2-1} &~:~ &
(i^k\phi^{1/4},1,i^l (\phi^2-1)^{1/4},0,0,1)& ,(k,l=0,...,3), \\
& & & (i^k \phi^{1/4},1,0,i^l (\phi^2-1)^{1/4},0,1)
&,(k,l=0,...,3),\\
& \mbox{or }\psi=\pm \sqrt{\phi^2-1 \over 2} &~:~ &
(i^k\phi^{1/4},1,0,0,0,1)& ,(k=0,...,3)\; , \\
\label{nodes2}
\end{array}
\end{equation}
where the $+/-$ signs are independent.

We proceed by restricting $\phi$ and $\psi$ to real
values in order to consider the $G_2$ orbifold
\begin{equation}
\G_5={\C_5\times S^1 \over w\I}\; ,
\end{equation}
where the involution $w\I$ is  defined as in Eq. (\ref{s}).
The orbifold point set $\Sigma \times \{0,\pi R\}$ is characterized by
the special Lagrangian 3-cycle $\Sigma$ of $w$-invariant points in
$\C_5$ given by:
\begin{equation}
x_6^4(x_1^8+x_2^8-2\phi x_1^4x_2^4)+(x_3^2-\psi x_6^2
x_2^4)^2+(x_4^2-\psi x_6^2 x_2^4)^2+x_5^4=0 \; ,
\label{fp2}
\end{equation}
where the unknowns are real and $x_{2}$, $x_6$ can be scaled to 1, thanks
to Eq. (\ref{forbid}). Solving for $x_1^4$, one obtains
\begin{equation}
x_1^4 =\phi \pm\sqrt{\phi^2-[1+
  (x_3^2-\psi)^2+(x_4^2-\psi)^2+x_5^4]}\; ,
\label{def3'}
\end{equation}
which implies $\phi\geq 1$ for having solutions. In the variables
\begin{equation}
X_1= x_1^4-\phi \;\; , \quad X_j=x_j^2-\psi \;  \; (j=3,4)\; , \quad
X_5=x_5^2~\mbox{sign}(x_5)~,
\label{maps}
\end{equation}
we find again a 3-sphere
\begin{equation}
X_1^2+X_3^2+X_4^2+X_5^2=\phi^2-1\; .
\label{s3}
\end{equation}
However, as for $\C_1$,  $X_5$ as a function of $x_5$ is one-to-one,
while $X_1$ as a function of $x_1$ is two-to-one so that
$\Sigma$ has again two disconnected components $\Sigma_+,\Sigma_-$.

Finally, we have to find the solutions for the variables $x_{3,4}$. From
the inequalities  $-\sqrt{\phi^2-1}\leq
x_{3,4}^2-\psi\leq\sqrt{\phi^2-1}$, one has:
\vskip .2in

$\bullet$ For $\psi > \sqrt{\phi^2-1}$:
There exist two disjoint sets of solutions for $x_{j}$  $(j=3,4)$
\begin{equation}
\begin{array}{lll}
&&0<(\psi-\sqrt{\phi^2-1})^{1/2}
\leq x_{j}\leq (\psi+\sqrt{\phi^2-1})^{1/2}\; \\ &\mbox{ or }&
\phantom{0}-(\psi+\sqrt{\phi^2-1})^{1/2}
\leq x_{j}\leq -(\psi-\sqrt{\phi^2-1})^{1/2}<0\; ,
\end{array}
\label{xx'}
\end{equation}
so that
\begin{equation}
\mbox{for $\psi > \sqrt{\phi^2-1}$} \quad : \qquad
\Sigma_+ = \bigcup_{i=1}^4 S^3 \; ,
\end{equation}
where the $S^3$'s are disconnected (see Figure 17(a)).
\begin{figure}[!h]
\centerline{\hbox{\psfig{figure=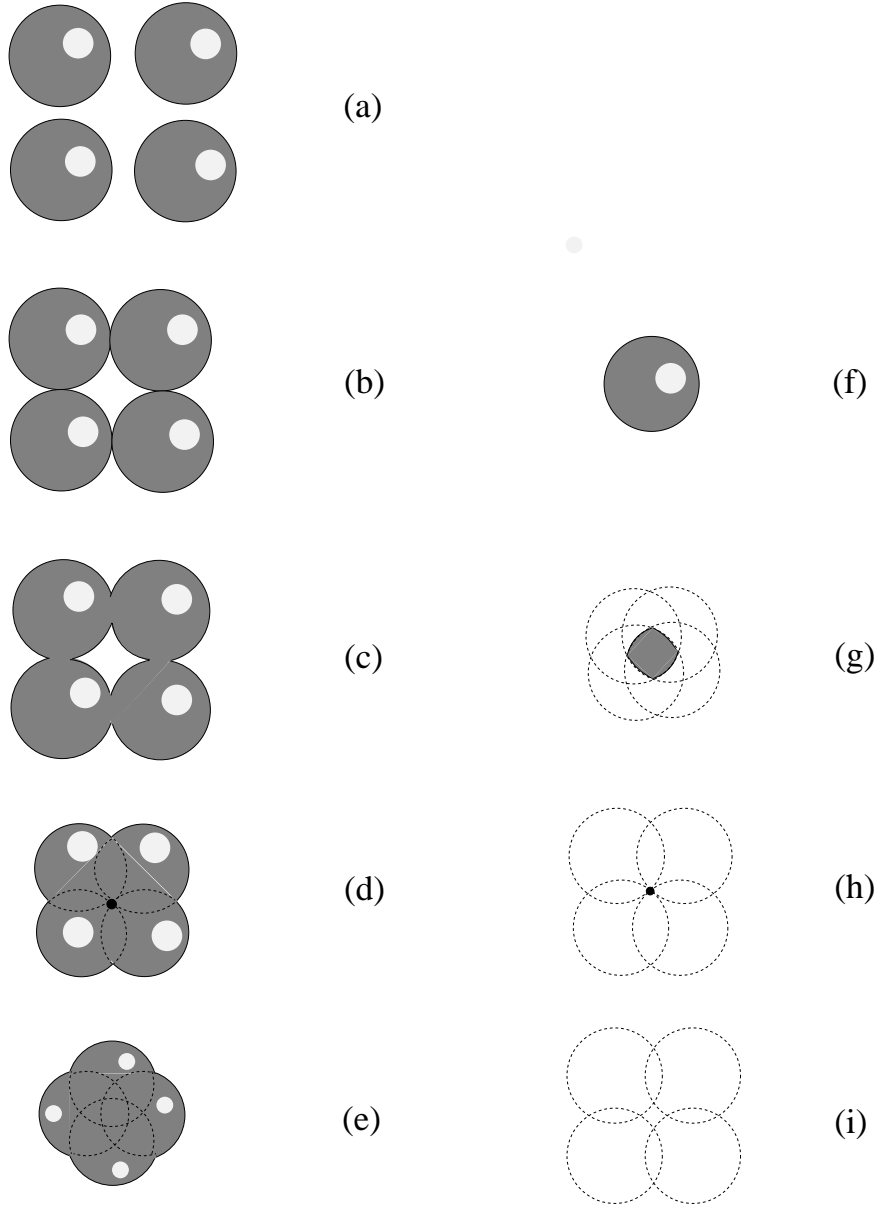,height=16cm}}}
\caption{{\bf Figure 17:}
{\footnotesize \em {\bf (a)}  $\Sigma_+$ is composed of four
  3-spheres that apprach each other. {\bf (b)}
  They intersect at four singular points of the manifold. {\bf (c)}
  They are connected so that topologically they are equivalent to a single
$S^2\times S^1$. {\bf (d)} the $S^1$ has shrunk to a singular point of
  the manifold. {\bf (e,f,g)} $\Sigma_+$ has become an $S^3$. {\bf
  (h)} This $S^3$ has shrunk to a singular point of the manifold. {\bf
  (i)} $\Sigma_+$ is empty.}}
\label{f7}
\end{figure}
\vskip .2in

$\bullet$ For $\psi = \sqrt{\phi^2-1}$: The four previous $S^3$'s intersect at
four points $(\phi^{1/4}, 1,\pm (\phi^2-1)^{1/4},0,0,1)$ and $(\phi^{1/4}, 1,0,
\pm (\phi^2-1)^{1/4},0,1)$ so that $\Sigma_+$ is connected with four singular
points (see Figure 17(b)). Note that the CY is also singular at these
points (see Eq. (\ref{nodes2})).
\vskip .2in

$\bullet$ For $\sqrt{\frac{\phi^2-1}{2}}<\psi < \sqrt{\phi^2-1}$:
The cyclic connected sum of the four 3-spheres has now become smooth. An
important remark is that the first Betti number of $\Sigma_+$ has
undergone a transition since we have now
$b_1(\Sigma_+)=1$. Topologically $\Sigma_+$ has become $S^2\times S^1$
(see Figure 17(c)):
\begin{equation}
\mbox{for $\sqrt{\frac{\phi^2-1}{2}}<\psi < \sqrt{\phi^2-1}$} \quad :
\qquad \Sigma_+ = \stackrel{4}{\mbox{\Large
    $\#$}}_{_{_{_{\mbox{\scriptsize \!\!\!\!\!\!\!\!\!$i{=}1$}}}}}
S^3\cong S^2\times S^1 \; .
\end{equation}
\vskip .2in

$\bullet$ For $\psi = \sqrt{\phi^2-1\over 2}$: The size of the $S^1$ vanishes
and $\Sigma_+$ can be recognized as a 3-sphere with its north and
south poles identified. The coordinates of this singular point are
$(\phi^{1/4}, 1, 0,0,0,1)$, where the CY is also singular (see Figure 17(d)).
\vskip .2in

$\bullet$ For $-\sqrt{\phi^2-1\over2}<\psi < \sqrt{\phi^2-1\over2}$:
The previously identified north  and south poles are now distinct.
As a result, we have now
\begin{equation}
\mbox{for $-\sqrt{\phi^2-1\over2}<\psi < \sqrt{\phi^2-1\over 2}$}
\quad : \qquad \Sigma_+ = S^3\; ,
\end{equation}
whose first Betti number is back to $b_1(\Sigma_+)=0$ (see Figures
17(e,f,g)).
\vskip .2in

$\bullet$ For $\psi =-\sqrt{\phi^2-1\over2}$: The size of the $S^3$ vanishes
and $\Sigma_+$ has collapsed to a point, so that
\begin{equation}
\mbox{for $\psi = -\sqrt{\phi^2-1\over 2}$} \quad : \qquad
\Sigma_+ = \{(\phi^{1/4}, 1, 0,0,0,1)\}\; ,
\end{equation}
where the CY is singular (see Figure 17(h)).
\vskip .2in

$\bullet$ Finally for $\psi <-\sqrt{\phi^2-1\over2}$: There is no solution for
$x_{3,4}$ and we have (see Figure 17(i))
\begin{equation}
\mbox{for $\psi < -\sqrt{\phi^2-1\over 2}$} \quad : \qquad
\Sigma_+ = \emptyset\; .
\end{equation}

 To summarize,  the topology of
the fixed point set $\Sigma_+$ takes the form, according to $\psi$ and
$\phi>1$,
\begin{equation}
\begin{array}{lll}
\psi>\sqrt{\phi^2-1},~~~ &
\Sigma_+=\bigcup_{i=1}^{4}S^3&\mbox{(disconnected sum)}\; , \\
\psi=\sqrt{\phi^2-1},~~~ &
\Sigma_+=\bigcup_{i=1}^{4}S^3&\mbox{(intersecting at four singular} \\
\sqrt{\phi^2-1\over 2}<\psi<\sqrt{\phi^2-1},~~~ &
\Sigma_+=\mbox{\large $\#$}_{i=1}^4
S^3\cong S^2\times S^1 &
\mbox{\phantom{----------------------------}points)}\; , \\
\psi=\sqrt{\phi^2-1\over 2},~~~ &
\Sigma_+=S^3 &\mbox{(north and south poles} \\
|\psi|<\sqrt{\phi^2-1\over 2},~~~ &
\Sigma_+=S^3 & \mbox{\phantom{------------------------}identified)}\; , \\
\psi=-\sqrt{\phi^2-1\over 2},~~~ &
\Sigma_+=\{(\phi^{1/4},1,0,0,0,1)\}&\mbox{(one singular point)}\; ,\\
\psi<-\sqrt{\phi^2-1 \over 2},~~~ & \Sigma_+= \emptyset \; .
\end{array} \label{ss22}
\end{equation}
Finally, we have represented in Figure 18 the associated phase
diagram. Let us determine now the massless spectrum.
\begin{figure}[!h]
\centerline{\hbox{\psfig{figure=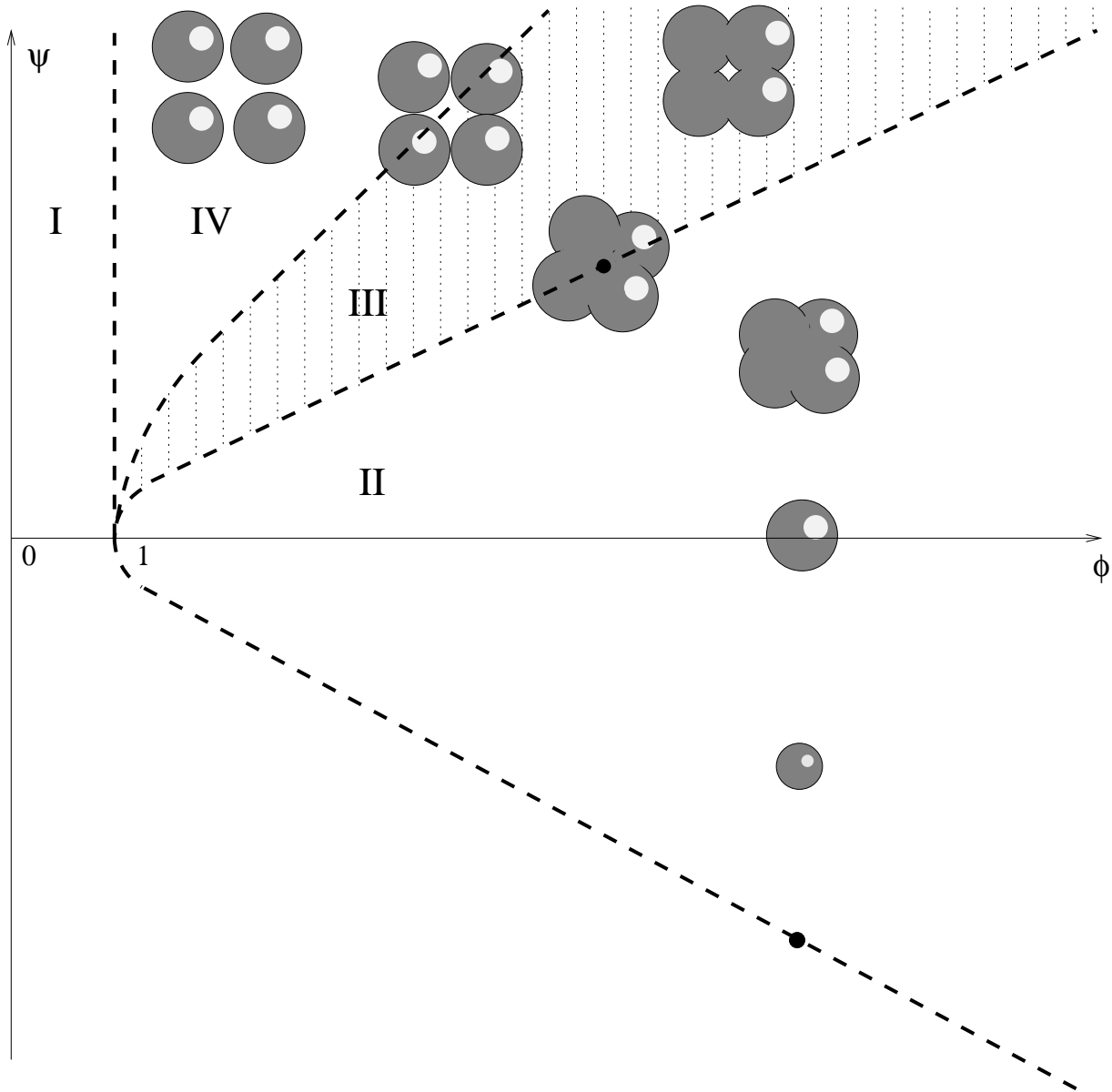,height=16cm}}}
\caption{{\bf Figure 18:}
{\footnotesize \em Phase diagram of the model based on
  $\G_5$. $\Sigma_+$ is represented for various values of
  $(\phi,\psi)$. It is composed of 3-spheres in phases II and IV. In
  Phase III, four 3-spheres are connected so that topologically
  $\Sigma_+=S^2\times S^1$. The transitions should occur at finite
  distance in moduli space. Hatches indicate the presence of a
  Coulomb branch parametrized by axes orthogonal to the figure.}}
\label{f18}
\end{figure}
\vspace{.3cm}

\noindent {\em \large -- Spectrum in phase I: $\phi<1$ or $\phi\geq 1,\,
\psi<-\sqrt{\phi^2-1\over 2}$ }
\vspace{.2cm}

\noindent
In this phase the orbifold is smooth and the spectrum is determined by
Eq. (\ref{b23}) where $h_{11}^+=0$. Thus, there are
\begin{equation}
\mbox{89 chiral multiplets } \qquad \mbox{and} \qquad
\mbox{no gauge group .}
\end{equation}

\vspace{.3cm}

\noindent {\em \large -- Spectrum in phase II: $\phi> 1,\,
|\psi|<\sqrt{\phi^2-1\over 2}$ }
\vspace{.2cm}

\noindent
This phase is similar to  phase II of the model based on $\G_1$.
In M-theory, there two copies of $\Sigma_+ \cup \Sigma_-=S^3\cup S^3$ of  $A_1$
 singularities, one at $x^{10}=0$ and the other at $x^{10}= \pi
R_{10}$. Hence, we have
\begin{equation}
\mbox{1 vector multiplet of }SU(2)^4 \qquad \mbox{and} \qquad
\mbox{89 neutral chiral multiplets .}
\end{equation}

\vspace{.3cm}

\noindent {\em \large -- Spectrum in phase III: $\phi> 1,\,
\sqrt{\phi^2-1\over 2}<\psi<\sqrt{\phi^2-1}$ }
\vspace{.2cm}

\noindent
As in phase II of the model based on $\G_4$, $\Sigma_+$ is composed of
a single connected component that gives rise to an $SU(2)$ gauge group
with one adjoint chiral field. In total, since we have four copies of
$S^2\times S^1$ in $\G_5$ and as we can go independently in the
Coulomb branch of each $SU(2)$ factor, the massless spectrum consists of
\begin{eqnarray}
&\mbox{an $SU(2)^{4-k}\times U(1)^k$ gauge group$ ~,~~~(k=0,...,4)\,
  ,$} &\nonumber  \\
&\mbox{with 1 adjoint chiral multiplet of $SU(2)^{4-k}$} & \label{spII'}\\
&\mbox{and $59+k$ neutral chiral multiplets .}&\nonumber
\end{eqnarray}

\vspace{.3cm}

\noindent {\em \large -- Spectrum in phase IV:  $\phi>1, \,
  \psi>\sqrt{\phi^2-1}$}
\vspace{.2cm}

\noindent
In this phase, $\Sigma_+$ is composed of four disconnected $S^3$'s so that
we have
\begin{equation}
\mbox{1 vector multiplet of }SU(2)^{16} \qquad \mbox{and} \qquad
\mbox{89 neutral chiral multiplets .}
\end{equation}
\vspace{.3cm}

\noindent {\em \large Partial interpretation}
\vspace{.2cm}

\noindent

As already said in footnote \ref{foo}, at each singularity of
Eq. (\ref{nodes2}), the determinant of second derivatives vanishes. In
fact, if one replaces $z_5^4$ in the defining polynomial $p_5$ in
Eq. (\ref{defp5}) by $(z_5^2-\psi_3z_6^2z_2^4)^2$ as in
Eq. (\ref{def333}), then the transitions become standard
conifold transitions. Explicitely, switching on $\psi_3$  has the
effect of separating the vanishing 3-spheres that are coincident when
$\psi_3=0$.  Thus, for $\psi_3\neq 0$ the transitions are
at finite distance in moduli space \cite{Candelas:1989di}, a fact that
should still
be valid in the case $\psi_3\equiv 0$ we studied in order to avoid irrelevant
complications in the discussion~{\footnote{As a particular case, if one
considers the transitions for $\psi_3\equiv \psi$, one finds that
there are in particular eight $S^3$'s centered at the corners of a
cube that are approaching each other before they are connected. As in our
case, the first Betti number of $\Sigma_+$ jumps to a non trivial
value $b_1>0$.}.

Actually, there is no new effect concerning the transitions II$\to$I
and IV$\to$I associated to confinement. As in Figure 9, they are
represented with dashed lines in Figure 18. To understand what is
happening at the transition IV$\to$III, it would be usefull to map
the M-theory geometry to a brane system of NS and D4
branes. Unfortunately, we have not been able to determine the dual
brane picture in details. However, for $\Sigma_+$ in phase IV, it
could involve four copies of the system of Figure 10, with a common O4
plane along a compact $x^6$ axis. The gauge group is then $SO(4)^4$ as
in M-theory. When $\psi$ approaches $\sqrt{\phi^2 -1}$, the NS brane of each
system approaches the NS' brane of the adjacent system till they
intersect. At this stage, there are four copies of intersecting orthogonal
Neveu-Schwarz fivebranes  along the compact $x^6$ direction. Then, each
pair of intersecting branes bend, as in Figure 4, and we are left with
two D4 branes and their mirrors along an O4 plane on $x^6$.
This bending causes the Higgsing of the gauge group
to a single $SO(4)$ due to non-perturbative effects.
However, in this brane picture, the
system is now locally $\N=4$, \ie with three massless chiral
multiplets in the adjoint of $SO(4)$, while the M-theory geometry
predicts only a local $\N=2$ system, \ie with one adjoint. Thus, there
must be some additional ingredients in the brane picture breaking
$\N=4 \to \N=2$. Also, these ingredients must be of first importance
for describing the transition III$\to$II, where only the massless adjoint
matter has disappeared. Clearly, such transitions where the first
Betti number of 3-cycles varies need to be further understood.

\vskip .5in

\noindent
{\bf Acknowledgments:}

\vskip .15in

\noindent

\noindent
We are grateful to C. Vafa for very usefull discussions
related to various points of this work. We also thank B. Acharya, C.
Angelantonj,  P. Mayr,
B. Pioline and
A. Uranga for discussions, and especially P. Kaste, who participated in
early stages of this paper.
We thank the CERN theory division where most of this
work has been done. H.P. would also like to thank the Physics
Department of NTU of Athens  for its kind hospitality.
The work of A.G. is supported in part by
the European RTN network HPRN-CT-2000-00122, the
BSF -- American-Israel Bi-National Science Foundation,
the Israel Academy of Sciences and Humanities --
Centers of Excellence Program, and
the German-Israel Bi-National Science Foundation.
The work of A.K. is partially supported by the
 $\Gamma\Gamma$ET grant E$\Lambda$/71, the ``Archimedes" NTUA
 programme and the
European RTN networks HPRN-CT-2000-00122, HPRN-CT-2000-00131. The work
of H.P. is supported in part by the PICS contract 
779 and the European RTN network
HPRN-CT-2000-00148.

\vskip .5in
\noindent
{\bf Note Added:}

\vskip .15in

\noindent

\noindent
As this article was being completed, we received the
paper \cite{Cachazo:2001sg}, which describes in detail
Seiberg duality from type IIB geometric set up.


\vskip .4in

\begin{thebibliography}{999}

\bibitem{gk}
A.~Giveon and D.~Kutasov,
``Brane dynamics and gauge theory,''
Rev.\ Mod.\ Phys.\  {\bf 71} (1999) 983
[arXiv:hep-th/9802067].

\bibitem{geometry}
W.~Lerche,
``Introduction to Seiberg-Witten theory and its stringy origin,''
Nucl.\ Phys.\ Proc.\ Suppl.\  {\bf 55B} (1997)  83
[Fortsch.\ Phys.\  {\bf 45} (1997) 293]
[arXiv:hep-th/9611190];
A.~Klemm,
``On the geometry behind N = 2 supersymmetric effective actions in four
dimensions,''
arXiv:hep-th/9705131;
P.~Mayr,
``Geometric construction of N = 2 gauge theories,''
Fortsch.\ Phys.\  {\bf 47} (1999) 39
[arXiv:hep-th/9807096].

\bibitem{Seiberg:1995pq}
N.~Seiberg,
``Electric - magnetic duality in supersymmetric nonAbelian gauge theories,''
Nucl.\ Phys.\ B {\bf 435} (1995) 129
[arXiv:hep-th/9411149].

\bibitem{egk}
S.~Elitzur, A.~Giveon and D.~Kutasov,
``Branes and N = 1 duality in string theory,''
Phys.\ Lett.\ B {\bf 400} (1997) 269
[arXiv:hep-th/9702014].

\bibitem{egkrs}
S.~Elitzur, A.~Giveon, D.~Kutasov, E.~Rabinovici and G.~Sarkissian,
``D-branes in the background of NS fivebranes,''
JHEP {\bf 0008} (2000) 046
[arXiv:hep-th/0005052].

\bibitem{Ooguri:1997ih}
H.~Ooguri and C.~Vafa,
``Geometry of N = 1 dualities in four dimensions,''
Nucl.\ Phys.\ B {\bf 500} (1997) 62
[arXiv:hep-th/9702180].

\bibitem{Sen:1997kz}
A.~Sen,
``A note on enhanced gauge symmetries in M- and string theory,''
JHEP {\bf 9709} (1997)  001 [arXiv:hep-th/9707123].

\bibitem{Ferrara:1998vf}
S.~Ferrara, A.~Kehagias, H.~Partouche and A.~Zaffaroni,
``Membranes and fivebranes with lower supersymmetry and their AdS
supergravity duals,''
Phys.\ Lett.\ B {\bf 431} (1998) 42
[arXiv:hep-th/9803109].

\bibitem{Townsend:1995kk}
P.~K.~Townsend,
``The eleven-dimensional supermembrane revisited,''
Phys.\ Lett.\ B {\bf 350} (1995) 184
[arXiv:hep-th/9501068].

\bibitem{Vafa:2000wi}
C.~Vafa,
``Superstrings and topological strings at large N,''
arXiv:hep-th/0008142.

\bibitem{Acharya:2000gb}
B.~S.~Acharya,
``On realising N = 1 super Yang-Mills in M-theory,''
arXiv:hep-th/0011089.

\bibitem{Atiyah:2000zz}
M.~Atiyah, J.~Maldacena and C.~Vafa,
``An M-theory flop as a large N duality,''
arXiv:hep-th/0011256.

\bibitem{Atiyah:2001qf}
M.~Atiyah and E.~Witten,
``M-theory dynamics on a manifold of $G_2$ holonomy,''
[arXiv:hep-th/0107177].

\bibitem{Dasgupta:2001um}
K.~Dasgupta, K.~Oh and R.~Tatar,
``Geometric transition, large N dualities and MQCD dynamics,''
Nucl.\ Phys.\ B {\bf 610} (2001) 331
[arXiv:hep-th/0105066].

\bibitem{Dasgupta:2001fg}
K.~Dasgupta, K.~Oh and R.~Tatar,
``Open/closed string dualities and Seiberg duality from geometric  transitions in M-theory,''
arXiv:hep-th/0106040.

\bibitem{Dasgupta:2001ac}
K.~Dasgupta, K.~h.~Oh, J.~Park and R.~Tatar,
``Geometric transition versus cascading solution,''
arXiv:hep-th/0110050.

\bibitem{Joyce2}
D. D. Joyce,
``Compact Riemannian 7-manifolds with $G_2$ holonomy, II,"
J. Diff. Geom. {\bf 43} (1996) 329.

\bibitem{Joyce:1999tz}
D.~Joyce,
``On counting special Lagrangian homology 3-spheres,''
arXiv:hep-th/9907013.

\bibitem{Kachru:2000vj}
S.~Kachru and J.~McGreevy,
``Supersymmetric three-cycles and (super)symmetry breaking,''
Phys.\ Rev.\ D {\bf 61} (2000) 026001
[arXiv:hep-th/9908135].

\bibitem{Kaste:2001iq}
P.~Kaste, A.~Kehagias and H.~Partouche,
``Phases of supersymmetric gauge theories from M-theory on $G_2$  manifolds,''
JHEP {\bf 0105} (2001) 058
[arXiv:hep-th/0104124].

\bibitem{Candelas:1990ug}
P.~Candelas, P.~S.~Green and T.~H{\"u}bsch,
``Rolling Among Calabi-Yau Vacua,''
Nucl.\ Phys.\ B {\bf 330} (1990) 49.

\bibitem{Partouche:2001uq}
H.~Partouche and B.~Pioline,
``Rolling among G(2) vacua,''
JHEP {\bf 0103} (2001) 005
[arXiv:hep-th/0011130].

\bibitem{Greene:1996dh}
B.~R.~Greene, D.~R.~Morrison and C.~Vafa,
``A geometric realization of confinement,''
Nucl.\ Phys.\ B {\bf 481} (1996) 513
[arXiv:hep-th/9608039].

\bibitem{ks}
I.~R.~Klebanov and M.~J.~Strassler,
``Supergravity and a confining gauge theory: Duality cascades and
$\chi$SB-resolution of naked singularities,''
JHEP {\bf 0008} (2000) 052
[arXiv:hep-th/0007191].

\bibitem{Gopakumar:1999ki}
R.~Gopakumar and C.~Vafa,
``On the gauge theory/geometry correspondence,''
Adv.\ Theor.\ Math.\ Phys.\  {\bf 3} (1999) 1415
[arXiv:hep-th/9811131].

\bibitem{Karch:1998yv}
A.~Karch, D.~Lust and D.~Smith,
``Equivalence of geometric engineering and Hanany-Witten via
fractional  branes,''
Nucl.\ Phys.\ B {\bf 533} (1998) 348
[arXiv:hep-th/9803232].

\bibitem{Sinha:2000ap}
S.~Sinha and C.~Vafa,
``SO and Sp Chern-Simons at large N,''
arXiv:hep-th/0012136.

\bibitem{Ooguri:2000bv}
H.~Ooguri and C.~Vafa,
``Knot invariants and topological strings,''
Nucl.\ Phys.\ B {\bf 577} (2000) 419
[arXiv:hep-th/9912123].

\bibitem{Acharya:2001hq}
B.~S.~Acharya,
``Confining strings from $G_2$-holonomy spacetimes,''
arXiv:hep-th/0101206.

\bibitem{Strominger:1996it}
A.~Strominger, S.~Yau and E.~Zaslow,
``Mirror symmetry is T-duality,''
Nucl.\ Phys.\ B {\bf 479} (1996) 243
[arXiv:hep-th/9606040].

\bibitem{Gomis:2001vk}
J.~Gomis,
``D-branes, holonomy and M-theory,''
Nucl.\ Phys.\ B {\bf 606} (2001) 3
[arXiv:hep-th/0103115].

\bibitem{Edelstein:2001pu}
J.~D.~Edelstein and C.~Nunez,
``D6 branes and M-theory geometrical transitions from gauged  supergravity,''
JHEP {\bf 0104} (2001) 028
[arXiv:hep-th/0103167].

\bibitem{Aganagic:2001nx}
M.~Aganagic, A.~Klemm and C.~Vafa,
``Disk instantons, mirror symmetry and the duality web,''
arXiv:hep-th/0105045.

\bibitem{Aganagic:2001jm}
M.~Aganagic and C.~Vafa,
``Mirror symmetry and a $G_2$ flop,''
arXiv:hep-th/0105225.

\bibitem{Brandhuber:2001yi}
A.~Brandhuber, J.~Gomis, S.~S.~Gubser and S.~Gukov,
``Gauge theory at large N and new G(2) holonomy metrics,''
Nucl.\ Phys.\ B {\bf 611} (2001) 179
[arXiv:hep-th/0106034].

\bibitem{Hernandez:2001bh}
R.~Hernandez,
``Branes wrapped on coassociative cycles,''
arXiv:hep-th/0106055.

\bibitem{Gomis:2001vg}
J.~Gomis and T.~Mateos,
``D6 branes wrapping Kaehler four-cycles,''
arXiv:hep-th/0108080.

\bibitem{Gibbons:1990er}
R. L. Bryant and S. M. Salamon, ``On the constrution of some complete metrics
with exceptional holonomy",  Duke Math. J. {\bf 58} (1989) 829;
G.~W.~Gibbons, D.~N.~Page and C.~N.~Pope,
``Einstein metrics on $S^3$,  $\R^3$ and $\R^4$ bundles,''
Commun.\ Math.\ Phys.\  {\bf 127} (1990) 529.

\bibitem{Kachru:2001je}
S.~Kachru and J.~McGreevy,
``M-theory on manifolds of $G_2$ holonomy and type IIA orientifolds,''
JHEP {\bf 0106} (2001) 027
[arXiv:hep-th/0103223].

\bibitem{Antoniadis:1996vb}
I.~Antoniadis, H.~Partouche and T.~R.~Taylor,
``Spontaneous breaking of N=2 global supersymmetry,''
Phys.\ Lett.\ B {\bf 372} (1996) 83
[arXiv:hep-th/9512006].

\bibitem{Partouche:1997yp}
H.~Partouche and B.~Pioline,
``Partial spontaneous breaking of global supersymmetry,''
Nucl.\ Phys.\ Proc.\ Suppl.\  {\bf 56B} (1997) 322
[arXiv:hep-th/9702115].

\bibitem{Taylor:2000ii}
T.~R.~Taylor and C.~Vafa,
``RR flux on Calabi-Yau and partial supersymmetry breaking,''
Phys.\ Lett.\ B {\bf 474} (2000) 130
[arXiv:hep-th/9912152].

\bibitem{Kiritsis:1997ca}
E.~Kiritsis and C.~Kounnas,
``Perturbative and non-perturbative partial supersymmetry breaking:  N
= 4 $\to$ N = 2 $\to$ N = 1,'' 
Nucl.\ Phys.\ B {\bf 503} (1997) 117
[arXiv:hep-th/9703059].

\bibitem{Mayr:2001hh}
P.~Mayr,
``On supersymmetry breaking in string theory and its realization in
brane  worlds,''
Nucl.\ Phys.\ B {\bf 593} (2001) 99
[arXiv:hep-th/0003198].

\bibitem{Papadopoulos:1995da}
G.~Papadopoulos and P.~K.~Townsend,
``Compactification of D = 11 supergravity on spaces of exceptional holonomy,''
Phys.\ Lett.\ B {\bf 357} (1995) 300  [arXiv:hep-th/9506150].

\bibitem{Giveon:1998sn}
A.~Giveon and O.~Pelc,
``M theory, type IIA string and 4D N = 1 SUSY SU(N(L)) x SU(N(R))
gauge  theory,''
Nucl.\ Phys.\ B {\bf 512} (1998) 103
[arXiv:hep-th/9708168].

\bibitem{Intriligator:1994uk}
K.~A.~Intriligator,
``'Integrating in' and exact superpotentials in 4-d,''
Phys.\ Lett.\ B {\bf 336} (1994) 409
[arXiv:hep-th/9407106].

\bibitem{Vafa:1996gm}
C.~Vafa and E.~Witten,
``Dual string pairs with N = 1 and N = 2 supersymmetry in four  dimensions,''
Nucl.\ Phys.\ Proc.\ Suppl.\  {\bf 46} (1996) 225
[arXiv:hep-th/9507050].

\bibitem{Hori:2000ck}
K.~Hori, A.~Iqbal and C.~Vafa,
``D-branes and mirror symmetry,''
arXiv:hep-th/0005247.

\bibitem{Aganagic:2000gs}
M.~Aganagic and C.~Vafa,
``Mirror symmetry, D-branes and counting holomorphic discs,''
arXiv:hep-th/0012041.

\bibitem{Mayr:2001xk}
P.~Mayr,
``N = 1 mirror symmetry and open/closed string duality,''
arXiv:hep-th/0108229.

\bibitem{Seiberg:1994rs}
N.~Seiberg and E.~Witten,
``Electric - magnetic duality, monopole condensation, and confinement
in N=2 supersymmetric Yang-Mills theory,''
Nucl.\ Phys.\ B {\bf 426} (1994) 19
[Erratum-ibid.\ B {\bf 430} (1994) 485]
[arXiv:hep-th/9407087].

\bibitem{Strominger:1995cz}
A.~Strominger,
``Massless black holes and conifolds in string theory,''
Nucl.\ Phys.\ B {\bf 451} (1995) 96
[arXiv:hep-th/9504090].

\bibitem{Greene:1995hu}
B.~R.~Greene, D.~R.~Morrison and A.~Strominger,
``Black hole condensation and the unification of string vacua,''
Nucl.\ Phys.\ B {\bf 451} (1995) 109
[arXiv:hep-th/9504145].

\bibitem{Candelas:1989di}
P.~Candelas, P.~S.~Green and T.~H{\"u}bsch,
``Finite distances between distinct Calabi-Yau vacua: (Other worlds
are just around the corner),''
Phys.\ Rev.\ Lett.\  {\bf 62} (1989) 1956.

\bibitem{Cachazo:2001sg}
F.~Cachazo, B.~Fiol, K.~A.~Intriligator, S.~Katz and C.~Vafa,
``A geometric unification of dualities,''
arXiv:hep-th/0110028.

\end{thebibliography}
\end{document}